\documentclass[aps,twocolumn,showpacs,showkeys,groupedaddress,nofootinbib]{revtex4}

\usepackage{graphicx}
\usepackage{bm}

\def\slashchar#1{\setbox0=\hbox{$#1$}
   \dimen0=\wd0 \setbox1=\hbox{/} \dimen1=\wd1
   \ifdim\dimen0>\dimen1 \rlap{\hbox to \dimen0{\hfil/\hfil}} #1
   \else  \rlap{\hbox to \dimen1{\hfil$#1$\hfil}} / \fi}

\def\w{\omega}

\newcommand{\tr}{{\textrm {tr}}}

\newcommand{\dlr}{\stackrel{\leftrightarrow}{\partial}}
\def\p{\slashchar{p}}
\makeatother
\input{epsf}
\begin{document}

\title{Chiral Solitons in the Spectral Quark Model}

\author{Enrique Ruiz Arriola}\email{earriola@ugr.es}
\affiliation{Departamento de F\'{\i}sica At\'omica, Molecular y Nuclear,  \\ 
Universidad de Granada, 
E-18071 Granada, Spain}
\author{Wojciech Broniowski} 
\email{Wojciech.Broniowski@ifj.edu.pl} 
\affiliation{The H. Niewodnicza\'nski Institute of Nuclear Physics,
PL-31342 Krak\'ow, Poland} 
\affiliation{Institute of Physics, \'Swi\c{e}tokrzyska Academy,
ul.~\'Swi\c{e}tokrzyska 15, PL-25406~Kielce, Poland} 
\author{Bojan Golli}\email{bojan.golli@ijs.si}
\affiliation{Faculty of Education,
              University of Ljubljana,
              1000 Ljubljana, Slovenia}
\affiliation{J.~Stefan Institute, 
              1000 Ljubljana, Slovenia}

\date{\today}

\begin{abstract}
Chiral solitons with baryon number one are investigated in the
Spectral Quark Model. In this model the quark propagator is a
superposition of complex-mass propagators weighted with a suitable
spectral function. This technique is a method of regularizing the 
effective quark theory in a way preserving many desired features
crucial in analysis of solitons.
We review the model in the vacuum sector,
stressing the feature of the absence of poles in the quark 
propagator. We also investigate in detail the analytic structure 
of meson two-point functions. We provide an appropriate prescription for 
constructing valence states in the spectral approach.
The valence state in the
baryonic soliton is identified with a saddle point of the Dirac
eigenvalue treated as a function of the spectral mass. Due to this feature the valence
quarks never become unbound nor dive into the negative spectrum, hence
providing stable solitons as absolute minima of the action. This is a
manifestation of the absence of poles in the quark propagator.
Self-consistent mean field hedgehog solutions
are found numerically and some of their properties are determined and compared 
to previous chiral soliton models. Our analysis
constitutes an involved example of a treatment of a relativistic complex-mass
system.
\end{abstract}



\maketitle 

\section{Introduction}

The original proposal by Skyrme of the early sixties foresaw that
baryons could be described as classical topological solitons of a
specific non-linear chirally symmetric Lagrangean in terms of meson
fields with non-trivial boundary
conditions~\cite{Skyrme:1961vq}. Within the accepted QCD framework the
large-$N_c$ analysis under the assumption of
confinement~\cite{'tHooft:1973jz,Witten:1979kh} supports many aspects
of this view. Moreover, the identification of the topological winding
number as the conserved baryon number of quarks from the occupied
Dirac sea in the background of large and spatially extended pion
fields~\cite{Goldstone:1981kk} suggested the underlying necessary
fermionic nature of the soliton at least for odd
$N_c$~\cite{Witten:1983tw}. The form of the Lagrangean remains
unspecified besides the requirement of chiral symmetry, leaving much
freedom on the assumed meson-field dynamics needed for practical
computations of low-energy baryonic
properties~\cite{Balachandran:1982cb,Adkins:1983ya}, usually organized
in terms of a finite number of bosons of increasing mass. One
appealing feature of the Skyrme model (for reviews see e.g.
\cite{Holzwarth:1985rb,Zahed:1986qz,Weigel:1995cz,Schechter:1999hg})
is that confinement appears to be explicitly incorporated, since the
baryon number topology of the meson fields cannot be changed with a
finite amount of energy and the soliton is absolutely stable against
decay into free quarks. This has also made possible the study of
excited baryons (see e.g. Ref.~\cite{Weigel:2004px} and references
therein).  However, when the partonic content of the baryon is
analyzed in deep inelastic scattering, in the Skyrme model the
Callan-Gross relation is violated, hence the partons turn out not to
be spin one-half objects~\cite{Chemtob:1987ut}. Therefore, making
credible non-perturbative estimates of high energy properties is out
of reach of the standard Skyrme model. This also suggests that the
relevant degrees of freedom should explicitly include constituent
quarks chirally coupled to
mesons~\cite{Birse:1983gm,Birse:1984js,Kahana:1984dx,Kaelbermann:1984xn,%
Broniowski:1984zd,Broniowski:1985kj,Diakonov:1987ty,Meissner:1988bg,Wakamatsu:1990ud}
motivating the use of chiral quark models with quarks and to search
for solitonic solutions to describe baryons. The approach is very much
in the spirit of the Skyrme model but with the important feature that
the partonic interpretation corresponds to spin one-half constituents,
despite the subtleties of the regularization~\cite{Weigel:1999pc}.
The considerable effort exerted to describe low-lying baryons as
solitons of effective chiral quark models has been described in detail
in the reviews
\cite{Birse:1991cx,Diakonov:1995zi,NJL:rev,Alkofer:rev,ripka:book}.

The chiral quark models that arise naturally in several approaches to
low-energy quark dynamics, such as the instanton-liquid model
\cite{Diakonov86} or the Schwinger-Dyson resummation of rainbow
diagrams \cite{Roberts94}, are {\em nonlocal}, {\em i.e.} the quark
mass function depends on the quark virtuality. For the derivations and
applications of these models see, {\em
e.g.},
\cite{Ball90,Ripka97,Cahill87,Holdom89,Krewald92,Ripka93,BowlerB,PlantB,%
coim99wb,bled99,Arrio,ArrSal,Plant00,Diak00,Gocke,Prasz,Dorokhov:2003kf}.
A major success was the finding of baryon hedgehog solitons in
nonlocal models \cite{Golli:1998rf,Broniowski:2001cx}, which turn out
to be stable also in the linear version of the model. Moreover, the
nonlocal models have nice features as compared to the local variants,
in particular they use a uniform regularization of both the normal and
abnormal parity processes, thus reproducing properly the anomalies in
the presence of regularization. In addition, the theory is made finite
to all orders of perturbation theory. The price to pay for the
non-locality is a complicated nature of interaction vertices with
currents, where the gauge invariance imposes non-local
contributions. Their presence leads to technical complications in the
treatment of non-local models.

A relevant issue for chiral quark solitons is related to the
confinement and stability of the mean field solution. The chiral quark
soliton model builds baryons as bound states of valence constituent
quarks in non-trivial meson fields. Indeed, such a model interpolates
between the non-relativistic quark model and the Skyrme model in the
limit of small solitons and large solitons,
respectively~\cite{Praszalowicz:1995vi}.  As a matter of fact, from
the view point of the chiral quark model, the apparent confinement in
the Skyrme model is reinterpreted as a strong binding effect, and
indeed deeply bound states are not sensitive to the confinement
properties of the interaction. Nevertheless, it is notorious that
depending on the details, the mean field soliton may in fact decay
into three free constituent quarks. Moreover, in local models for
small enough solitons the valence state becomes an unbound free quark
at rest. Actually, to our knowledge there is no satisfactory
calculation of excited baryon states in the chiral quark soliton model
precisely because of the lack of confinement~\cite{BC86}. These
features happen for realistic parameter values and despite the alleged
compatibility of the chiral quark model with the large-$N_c$ limit. As
already mentioned the soliton description of baryons in the 
large-$N_c$ limit is based on the assumption of confinement.


Let us define the scope of the present work.  After the many
years, the issue of color confinement remains a crucial and intriguing
subject for which no obvious solution exists yet. In particular, its
realization implies two relevant consequences. In the
first place, there exists a spectrum of color singlet excited
states. Secondly, quarks cannot become on shell and hence quark
propagators merely cannot have poles on the real axis. This restrained
meaning of confinement is often called {\it analytic confinement} in
the literature and in that sense is adopted in the present work. We
hasten to emphasize that although our work obviously {\it does not} suggest how
the problem of color confinement might be resolved, we manage to set
up a framework where the calculation of color singlet states such as
excited baryons becomes possible as a matter of principle within
a chiral quark soliton approach. 

In this paper we show how a recently proposed version of the
chiral quark model, the Spectral Quark Model
(SQM)~\cite{RuizArriola:2001rr,RuizArriola:2003bs,RuizArriola:2003wi,Megias:2004uj}
(see also \cite{Efimov:1993zg} for the original insights on the
spectral problem), not only allows for solitonic solutions but due to
its unconventional and indeed remarkable analytic properties yields a
valence eigenvalue which never becomes unbound. Thus the soliton is
absolutely stable.  The model is based on a generalized Lehmann
representation where the spectral function is generically complex,
involving a continuous superposition of complex masses. The subject of
defining a well founded quantum field theory of fixed complex masses
is an old story \cite{Lee:1969fy,Cutkosky:1969fq,Nakanishi:1972pt,%
Gribov:book,Kleefeld:2004jb}. Our approach differs from these works,
since our complex mass is an integration variable so that many
standard objections are sidestepped by choosing an appropriate
integration contour.  In practical terms the spectral function acts as
a finite regulator fulfilling suitable spectral conditions but with
many desirable properties, in particular the simple implementation of
chiral symmetry and gauge invariance. One of its outstanding features
is the uniform treatment of normal and abnormal parity processes
ensuring both finiteness of the action and a simultaneous
implementation of the correctly normalized Wess-Zumino-Witten
term\footnote{In the standard treatment of local models the somewhat
artificial and certainly asymmetric prescription of regularizing the
real part of the Euclidean action and not regularizing the imaginary
part has been used.}. This uniform treatment of regularization has
some impact on soliton calculations even in the SU(2) case (where the
Wess-Zumino-Witten term vanishes) since the valence and sea
contributions to the soliton energy are indeed related to the abnormal
and normal parity separation of the effective action,
respectively. The basics of SQM are described in detail in
\cite{RuizArriola:2003bs,Megias:2004uj} and the reader is referred
there for the description of the method and numerous applications to
the pion phenomenology. The SQM approach bares similarities to
nonlocal models, however the construction of interaction vertices with
currents is very simple in SQM, as opposed to almost prohibitive
complications of the nonlocal approach.

The purpose of this paper is twofold: Firstly, we show that absolutely
stable baryon solitons in SQM exist and discuss their
properties. Secondly and more generally, we study an instance of an
involved complex-mass system, and show how to treat valence states.
Despite the complexity, the resulting prescription turns out to be very simple
and easy to implement in practical calculations. It amounts to locating the 
saddles of the valence eigenvalue, $\epsilon_0$,  
as a function of the complex mass $\w$, based on 
the condition 
\begin{equation}
\frac{d\epsilon_0(\w)}{d \w}=0.
\end{equation}

The outline of our paper is as follows: In 
Sect.~\ref{sec:cons-sqm} we provide an operational justification
of the need of some uniform regularization for the full action without
an explicit separation between normal and abnormal parity
contributions. Certain
{\em a priori} field-theoretic 
consistency conditions are discussed. SQM is shown to
fulfill the {\em one-body} consistency condition in contrast to previous
local versions of chiral quark models. We also discuss in some detail
the analytic structure of the meson correlators, showing that despite
analytically-confined quarks they possess cuts at large values of
$q^2$, as requested by (asymptotic) quark unitarity. However, the
meromorphic structure expected from general large $N_c$
considerations is violated. This point is analyzed in the light of a
large-$N_c$ Regge model. Next, we pass to the analysis of solitons.
First, we discuss the connection between the baryon and topological
currents in the limit of large solitons, which holds in the presence
of the spectral regularization. Following the standard field-theoretic
approach the construction of the baryon state is pursued in
Sect.~\ref{sec:formal}. The unconventional appearance of
complex-masses requires a demanding mathematical treatment of both the
valence and sea contributions.  Nevertheless, ready-to-use formulas
for the total soliton mass are derived and analyzed for several
soliton profiles showing the existence of chiral solitons with baryon
number one. An alternative derivation is provided in
Appendix~\ref{sec:spec-first} based on computing the spectral integral
exactly. Dealing with complex mass Dirac operators both for bound
states and continuum states is involved and some of the features may
be studied in a somewhat comprehensive toy model in
Appendix~\ref{sec:toy}. The techniques are close to the more familiar
complex potentials in non-relativistic quantum mechanics which are
reviewed for completeness in Appendix~\ref{qm}.  In
Appendix~\ref{exlin} we show that a linear extension of SQM leads to
instability of the vacuum, hence SQM can only be constructed in the
originally proposed nonlinear realization. In Sect.~\ref{sec:self} we
look for self-consistent hedgehog solutions and determine their
properties both in the chiral limit as well as for finite pion
masses. Finally, in Sect.~\ref{sec:concl} we come to the conclusions.

\section{The Spectral Quark Model and consistency relations}
\label{sec:cons-sqm} 

We begin with some basic expressions of the general field-theoretic
treatment of chiral quark models, which are the groundwork for our
treatment of solitons and the method of including valence states in
SQM discussed in Sect.~\ref{sec:formal}. In this section we highlight an important
consistency condition which was not fulfilled in the hitherto
extensively used chiral quark soliton models based on local
interactions. Remarkably, this condition happens to be satisfied in
SQM. The analytic properties of the quark propagator and in
particular the lack of poles are reviewed.  We also analyze some
important aspects of the meson sector, and in particular the
analytic structure of two-point correlators in the complex
$q^2$-plane. Finally, we show how the correct topological current
arises in the presence of regularization.

\subsection{A consistency condition \label{sect:consist}}

The vacuum-to-vacuum transition amplitude in the presence of {\em
external} bosonic $(s,p,v,a)$ and fermionic $(\eta, \bar\eta) $ fields
of a chiral quark model Lagrangian can be written in the path-integral
form as
\begin{eqnarray} 
\!\!\!\!\!\! Z[j,\eta,\bar\eta] &=& \langle 0 | {\rm T} \exp \Bigl\{ i \int
d^4 x \Big[\bar q  j q + \bar \eta q + \bar q \eta \Big] \Big\} |0
\rangle, \label{eq:Z} 
\end{eqnarray} 
where the compact notation 
\begin{eqnarray} 
j &=& {\slashchar v}+{\slashchar a} \gamma_5-(s+i\gamma_5 p)  
\end{eqnarray}
has been introduced. The symbols $s$, $p$, $v_\mu$, and $a_\mu $
denote the external scalar, pseudoscalar, vector, and axial flavor
sources, respectively, given in terms of the generators of the flavor
$SU(3)$ group,
\begin{eqnarray} 
s = \sum_{a=0}^{N_F^2-1} s_a \frac{\lambda_a}2, \qquad \dots  
\end{eqnarray}   
with $\lambda_a$ representing the Gell-Mann matrices. Any physical matrix
element can be computed by functional differentiation with respect to
the external sources. 

Let us consider the calculation of a bilinear quark operator, such as,
{\em e.g.}, the quark condensate (for a single flavor) $\langle \bar q
q \rangle $. We can think of two possible ways of making such a
computation, namely via coupling of an external scalar source $s(x)$
(a mass term) and differentiating with respect to $s(x)$, or by
calculating the second functional derivative with respect to the
Grassmann external sources $\eta(x)$ and $\bar \eta(x)$ taken at the same
spatial point. The consistency of the calculation requires the
following trivial identity for the generating functional:
\begin{eqnarray} 
\langle \bar q(x) q(x) \rangle &=& i \frac1{Z}\frac{\delta Z}{\delta
s(x)} \Big|_0 \nonumber \\ = 
\lim_{x' \to x} \langle \bar q(x') q(x) \rangle &=& 
\lim_{x' \to x} (-i)^2 \frac1{Z} \frac{\delta^2
Z}{\delta \eta(x) \bar \eta (x')} \Big|_0,
\label{eq:cons-qq}
\end{eqnarray} 
where $|_0$ means all external sources set to zero. This requirement
can be generalized to any quark bilinear with any bosonic quantum
numbers and thus we call it the {\em one-body consistency
condition}. In
the traditional treatment of local chiral quark models the l.h.s. of
the above formula corresponds to a closed quark line and is divergent,
calling for regularization, whereas for $x' \neq x $ the r.h.s. is
finite and corresponds to an open quark line, thus no regularization
is demanded. This poses a consistency problem which actually becomes
crucial in the analysis of high-energy processes and introduces an
ambiguity in the partonic interpretation as well as conflicts with
gauge invariance and energy-momentum conservation (see, {\em e.g.},
Ref.~\cite{Ru02} for a further discussion on these subtle but relevant
issues). Obviously, the previous argument could equally be applied to
sources with any quantum numbers suggesting that any {\em consistent}
regularization should be applied to the full action. As already
mentioned in the Introduction, in the traditional approach to local
models the treatment of singularities requires first to separate the
normal and abnormal parity contributions and to regularize only the
normal parity piece. In the Euclidean space this separation corresponds
to the real and imaginary parts of the action.

In the next section we show that SQM fulfills the one-body consistency
condition. More generally, one might want to extend the condition
(\ref{eq:cons-qq}) to any number of quark bilinears contracted to
bosonic quantum numbers, namely the $N-$body consistency condition
\begin{eqnarray}
&& \langle \bar q(x_1) \Gamma_1 q(x_1)
\cdots \bar q(x_N) \Gamma_1 q(x_N) \rangle \nonumber \\ &=& 
\lim_{x'_i \to x_i } \langle \bar q(x_1') \Gamma_1 q(x_1) \cdots \bar
q(x_N') \Gamma_1 q(x_N) \rangle 
\label{eq:N-body} 
\end{eqnarray} 
where the l.h.s. is evaluated after functional derivatives with
respect to bosonic sources and the r.h.s. with respect to fermionic
sources respectively. Here $\Gamma_i$ are general spin-flavor 
matrices. We have found that this is not possible in the SQM scheme (see the discussion in Sect.~\ref{sec:def-SQM} below).

\subsection{The Spectral Quark Model and the quark propagator}
\label{sec:def-SQM}

In SQM the regularization is imposed already at the level of one open 
line through the use of a generalized Lehmann representation for the
full quark propagator in the absence of external sources,
\begin{eqnarray}
S ( \slashchar{p} ) = \int_C d \w \frac{\rho(\w)}{ \slashchar{p}-\w}= 
\p A(p^2 ) + B(p^2) 
\label{eq:Sp}
\end{eqnarray}
where $C$ is a suitable contour in the complex
mass-plane~\cite{RuizArriola:2003wi} (see also Fig.~1 below). Chiral and 
electromagnetic gauge invariance can be taken care of through the use of the gauge technique
of Delbourgo and West~\cite{Delbourgo:1977jc,Delbourgo:1978bu}, which
provide particular solutions to chiral Ward-Takahashi identities, or
through the use the standard effective action approach, applied in
this paper.  As a result, one can ``open the quark line'' from one
closed loop and compute high-energy processes with a partonic
interpretation, such as, {\em e.g.}, the structure function and the
light-cone wave function of the pion, {\em
etc.}~\cite{RuizArriola:2003wi}, or the photon and $\rho$-meson light
cone wave functions \cite{Dorokhov:2006qm}.  The generalization of the
(one body) consistency condition for scalar sources,
Eq.~(\ref{eq:cons-qq}), to all possible bosonic quantum numbers can be
achieved by taking the generating functional to be
\begin{eqnarray}
Z[\eta, \bar \eta, s , p, \dots ] = \int DU e^{ - i \langle \bar \eta ,
S[U,s,p,v,a] \eta \rangle } e^{i \Gamma [U,s,p,v,a]},
\label{eq:pathint}
\end{eqnarray} 
where the quark propagator and the effective action are given by 
\begin{eqnarray} 
\langle x' | S[U,s,p,v,a]_{a a'} | x \rangle  = \int_C d \w \rho(\w) 
\langle x| ( {\bf D}  )^{-1}_{a a'}|x'\rangle ,
\label{eq:eff_prop} 
\end{eqnarray} 
and 
\begin{eqnarray}
\Gamma[U,s,p,v,a] =-i N_c \int_C d \omega \rho(\omega) {\rm Tr} \log
\left( i {\bf D} \right),
\label{eq:eff_ac} 
\end{eqnarray} 
respectively. The Dirac operator has the form
\begin{eqnarray}
i {\bf D} &=& i\slashchar{\partial} - \omega U^5 - {\hat m_0} + \left(
\slashchar{v} + \slashchar{a} \gamma_5 - s - i \gamma^5 p \right).
\nonumber 
\label{eq:dirac_op} 
\end{eqnarray} 
For a bilocal (Dirac- and flavor-matrix valued) operator $A(x,x')$ we
use the notation
\begin{eqnarray}
{\rm Tr} A = \int d^4 x \,{\rm tr} \langle A(x,x) \rangle,
\end{eqnarray} 
with ${\rm tr}$ denoting the Dirac trace and $\langle \, \rangle $
the flavor trace.  The matrix 
\begin{eqnarray}
U^5 &=& e^{ i \gamma_5 \sqrt{2}\Phi /f} \nonumber \\ &=&  \frac12 ( 1 +
\gamma_5 ) U + \frac12 ( 1 - \gamma_5 ) U^\dagger \, ,
\end{eqnarray}
while $U = u^2 = e^{ { i} \sqrt{2}\Phi / f }$ is the flavor matrix
representing the pseudoscalar octet of mesons in the nonlinear
representation,
\begin{eqnarray} 
        \Phi = \left( \matrix{ \frac{1}{\sqrt{2}} \pi^0 +
        \frac{1}{\sqrt{6}} \eta & \pi^+ & K^+  \cr  \pi^- & -
        \frac{1}{\sqrt{2}} \pi^0 + \frac{1}{\sqrt{6}} \eta & K^0  \cr 
        K^- & \bar{K}^0 & - \frac{2}{\sqrt{6}} \eta }
        \right) .
\end{eqnarray}
The matrix ${\hat m}_0={\rm diag}(m_u, m_d, m_s)$ is the current quark
mass matrix and $f=86~{\rm MeV}$ denotes the pion weak-decay constant
in the chiral limit, ensuring the proper normalization condition of
the pseudoscalar fields. For a two-flavor model it is enough to
consider \mbox{$\Phi=\vec{\tau}\cdot \vec{\pi}/\sqrt{2}$}.

In Eq.~(\ref{eq:pathint}) the Dirac operator appears both in the fermion determinant
as well as in the quark propagator and could, in principle, be treated
independently. The one-body consistency condition is fulfilled {\it
precisely} because we use the same spectral function $\rho(\w) $ for
both.  We have refrained on purpose from writing a Lagrangean in terms
of quarks explicitly since anyhow chiral quark models are defined in
conjunction with the regularization and the approximation used. In our
case we work in the leading order of the large-$N_c$ expansion, which
amounts to a saddle point approximation in the bosonic $U$-fields and
use the spectral regularization which is most explicitly displayed in
terms of the generating functional presented above.  The new ingredient of SQM
compared to earlier chiral quark models is the presence of the quark
spectral function $\rho(\omega)$ in
Eq.~(\ref{eq:eff_prop},\ref{eq:eff_ac}) and the integration over
$\omega$ along a suitably chosen contour $C$ in the complex-$\omega$
plane.  Our approach extends the early model of Efimov and Ivanov
\cite{Efimov:1993zg} by including the gauge invariance, the chiral
symmetry, and the vector meson dominance, as well as applying the
model to both low- and high-energy processes.

The above SQM construction implements the one-body consistency
condition, as follows
\begin{eqnarray} 
i \frac{\delta Z}{\delta j_\alpha (x)} \Big|_{\eta=\bar
\eta=0} = \lim_{x' \to x} (-i)^2  \frac{\delta}{\delta
\eta(x)} \Gamma_\alpha \frac{\delta}{ \bar \eta (x')} Z
\Big|_{\eta=\bar \eta=0},  \nonumber \\ 
\label{eq:consistency}
\end{eqnarray} 
which is more general than Eq.~(\ref{eq:cons-qq}) since it is valid in
the presence of non-vanishing external bosonic sources with any
quantum numbers and non-trivial background pion field, $U$. This
represents a clear improvement on the previous {\it local} chiral
quark models where this requirement was violated.
Eq.~(\ref{eq:consistency}) has direct applicability on the soliton
sector as we will discuss in Sect.~\ref{sec:formal}. It should be
mentioned, however, that similarly to many other chiral quark models,
the two-body and higher consistency conditions, Eq.~(\ref{eq:N-body})
are not satisfied in SQM. For instance, for the two body consistency
condition one gets  
\begin{eqnarray} 
&& \langle \bar q(x_1) \Gamma_1 q (x_1) \bar q(x_2) \Gamma_2 q(x_2)
\rangle = \frac1Z \int DU \int d \w \rho (\w) \nonumber \\ &\times& 
\tr \left[ \Gamma_1 
\langle x_1 | {\bf D}^{-1} (\w) | x_1 \rangle \right] \tr \left[ \Gamma_2 \langle x_2 | {\bf
D}^{-1} (\w) | x_2 \rangle \right] \, , \nonumber \\
\end{eqnarray} 
while 
\begin{eqnarray} 
&& \lim_{x_i'\to x_i} \langle \bar q(x_1) \Gamma_1 q (x_1') \bar
q(x_2) \Gamma_2 q(x_2') \rangle = \lim_{x_i'\to x_i}
\frac1Z \int DU \nonumber \\ &\times& \int d\w_1 \rho (\w_1) \tr \left[ \Gamma_1 \langle x_1' |
{\bf D}^{-1} (\w_1) | x_1 \rangle \right] \nonumber \\ &\times& \int d\w_2
\rho (\w_2) \tr \left[ \Gamma_2 \langle x_1' | {\bf D}^{-1} (\w_2) | x_1
\rangle \right] \nonumber \\ &\neq& \langle \bar q(x_1) \Gamma_1 q (x_1) \bar
q(x_2) \Gamma_2 q(x_2) \rangle \, . 
\end{eqnarray}
Likewise, we do not know if this one body
consistency condition is fulfilled beyond the leading large-$N_c$
approximation. At present the only known way to fulfill all
consistency conditions is by returning to nonlocal versions of the
chiral quark model.  Therefore, we must provide a prescription of how
higher functional derivatives should be handled in case were more than
a single quark line could be opened. Eq.~(\ref{eq:consistency})
suggests to use {\it always} the method based on the bosonic sources
for operators involving any number of quark bilinear and local
operators. For operators involving one single bilocal and bilinear
operator $q(x) \bar q (x') $ we may use the method based on fermionic
sources, since Eq.~(\ref{eq:consistency}) guarantees the consistency
in the coincidence limit $x' \to x$.  Of course, this prescription
does not yield a unique result when more then one bilocal and bilinear
quark operator is involved or equivalently when more than one quark line
is opened.

Using the standard variational differentiation for the generating
functional the Feynman rules for SQM follow from the
action~(\ref{eq:eff_ac}). They have the form of the usual Feynman
rules for a local theory, amended with the spectral integration
associated to each quark line, according to
Eq.~(\ref{eq:eff_ac}). This resembles very much the well known
Pauli-Villars regularization method (however with a continuous
superposition of complex masses) and allows for very efficient
computations, see
\cite{RuizArriola:2001rr,RuizArriola:2003bs,RuizArriola:2003wi,Megias:2004uj}.

\begin{figure}[tb]
\begin{center}
\includegraphics[width=7cm]{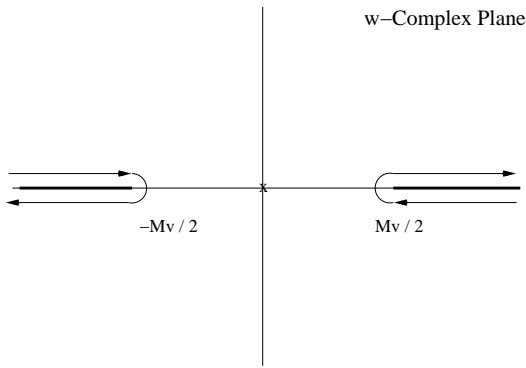}
\end{center}
\caption{The contour $C$ in the complex-$\omega$ 
plane for the calculations in the vacuum sector in
the meson dominance variant of SQM. $M_V$ denotes the $\rho$-meson
mass. The two segments shown in the figure are connected at $\pm$ infinity
with semi-circles, not displayed.
\label{contour0}}
\end{figure}

The basic paper~\cite{RuizArriola:2003bs} explains the general
construction and the conditions for the moments of the spectral
function $\rho(\omega)$ coming from physics constraints. In
particular, normalization requires
\begin{eqnarray}
\int_C d\omega \rho(\omega)=1, 
\label{normrho}
\end{eqnarray}
while observables are related to the log-moments and inverse moments
of $\rho(\omega)$~\cite{RuizArriola:2003bs}.  The full spectral
function consists of two parts of different parity under the change
$\omega \to -\omega$, {\em i.e.} $\rho(\omega)=
\rho_V(\omega)+\rho_S(\omega)$, with the scalar part, $\rho_S$, even
and the vector part, $\rho_V$, odd.  A particular implementation of
SQM is the {\em meson dominance model}, where one requests that the
large-$N_c$ pion electromagnetic form factor has the monopole form of
the vector-meson dominance (VMD), 
\begin{eqnarray}
F_\pi(q^2)=\frac{M_V^2}{M_V^2-q^2},
\label{eq:pion-formfac}
\end{eqnarray}
where $M_V$ denotes the $\rho$-meson mass.\footnote{This example shows
in a transparent way a peculiar feature of the model. In the standard
constant mass case the pion form factor, a three point function, due
to Cutkosky's rules, displays a cut in the form factor for
sufficiently large energy due to a superposition of poles in the quark
propagator. In SQM the mechanism is just the opposite; the cuts
conspire to build a pole in the form factor.} The matching of the
model predictions to this form yields the rather unusual spectral
function
\begin{eqnarray}
\rho_V (\omega) &=& \frac{1}{2\pi i} \frac{1}{\omega}
\frac{1}{(1-4\omega^2/M_V^2)^{5/2}},  \label{rhov}
\end{eqnarray}
with the pole at the origin and cuts starting at $\pm M_V/2$, where
$M_V$ is the mass of the vector meson. Similar considerations for the
photon light-cone wave function~\cite{Dorokhov:2006qm} and matching to
VMD yield for the scalar part an analogous form,
\begin{eqnarray}
\rho_S (\omega) &=& -
\frac{1}{2\pi i} \frac{48 \pi^2 \langle \bar q q \rangle } {N_c M_S^4
(1-4\omega^2/M_S^2)^{5/2}}. \label{rhos}
\end{eqnarray} 
where $\langle \bar q q \rangle $ is the single flavor quark
condensate, and $M_S=M_V$ \cite{Megias:2004uj}. The contour $C$ for
the $\omega$ integration to be used with formulas
(\ref{rhov},\ref{rhos}) is shown in Fig.~\ref{contour0} for $\rho_V$ 
(for $\rho_S$ one has the same contour with $M_V \to M_S$). This contour
is applicable in the vacuum ({\em i.e.}, no baryon number) sector of
the model. The extension to baryons is described in the next
section. From Eq.~(\ref{rhov}) and (\ref{rhos}) one gets the quark
propagator functions from Eq.~(\ref{eq:Sp}),
\begin{eqnarray}
A(p^2) &=& \int_C d\w \frac{\rho_V (\w)}{p^2-\w^2} \nonumber \\ &=&  
\frac1{p^2} \left[ 1 - \frac1{(1-4p^2/M_V^2)^{5/2}} \right],
\nonumber \\ 
B(p^2) &=& \int_C d\w \frac{\w \rho_S (\w)}{p^2-\w^2} \nonumber \\ &=&    
\frac{48 \pi^2 \langle \bar q q\rangle }
{M_S^4 N_c (1-4p^2/M_S^2)^{5/2}} \label{AB}.
\end{eqnarray} 
These functions do not possess poles (the alleged pole at $p^2=0$ in
$A(p^2)$ is canceled), but only cuts starting at $p^2=M_V^2/4$,
reflecting the structure of $\rho(\omega)$. Actually, the
$\w$-integral can be evaluated using the integral transformations
displayed in the first lines of Eqs.~(\ref{AB}). For instance, taking
the limit $p^2 \to \infty $ one gets the moments 
\begin{eqnarray}
\delta_{k,0} &=& \int_C d\w \, \rho_V (\w) \w^{2k} \qquad , \,
k=0,1,2, \dots \\ 0 &=& \int_C d\w \, \rho_S (\w) \w^{2k+1} \qquad ,
\, k=0,1,2, \dots
\label{eq:moments}
\end{eqnarray} 
which imply in particular the vanishing of all positive moments for the 
spectral function $\rho(\w)$. 

The pion weak-decay constant
in the chiral limit comes out to be~\cite{RuizArriola:2003bs}
\begin{eqnarray}
f^2 = \frac{ N_c M_V^2}{24 \pi^2}, 
\label{eq:fpi-MV}
\end{eqnarray} 
a relation which works well phenomenologically. This relation will be
used as an identity in the rest of the paper. Compared to the standard
field-theoretic case, each quark line is supplemented with a spectral
integration $\int_C d\omega \rho(\omega)$. This makes calculations
very straightforward and practical for numerous hadronic processes
involving mesons and photons.  We stress that despite a rather
``exotic'' appearance of the quark spectral function, SQM leads to
proper phenomenology for the soft pion, including the Gasser-Leutwyler
coefficients \cite{Megias:2004uj}, the soft matrix elements for hard
processes, such as the distribution amplitude, transition form factor,
or structure functions \cite{RuizArriola:2003bs}, the generalized
forward parton distribution of the pion \cite{Broniowski:2003rp}, the
photon distribution amplitude and light-cone wave
functions~\cite{Dorokhov:2006qm}, or the pion-photon 
transition distribution amplitude~\cite{TDA}. In addition, the
calculations are straightforward, leading to simple analytic
results. Interestingly, the quark propagator corresponding to
(\ref{rhov},\ref{rhos}) has no poles, only cuts, in the momentum
space. The evaluated mass function of the quark, $M(Q^2)$, displays a
typical dependence on the virtuality $Q^2$ in the Euclidean region and
at the qualitative and quantitative level compares favorably to the
non-local quark models and to the lattice calculations
\cite{Bowman:2002bm}. As a sample calculation, we extend in the next subsections the previous
considerations to the calculation of vacuum properties and two-point correlators.

\subsection{The vacuum sector\label{sec:vac}}

As mentioned above the functional (one-body) consistency conditions
guarantee the unambiguous calculation of observables obtained from
quark fields, either as double Grassmann functional derivatives or as
single bosonic ones. These identities are generically formally
satisfied but become tricky under regularization.  Here we use the
first possibility for the quark condensate and the vacuum energy
density. The second method based on the effective action and yielding
identical results is outlined in Appendix~\ref{sec:eff2}.

The (single flavor) quark condensate $\langle \bar q q \rangle $ can
directly be computed from the quark propagator
\begin{eqnarray}
N_f \langle \bar q q \rangle &=& -i N_c \int \frac{d^4 p}{(2\pi)^4}
{\rm Tr} S (\p) \nonumber \\ &=& 4 N_c N_f \frac1{i} \int \frac{d^4 p}{(2\pi)^4} B(p^2),
\end{eqnarray} 
which through the use of Eq.~(\ref{AB}) becomes an identity. The vacuum energy
density can be computed by using the trace of the energy momentum
tensor, yielding
\begin{eqnarray} 
\langle \theta^{\mu \nu} \rangle &=& \langle \bar q (x)
\frac{i}{4}\left[ \gamma^\nu  {\dlr}{}^\mu + \gamma^\mu {\dlr}{}^\nu 
\right] q(x) \rangle \nonumber \\ &=&
-i N_c \int \frac{d^4
p}{(2\pi)^4} {\rm Tr} \left[ \frac12 ( p^\mu \gamma^\nu + p^\nu \gamma^\mu ) S (\p) \right] \nonumber \\ &=& -4i N_c N_f \int \frac{d^4
p}{(2\pi)^4}  p^\mu p^\nu A (p^2)  \nonumber \\ &=&
\frac14 g^{\mu \nu} \epsilon  + \langle \theta^{\mu \nu} \rangle_0  ,
\end{eqnarray} 
where the subscript $0$ refers to the trivial vacuum ($A=1$ and
$B=0$).  The bag constant is defined by the difference $ \epsilon =
\frac14 \langle \theta^\mu_\mu \rangle - \frac14 \langle
\theta^\mu_\mu \rangle_0 $ and after computing the integral becomes
\begin{eqnarray} 
\epsilon &=& -\frac{M_V^4 N_f N_c }{192 \pi^2} . \label{enden}
\end{eqnarray} 
The formula implies that it costs energy to dig a hole in the vacuum
with a non-vanishing pion weak decay constant (see
Eq.~(\ref{eq:fpi-MV})), as one might expect for the chirally symmetric
broken phase.

\subsection{Two-point correlators in the meson sector} 

In SQM the quark propagator has no poles but cuts at $p^2=M_V^2/4$
preventing the occurrence of quarks on the mass shell. This complies
to the notion of analytic confinement, a necessary but certainly not
sufficient condition for color confinement. This non-standard behavior
already suggests a possible departure from the standard
treatment. However, to justify the claim of confinement one should
check, in addition, the absence of cuts in physical correlators. Note
that this question is in principle unrelated to the analytic
confinement of bound quarks in a soliton with baryon number one which
will be discussed in further sections below, but it is still
interesting to see the analytic properties of the model. We analyze
this issue below in some detail for the two-point mesonic correlators
defined as 
\begin{eqnarray}
 \Pi_{AB} (q) = i \int d^4 x e^{-i q \cdot x}
\langle 0 | T \left\{ J_A (x) J_B  (0) \right\} | 0 \rangle ,
\end{eqnarray}
where $J_A (x)$ and $J_B (x)$ are interpolating currents with the
relevant meson quantum numbers. In the standard path
integral approach (see Eq.~(\ref{eq:pathint})) the time-ordered
products of currents in the vacuum can be evaluated by suitable
functional derivatives of the generating functional (\ref{eq:eff_ac})
with respect to external bosonic currents, and in the large-$N_c$
limit the path integral is driven to the saddle point of the path
integral in the presence of those currents. On the other hand, at
large $N_c$ any two-point mesonic correlation function should have the general
structure~\cite{'tHooft:1973jz,Witten:1979kh}
\begin{eqnarray}
\Pi_J (q^2 ) = \sum_n \frac{f_{J,n}^2}{M_{J,n}^2-q^2}
\label{eq:largeNc}
\end{eqnarray}  
due to confinement. This is a very stringent test, since it implies in
particular a meromorphic analytic structure in the $q^2$ complex
plane. Amazingly, this rather simple requirement has never been
accomplished in chiral quark models, either local or
nonlocal\footnote{Instead, calculations at large values of $Q^2$ have
only been confronted to QCD sum rules (see, {\em e.g.},
Ref.~\cite{Dorokhov:2003kf} and references therein.)}. 

A straightforward consequence of the large-$N_c$ representation of the
two point correlator in Eq.~(\ref{eq:largeNc}) is {\it positivity},
$\Pi_J( q^2 ) > 0 $, in the Euclidean region $q^2=-Q^2< 0$ and {\it
quark unitarity} at large values of $q^2$.  This ensures that, for
instance, the inclusive $e^+ e^- \to {\it hadrons} $ total cross
section is proportional to the imaginary part of the polarization
operator in the vector channel.~\footnote{This asymptotic quark
unitarity holds for external currents and should not be confused with
the S-matrix hadron unitarity which is violated at any finite order of
the large $N_c$ expansion due to the $1/\sqrt{N_c} $ suppressed
behavior of the meson-meson couplings.}

In our previous work~\cite{RuizArriola:2003bs} we have evaluated the $VV$
and $AA$ correlators corresponding to the conserved vector and axial
currents
\begin{eqnarray}
J_V^{\mu,a} (x) &=& \bar q(x) \gamma^\mu {\lambda_a \over 2} q(x) ,\\
J_A^{\mu,a} (x) &=& \bar q(x) \gamma^\mu \gamma_5 {\lambda_a \over 2} q(x) \, 
\end{eqnarray} 
for a general spectral function $\rho(\w)$ using solutions to the
Ward-Takahashi identities based on the gauge technique. 
In this paper we use the
effective action (\ref{eq:eff_ac}) to evaluate the correlation functions. 
Details of the calculation are provided in Appendix~\ref{sec:eff2}. We find
\begin{eqnarray} 
\Pi_{VV}^{\mu a , \nu b } (q) = \frac12 
\delta_{ab} \left( g^{\mu\nu } q^2 -  q^\mu q^\nu  \right)
\Pi_V^T (q^2),
\end{eqnarray} 
with\footnote{We correct here a typo in our previous
work~\cite{RuizArriola:2003bs}, Eq.~(4.3). Our conventions here are 
$\Pi^T_V (q^2) = - 2 \Pi (q^2) /q^2 $.}
\begin{eqnarray}
&&\Pi_V^T (q^2 ) = \frac{2 N_c}{3 q^2} \int \rho(\omega) d\omega \times  \\
&&\;\; \left\{ 2 \omega^2 (I(q^2,\omega)-I(0,\omega)) - q^2 (\frac{1}{48 \pi^2}-I(q^2,\omega)) \right\} , \nonumber
\end{eqnarray} 
and 
\begin{eqnarray} 
\Pi_{AA}^{\mu a , \nu b } (q) = \frac12 \delta_{ab} \left(
g^{\mu\nu } q^2 - q^\mu q^\nu  \right) \Pi_A^T (q^2),
\end{eqnarray} 
with
\begin{eqnarray}
\Pi_A^T (q^2) = \Pi_V^T (q^2 ) + 4 N_c \int d\omega \omega^2 \rho(\omega) I (q^2,\omega).
\end{eqnarray}
In the particular case of the meson dominance model for the spectral
function, Eq.~(\ref{rhov}), up to an (infinite) constant one has a remarkably
simple result
\begin{eqnarray}
\Pi_V^T (q^2) &=& \frac{f_V^2}{M_V^2-q^2} - \frac{N_c}{24 \pi^2} \log\left( 1- \frac{q^2}{M_V^2}\right), \nonumber \\  
\Pi_A^T (q^2) &=& -\frac{f^2}{q^2} - \frac{N_c}{24 \pi^2} \log\left( 1- \frac{q^2}{M_V^2}\right), 
\label{eq:VV+AA} 
\end{eqnarray} 
where $f_V^2 = f^2$. The above formulas clearly display the $\rho$
meson pole in the vector channel and the pion pole in the axial
channel. Also, we see that $f_A^2 =0 $, {\em i.e.}, there is no axial
meson dominance. Our expressions fulfill the first Weinberg sum rule,
\begin{eqnarray} 
\lim_{q^2 \to 0 } q^2\left[ \Pi_V^T (q^2) - \Pi_A^T (q^2) \right]=f^2.
\end{eqnarray}  
Similarly as in other local quark models, the second Weinberg sum rule is not satisfied,
\begin{eqnarray} 
\lim_{q^2 \to -\infty } q^4\left[ \Pi_V^T (q^2) - \Pi_A^T (q^2)
\right]= -M_V^2 f^2 \neq 0,
\end{eqnarray}  
as noted in our previous work~\cite{RuizArriola:2003bs}.  Since the
second Weinberg sum rule is a high-energy feature of the theory, one
hopes that it is not crucial for the low-energy phenomena studied in
this paper and in particular for the soliton properties which probe
Euclidean momenta corresponding to a soliton size $\sim \sqrt{6}/M_V$
(see, {\em e.g.}, Eq.~(\ref{eq:TPA}) below). This assumption has
implicitly been made in state-of-the-art chiral quark soliton
models~\cite{Birse:1991cx,Diakonov:1995zi,NJL:rev,Alkofer:rev,ripka:book}.

The presence of the log pieces in the correlators guarantees the
fulfillment of quark unitarity since\footnote{We define the discontinuity as
$$ {\rm Disc} \Pi (q^2) = \Pi (q^2 + i 0^+) - \Pi (q^2 - i 0^+) = 2 i
{\rm Im} \Pi (q^2).$$}
\begin{eqnarray} 
{\rm Im} \Pi_V^T (q^2) = {\rm Im} \Pi_A^T (q^2) = 
\frac{N_c}{24 \pi} ,   \qquad q^2 > M_V^2 .  
\end{eqnarray} 
The appearance of these quark unitarity cuts can be inferred from the
general structure of the quark propagator (see Eq.~(\ref{AB})) for any
quark momentum $p^2 > M_V^2/ 4 $, in spite of the fact that there are
no poles in the quark propagator. A more detailed analysis is
presented in Appendix~\ref{sec:integrals}. Note that in QCD one
obtains these parton-like relations for $q^2 \to \infty $. We stress
that the coefficients of the log terms are precisely such as in the
one-loop QCD calculation, complying to the parton-hadron
duality. Moreover, despite the unconventional features of SQM
involving complex masses, {\em positivity} is preserved both for the
pole and for the cut contributions to the imaginary parts of the
considered correlators, as can be seen from Eq.~(\ref{eq:VV+AA}). This
is a virtue of the spectral model, not easily fulfilled in other
chiral quark models\footnote{For instance, the Pauli-Villars
regularization spoils positivity due to subtractions, while dispersion
relations are fulfilled. The proper-time regularization preserves
positivity but does not fulfill dispersion relations due to a plethora
of complex poles \cite{Broniowski:1995yq}.}. It is important to
realize that this quark unitarity relation is hidden in the
large-$N_c$ meromorphic representation of Eq.~(\ref{eq:largeNc}) at
large $q^2$ through the asymptotic density of $\bar q q $
states~\cite{Arriola:2006sv}. 

To see this we may compare to the
large-$N_c$ Regge models (see e.g. \cite{RuizArriola:2006gq} and
references therein), where the meson spectrum is chosen to be a tower
of infinitely many radially excited states with masses $M_{n,V}^2 =
M_V^2 + 2 \pi \sigma n $, where $\sigma $ is the string tension and
the residue is taken to be constant $f_{n,V}^2 = N_c \sigma / (12 \pi)
$ precisely to implement quark-hadron duality at large $q^2$. As we
see, SQM corresponds to keeping one pole in the vector channel and
approximating the higher states by a logarithm.
More explicitly, one has
\begin{eqnarray} 
&&\Pi_V (q^2 ) - \Pi_V(0)  \nonumber \\
&& =\; \frac{N_c \sigma}{12 \pi} \sum_{n=0}
\left[ \frac{1}{M_V^2 + 2 \pi \sigma n -q^2}- \frac{1}{M_V^2 + 2 \pi
\sigma n} \right] \nonumber \\ && = 
\frac{N_c}{24\pi^2} \left[
\Psi \left( \frac{M_V^2 -q^2}{2\pi\sigma} \right) - \Psi \left(
\frac{M_V^2}{2\pi\sigma} \right) \right],  \nonumber
\end{eqnarray} 
where the digamma function, $\Psi'(z)= \Gamma'(z)/\Gamma(z)$ has been
introduced.  SQM corresponds to representing the infinite Regge sum
\begin{eqnarray}
\Psi(1+z) - \Psi (1) = - \sum_{n=1}^\infty
\left[\frac{1}{n+z}-\frac1{n} \right],
\end{eqnarray}
where $-\Psi(1) = \gamma$ is the Euler-Mascheroni constant, by the
approximation of taking explicitly the first pole. The higher poles
lead to the asymptotic behavior
\begin{eqnarray}
 -1/(z+1)+1 + \log (1+z),
\end{eqnarray}
approximated by the cut in SQM.  The accuracy is better than 20 \% for
$0 < z < \infty$.  We also recall that the large-$N_c$ analyses
restricted to a finite number of
resonances~\cite{Ecker:1988te,Pich:2002xy} provide a meromorphic
structure but fail to give the large-$q^2$ parton-hadron duality conditions.

For completeness we also discuss the scalar and pseudoscalar channels,
where the currents are given by
\begin{eqnarray}
J_S^{a} (x) &=& \bar q(x) {\lambda_a \over 2} q(x) ,\\
J_P^{a} (x) &=& \bar q(x) i \gamma_5 {\lambda_a \over 2} q(x).
\end{eqnarray} 
No Ward-Takahashi identities may be written for these currents and
thus they are not directly amenable to the gauge technique. The
effective action approach used here is superior to the gauge technique
since it also allows to compute the two-point correlators not directly
related to conserved currents.  The result is given up to two
subtraction terms as follows (see Appendix~\ref{sec:eff2}). Defining 
\begin{eqnarray}
\Pi_{SS}^{ab} (q) &=& \frac12 \delta^{ab} \Pi_S (q) \, , \nonumber \\   
\Pi_{PP}^{ab} (q) &=& \frac12 \delta^{ab} \Pi_P (q) \,  
\end{eqnarray} 
one gets 
\begin{eqnarray}
\Pi_{S} (q^2) &=& \frac{ N_c}{8 \pi^2} q^2 \log\left( 1-
\frac{q^2}{M_V^2}\right), \nonumber \\ \Pi_{P} (q^2) &=& \frac{
N_c}{8 \pi^2} q^2 \log\left( 1- \frac{q^2}{M_V^2}\right) -\frac{2 f^2
M_V^2}{M_V^2-q^2} \nonumber \\ &+& \frac{2 \langle \bar q q
\rangle^2}{f^2} \frac1{q^2} \frac{M_S^4 (M_V^2 - q^2) }{M_V^2
(M_S^2-q^2)^2},
\label{eq:SS+PP}
\end{eqnarray} 
where the residue of the pion pole is $2 f^2 B_0^2$, with 
$B_0 =- \langle \bar q q \rangle / f^2 $. 
We see that $f_S=0$, {\em i.e.}, there is no excited pseudoscalar
meson dominance. Again, the emergence of the quark unitarity cuts is
manifest. Similarly to the vector and axial correlators, the first $SS-PP$
Weinberg-like sum rule is verified
\begin{eqnarray} 
\lim_{q^2 \to 0 } q^2 \left[ \Pi_{S} (q^2) - \Pi_{P} (q^2) \right]= - 2
B_0^2 f^2 ,
\end{eqnarray}  
while the second  $SS-PP$ sum rule is not,
\begin{eqnarray} 
\lim_{q^2 \to -\infty } q^2\left[ \Pi_{S} (q^2) - \Pi_{P} (q^2)
\right]= -2 M_V^2 f^2 \neq 0. 
\end{eqnarray}  
Our analysis of both the second Weinberg sum rules for $VV-AA$ and
$SS-PP$ correlators agrees with the observed mismatch of the 
low-energy coefficients $L_{8}$ and $L_{10}$ between
SQM~\cite{Megias:2004uj} and the large-$N_c$ evaluation in the single-resonance
approximation (SRA)~\cite{Ecker:1988te,Pich:2002xy} (where both sum
rules are enforced). In \cite{Megias:2004uj} it was also shown that
matching $L_3$ of both SQM and the large-$N_c$ SRA yields the identity
between the scalar and meson masses, $M_S=M_V$ (as we assume for the
rest of the paper). It is interesting that the same condition also follows
from the requirement of having a single pole in the $PP$ correlator,
Eq.~(\ref{eq:SS+PP}). Moreover, at small $q^2 $ one has
\begin{eqnarray}
\Pi_{V} (q) - \Pi_{A} (q) &=& \frac{f^2}{q^2}- 4 L_{10} + \dots \, , \\
\Pi_{S} (q) - \Pi_{P} (q) &=& -2 B_0^2 \left[ 
\frac{f^2}{q^2} + 16 L_8 + \dots \right] \, . 
\end{eqnarray} 
Using Eqs.~(\ref{eq:VV+AA}) and (\ref{eq:SS+PP}) and taking $M_S =
M_V$ yields 
\begin{eqnarray} 
L_8 &=& \frac{N_c}{384 \pi^2}- \frac{f^6 }{16 \langle \bar q q
\rangle^2} \, , \nonumber \\ L_{10} &=& - \frac{N_c}{92 \pi^2} \, , 
\end{eqnarray} 
in agreement with results from the derivative expansion carried out in
\cite{Megias:2004uj}. The large-$N_c$ SRA yields $L_{8}= 3 f^2 / (32
M_S^2) $ and $L_{10}= - 3 f^2 / ( 8 M_V^2)$ \cite{Ecker:1988te,Pich:2002xy}, while the large-$N_c$ Regge models
produce $L_{10} = - N_c / (96 \sqrt{3} \pi )$.%
\footnote{At the mean field level in a gradient expansion the $L_8$ 
  term corresponds to ${\cal O}(m^4)$ corrections to the soliton energy, while $L_{10}$ 
   couples to external axial and vector currents. So a slight mismatch in 
   these values should not influence strongly the mean field soliton 
   properties. We estimate the corresponding correction to the energy as 
$\Delta E_8 = \Delta L_8 4 m_\pi^4 \int d^3 x [ \cos(2\theta(r)) -1] \sim  -3 {\rm MeV}$, with 
$\Delta L_8 \sim 10^{-3}$ -- a negligible number.}

To summarize this section, SQM provides meson two-point correlators
which carry poles as well as cuts, while strictly speaking large $N_c$
requires only having poles. Thus, despite the quark propagator not
having poles, the meson correlations do not exhibit true
confinement. Although we cannot prove it in general for any
correlation function, all two-point correlators we have considered
obey positivity and analyticity, {\em i.e.} dispersion relations. This is
required by the (asymptotic) quark unitarity and causality, {\em i.e.} the current
commutators must vanish outside the causal cone, $[j(x),j(0)] =0 $ for $x^2
< 0$. The coefficients of the log terms are in agreement with the
parton-hadron duality. On the other hand, the second Weinberg sum rule
is not satisfied, as in other local chiral quark models. This calls
for caution, in particular when analyzing processes sensitive to
high-momenta.~\footnote{We remind that nonlocal models do indeed
fulfill this high energy constraint~\cite{coim99wb} and the violation
of the second Weinberg sum rule in local models is probably related to
the violation of the two body consistency conditions discussed above
(see Sects.~\ref{sect:consist} and Sect.~\ref{sec:def-SQM}). On the
other hand it is uncertain if nonlocal models do indeed satisfy
analyticity.}

\subsection{The topological current}

In a previous work~\cite{Megias:2004uj} it was shown how the
Wess-Zumino-Witten term arises for SQM in the {\em presence} of the
spectral regularization. As a sample calculation illustrating the
consistency condition of Sect.~\ref{sect:consist} we compute the
baryon current in the limit of spatially large backgrounds. This also
shows how the regularization effects cancel in the final result,
yielding the well known Goldstone-Wilczek
current~\cite{Goldstone:1981kk}.  Taking the appropriate functional
derivative in Eq.~(\ref{eq:Z}) we find
\begin{eqnarray} 
\!\!\!\!\!\!\!\!\! \langle \bar q(x) \gamma^\mu q(x) \rangle &=& i
\frac1{Z}\frac{\delta Z}{\delta v_\mu (x)} \Big|_0 \nonumber \\ &=&
\lim_{x' \to x} (-i)^2 \frac1{Z} \frac{\delta}{\delta \eta(x)}
\gamma^\mu \frac{\delta Z}{\delta  \bar \eta (x')} \Big|_0
\nonumber \\ &=& \int_C d \w \rho(\w) \tr \left[ \gamma^\mu \langle x
| \frac{-i}{ i\slashchar{\partial} - \omega U^5 } |x \rangle \right].
\label{eq:bc-sea}
\end{eqnarray} 
The first two lines display the consistency condition. The
derivative expansion of the Dirac operator can be neatly done with the
help of the identity
\begin{eqnarray}
 \langle x | \frac1{i \slashchar{\partial} - \omega U^5 } | x
\rangle = {\rm tr}\int \frac{d^4 k}{(2 \pi)^4 } \frac1{\slashchar{k} + i
\slashchar{\partial}  - \omega U^5 },  
\label{eq:bas1}
\end{eqnarray} 
where the differential operator acts to the right on the function
$f(x)=1$. This formula can be justified through the use of an
asymmetric version of the Wigner transformation presented in
Ref.~\cite{Salcedo:1994qy}.
Formal expansion in powers of
$\partial$ 
yields
\begin{eqnarray} 
 \langle x | \frac1{i \slashchar{\partial} - \omega U } | x
\rangle &=& \sum_{n=0}^\infty {\rm tr} \int \frac{d^4 k}{(2 \pi)^4 }
\left[\frac{1}{k^2-\omega^2} \right]^{n+1} \nonumber \\ &&
\!\!\!\!\!\!\!\!\!\!\!\!\left(\slashchar{k} + \omega U^{5\dagger} \right)
\left[ \left(-i \slashchar{\partial}  \right) \left(\slashchar{k} +
\omega U^{5\dagger} \right) \right]^n.
\label{eq:bas2}
\end{eqnarray} 
The leading non-vanishing term is 
\begin{eqnarray} 
\langle \bar q(x) \gamma^\mu q(x) \rangle &=& \int_C d \w \rho(\w)
\int \frac{d^4 k}{(2 \pi)^4} \times \\ & \times & \frac{\w^4}{[k^2-\w^2]^4} 
\tr \left [ \gamma^\mu U^{5\dagger} \left( i\slashchar{\partial} U^5 \right)^3 \right ] =\rho_0 B^\mu,\nonumber 
\end{eqnarray} 
where $B_\mu (x) $ is the topological Goldstone-Wilczek current, 
\begin{eqnarray}
B_\mu = \frac{1}{24 \pi^2} \epsilon^{\alpha \beta \mu \nu} \langle \left(
U^\dagger \partial_\alpha U \right) \left( U^\dagger \partial_\beta U
\right) \left( U^\dagger \partial_\nu U \right) \rangle,
\end{eqnarray} 
and 
\begin{eqnarray}
\rho_0 &=& (- 96 \pi^2 i)  \int_C d \w \rho(\w) \int \frac{d^4 k}{(2
\pi)^4} \frac{\w^4}{[k^2-\w^2]^4} \nonumber \\ 
&=& \int_C d \w \rho(\w) =1, 
\end{eqnarray} 
where in the first line we have computed the momentum integral 
and in the second line used the normalization condition~(\ref{normrho}). 
Thus the proper normalization of the spectral function is responsible for the 
correct topological properties in SQM, preservation of anomalies, {\em etc.}
In fact, we view the uniform treatment of the anomalous processes as one of the main 
advantages of SQM over other local chiral quark models.

\section{Building the baryon \label{sec:formal}} 

In this section we show how valence orbits are constructed in SQM.
The issue is far from trivial, as unlike the case of local models,
there is no obvious Fock space representation. Already the experience
of nonlocal models showed that the construction of valenceness is an
involved issue \cite{Golli:1998rf,Broniowski:2001cx}.  There a
bound-state pole of the propagator was found in the background of
hedgehog chiral fields, and this state was occupied with $N_c=3$
valence quarks. The full contribution to the baryon current (local and
nonlocal) yielded a correct (and quantized) baryon number of the
soliton. In the present case we face a different situation, with the
spectral density and inherent complex masses present.  Thus we start
our derivation from very basic field-theoretic foundations, arriving
in the end at a very simple prescription holding under plausible
assumptions concerning the analyticity properties of the valence
eigenvalue as a function of the complex mass.

\subsection{Interpolating baryon fields}

In the standard field-theoretic approach a baryon can be described in
terms of the corresponding correlation function
\begin{eqnarray} 
\Pi_B (x,x') = \langle 0 | T \Bigl\{ B(x) \bar B(x') \Bigl\} | 0
\rangle 
\label{eq:baryon-corr}
\end{eqnarray} 
with $B(x)$ denoting an interpolating baryonic operator in terms of
anti-commuting quark fields. We take the simplest combination
\begin{eqnarray} 
B(x) = {1\over N_c !} \epsilon^{\alpha_1, \dots, \alpha_{N_c}}
\Phi^{a_1,\dots, a_{N_c}} q_{\alpha_1 a_1}(x) \cdots q_{\alpha_{N_c}
a_{N_c}}(x) \, , \nonumber \\ 
\label{eq:baryo-field}
\end{eqnarray} 
where $(\alpha_1, \dots , \alpha_{N_c})$ are the color indices, $(a_1,
\dots, a_{N_c})$ the spinor-flavor indices, and 
$\Phi^{a_1,\dots,a_{N_c}}$ is the appropriate completely symmetric spinor-flavor
amplitude. Inserting a complete set of baryon eigenstates gives
\begin{eqnarray} 
&&\Pi_B (x,x') = \\ && \theta(t-t^\prime)\sum_n\langle
0|B(0)|B_n,\vec{k}\rangle \langle B_n,\vec{k}|\bar{B}(0)|0\rangle
e^{-i(x-x^\prime)k}  \nonumber \\ && + (-1)^{N_c}  \theta(t^\prime-t)
\sum_n\langle 0|\bar{B}(0)|\bar{B}_n,\vec{k}\rangle \langle
\bar{B}_n,\vec{k}|B(0)|0\rangle \nonumber \\ 
&& \qquad \times e^{+i(x-x^\prime)k}, \nonumber
\label{eq:spec_th}
\end{eqnarray} 
where $B_n$ ($\bar{B}_n$) are the baryon (antibaryon) states with
momentum $\vec{k}$. Next we 
take
the limit $t-t' = T \to - i\infty $. That way the lightest baryon at rest
is selected in the sum,
\begin{eqnarray} 
\Pi_B (x,x') = \langle 0|B(0)|B\rangle
\langle B|\bar{B}(0)|0\rangle e^{-iTM_B}.\label{barbar}
\end{eqnarray} 
In the large-$N_c$ limit, we first rewrite the time ordered product as
a path integral over the fermionic degrees of freedom with the weight 
${\rm exp}(iS_{\rm SQM})$. The resulting expression, in turn, can be
obtained by the appropriate functional differentiation of the generating
functional $Z[s,p,v,a,\eta,\bar\eta] $ with respect to the external
quark fields $\eta(x)$ and $\bar\eta(x)$, yielding
\begin{eqnarray} 
\Pi_B (x,x') &=& \Phi^{a_1,\dots, a_{N_c}} \bar \Phi^{a_1',\dots,
a_{N_c}'} \nonumber \\ &\times& 
\int DU  e^{i \Gamma[U]} \prod_{i=1}^{N_c} iS_{a_i a'_i}(x,x') .
\label{eq:sol-corr}
\end{eqnarray} 
Here the one-particle Green function in SQM is given by
\begin{eqnarray} 
S_{a a'} (x,x') = \int_{C'} d \w \rho(\w) \langle x|
\left(i\slashchar{\partial} - \omega U^5\right)^{-1}_{a a'}|x'\rangle
\label{eq:eff_prop_0} 
\end{eqnarray} 
and the one loop effective action has the form 
\begin{eqnarray}
\Gamma[U] =-i N_c \int_{C'} d \omega \rho(\omega) {\rm Tr} \log
\left( i\slashchar{\partial} - \omega U^5  \right).
\label{eq:eff_ac_0} 
\end{eqnarray} 
Note the presence of the spectral integration, however, the contour
$C'$ to be used in the baryon sector is yet to be determined. 
There is no a priori reason why it should be equal to the vacuum 
contour $C$\footnote{Note by analogy that in the standard many-body quantum field
theory the contour in the complex {\em energy} variable selects the
orbits to be occupied and is clearly different for states with
different baryon number, or baryon density, where it crosses the real
axis at the value of the chemical potential.}.  The limit $N_c \to
\infty$ drives the functional integral over the bosonic $U$ fields
into a saddle point. In the vacuum sector one then obtains the
mean-field vacuum. In the baryon sector, due to the presence of $N_c$
factors $S_{a,a^\prime}$ in the numerator, the dominating saddle point
configuration is of course different from that of the vacuum and
depends non-trivially on $x$ and $x^\prime$.  The limit of a large
evolution time $T$ selects the minimum-energy stationary
configuration. For stationary configurations the Dirac operator can be
written as
\begin{eqnarray} 
i\slashchar{\partial} - \omega U^5  = \gamma_0 (i \partial_t -H(\w) ),
\end{eqnarray} 
with the Dirac Hamiltonian equal to
\begin{eqnarray} 
H(\w) = -i \alpha \cdot \nabla + \w U^5.
\label{eq:Hdirac}
\end{eqnarray} 
Note that the spectral mass $\omega$ appearing here is {\em complex}
and the situation is unconventional, since $H(\w)$ is {\em not
Hermitean}. In such a case one must distinguish between the right and
left eigenvectors, $ H \psi^R_n = \epsilon_n \psi^R_n $ and $
H^\dagger \psi^L_n = \epsilon_n \psi^L_n $, not related by the
Hermitean conjugation, {\em i.e.}  $(\psi^R_n )^\dagger \neq \psi^L_n$. 
The orthogonality relation is $ \langle \psi^L_n , \psi^R_m \rangle = \delta_{nm}$ 
and completeness is given in terms of the
left-right identity $ \sum | \psi^R_n \rangle \langle \psi^L_n |= 1$. 
We will not use this fancy notation, but will implicitly understand
that $\psi$ is the right eigenvector, while $\psi^\dagger $ is in fact
the complex-conjugated left eigenvector.
The corresponding eigenvalue problem is then
\begin{eqnarray}
H(\w) \psi_{n}(\vec x; \w) = \epsilon_n(\w) \psi_{n}(\vec x; \w).
\label{eq:direq}   
\end{eqnarray}

For our particular Hamiltonian one has the following useful properties
\begin{eqnarray}
H(\w)^\dagger &=& H (\w^*), \nonumber \\ \gamma_5 H(\w) \gamma_5^{-1}
&=& H (-\w), \nonumber \\ \gamma_0 H(\w) \gamma_0^{-1} &=& -H
(-\w^*)^\dagger, \nonumber \\ ( \gamma_0 \gamma_5 ) H(\w) (\gamma_0
\gamma_5 )^{-1} &=& -H (\w) , \nonumber \\ 
{\rm tr} H(\w) &=& 0  
. \label{Hp}
\end{eqnarray} 
where the trace is in the Dirac sense. Some properties of the
eigenvalues deduced from the properties (\ref{Hp}) are
\begin{eqnarray}
\epsilon_n (\w)^* &=& \epsilon_n (\w^*), \label{eq:reflexion} \\ 
\epsilon_n (\w )  &=& \epsilon_n (-\w). \nonumber   
\end{eqnarray}
One can now use the spectral representation of the propagator 
\begin{eqnarray} 
&&i S_{a a'} ( \vec x , t ; \vec x' , t' ) = \int_{C'} d\w \rho(\w) \label{Sspec}
\\ &\times&
\int_\gamma \frac{d \nu}{2\pi}  \sum_n \frac{ e^{i \nu (t-t') } }{\nu -
\epsilon_n (\w) } \psi_{n a}(\vec x; \w) \bar
\psi_{na'}(\vec x';\w) \nonumber 
\end{eqnarray} 
with $\psi_{n a}(\vec x; \w)$ ($\bar \psi_{n a}(\vec x; \w)$) and $\epsilon_n
(\w) $ denoting the right (left conjugated) eigenfunctions and
eigenvalues of $H(\w)$ evaluated at the stationary bosonic configuration.
The contour for the energy integration, discussed in the following, is denoted by $\gamma$.
For the calculation of the baryon state note that the propagator given
by Eq.~(\ref{Sspec}) and the fermion determinant given by
Eq.~(\ref{eq:eff_ac_0}) have to be evaluated at large Euclidean times.

\subsection{Hedgehog ansatz}

The stationary solutions of chiral quark models have the familiar 
hedgehog form,
\begin{eqnarray}
U^{5}(\vec x)=\exp(i \gamma_5 {\vec \tau} \cdot {\vec \theta}(\vec x)) ,
\label{U5}
\end{eqnarray}
with ${\vec \theta}$ denoting the chiral phase field. In the hedgehog ansatz
\begin{eqnarray}
{\vec \theta}=\hat{ r} \theta(r), \label{hedgehog} 
\end{eqnarray}
with $\theta(r)$ being a radial function. 
In the following two subsections we analyze the Dirac problem (\ref{eq:direq})
for the exponential profile as an example,
\begin{eqnarray}
\theta (r) = \pi e^{-r/R},  
\label{eq:exp-prof}
\end{eqnarray} 
In particular our figures are obtained for that case. Other popular 
cases include the linear profile
\begin{eqnarray}
\theta (r) = \pi (1-r/R) \theta(R-r)
\label{eq:lin-prof}
\end{eqnarray}
and the ${\rm arctan}$ profile \cite{Diakonov:1987ty}
\begin{eqnarray}
\theta (r) = 2 {\rm arctan}(R/r)^2. 
\label{eq:atan-prof}
\end{eqnarray} 
The parameter $R$ is a generic size scale of the 
soliton
proportional to the baryon rms radius.
The profile (\ref{eq:exp-prof}) is in fact a fairly good approximation to 
the fully self-consistent profile found numerically in Sect.~\ref{sec:self}.

In the treatment of hedgehog systems it is relevant to consider symmetries 
such as the grand spin $G=I+J$, the sum of isospin and spin.
The reader is referred to 
Refs.~\cite{Birse:1991cx,Diakonov:1995zi,NJL:rev,Alkofer:rev} for necessary details.

\subsection{The valence contribution}
\label{sec:val} 

\begin{figure}[tb]
\begin{center}
\includegraphics[width=9cm]{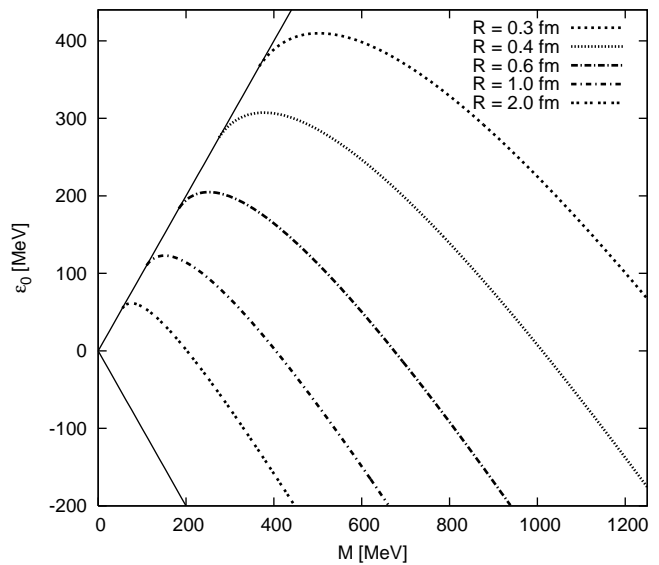}
\end{center}
\caption{The lowest $J^P=0^+$ eigenvalue $\epsilon_0$ as a function of mass $M$ 
for several soliton size parameters $R$ in the profile (\ref{eq:exp-prof}).
The curves are arranged with the lowest value of $R$ at the top. The two straight solid 
lines indicate the boundaries of the gap, $M$ and $-M$. 
Each curve leaves the positive continuum
at some critical value of $M$, assumes a maximum, and then decreases, 
asymptotically becoming parallel to the $-M$ line.   
\label{fig:valence}}
\end{figure}
 
Now we resort to the numerical calculation of the spectrum in the
chiral soliton, described in detail in the proceeding sections. What
we need for the present discussion is the behavior of the spectrum
when we vary the spectral mass $\omega$ as an independent variable.  We focus
on the grand-spin $0$ parity $+$ state, $G^P=0^+$, where $G=I+J$ is
the sum of isospin and spin, appropriate to classify states in the
hedgehog background (\ref{hedgehog}).  In Fig.~\ref{fig:valence} we
show the dependence of the lowest $0^+$ eigenvalue on the mass
parameter $M$ for {\em real masses} (in our notation $M$ is the same
as real $\w$).  Subsequent curves are obtained for the
profile~(\ref{eq:exp-prof}) with a fixed radius $R$.  In fact, all the
displayed curves are related to each other via simple scaling. Indeed,
introducing ${\vec \xi}= {\vec r} /R$ and $\Omega=\w R$ we can rewrite
the Dirac equation in the form
\begin{eqnarray}
\!\!\!\!\!\! \left ( -i \alpha \cdot \nabla_{\vec \xi} + \Omega U^5({\vec \xi}) \right )
\Psi_{n}(\vec \xi; \Omega) = E_n(\Omega) \Psi_{n}(\vec \xi; \Omega),
\label{eq:direqsc}   
\end{eqnarray}
which depends on a single scale $\Omega$.
By identification 
\begin{eqnarray}
\epsilon_n(\w)=\frac{E_n(\w R)}{R}. \label{eq:scaled}
\end{eqnarray}

As can be promptly seen from Fig.~\ref{fig:valence}, for any value of
$R$ we have for the valence $0^+$ level the limiting behavior
\begin{eqnarray}
\epsilon_0 (M) &\sim & M \qquad \qquad \hspace{2mm} {\rm at} \qquad M
\to 0, \\ \epsilon_0 (M) &\sim & -M + a \qquad {\rm at} \qquad M \to
\infty, \label{dive}
\end{eqnarray} 
where $a$ is a constant.  Correspondingly, at fixed $M$ in the limit
of low $R$ (small solitons) we have $\epsilon_0 (M) \simeq M$, {\em
i.e.} the level goes to the top of the gap, whereas in the limit of
high $R$ (large solitons) the behavior $\epsilon_0 (M) \simeq -M + a$
shows that the level goes to the bottom of the gap.  The small $M$
behavior, together with the reflexion property
Eq.~(\ref{eq:reflexion}) suggests a branch point behavior at the
origin
\begin{eqnarray}
\epsilon_0 (\w) &\to & \sqrt{\w^2} \qquad \qquad \w  \to 0  
\end{eqnarray} 
which is no mystery, since it corresponds to a free particle with the
complex mass and with energy $ \sqrt{k^2 + \w^2 }$ at zero momentum.
Unfortunately, in our study we only have access to the chiral soliton
spectrum for real $M$, thus we do not possess the information on
analyticity properties of the eigenvalues as functions of $\omega$,
which is fully accessible only in exactly soluble problems.  

Indeed, due to the inherent
numerical complications, finite-size discrete basis used, {\em etc.},
such information would  be difficult to obtain numerically in a reliable
way. Thus, we proceed motivated by the real mass results and the
analogy to a similar exactly-solvable model presented in
Appendix~\ref{sec:toy}. In other words, our assumptions made in the
general hedgehog case appear to be justified.

Let us now review the calculation of the Dirac propagator for the
standard case of a {\em real} mass. If we first compute the $\nu$ integral in 
Eq.~(\ref{Sspec}), we have to consider the $\gamma$ contour in the complex-$\nu$ space which
for real masses and hence real eigenvalues has the standard form of
going below the real axis for states with energy below the gap and
going above the real axis for states above the gap. This yields
\begin{eqnarray} 
&& \int_\gamma \frac{d \nu}{2\pi i } \frac{ e^{i \nu (t-t') } }{\nu -\epsilon_n (M) } = 
\label{eq:nu-contour}
\\ && \!\!\!\!\!\! \left[ \theta(t-t')\theta(\epsilon_n(M) )
-\theta(t'-t)\theta(-\epsilon_n (M) ) \right] e^{-i \epsilon_n (M)
(t-t')}, \nonumber 
\end{eqnarray} 
hence the positive (negative) energy states propagate forward
(backward) in time. This is equivalent to a damped (exploding)
imaginary-time behavior for $ t-t' \to \pm i \infty $. This real mass
result coincides with the standard one. If we change $M$ the valence
level may change sign (because it is equivalent to changing the
soliton size $R$ and the level dives into the sea according to
(\ref{dive})).  Thus, according to Eq.~(\ref{eq:nu-contour}) one keeps
the contour.  If we deformed the contour deciding that we occupy
the negative eigenvalue then the imaginary-time behavior implying the
standard particle (antiparticle) interpretation would be 
violated.

Let us now analyze the case of a {\em complex} mass. For that purpose we go
slightly off the real axis, $\w = M + i \Gamma $, and expand
\begin{eqnarray}
\epsilon_n ( M + i \Gamma ) = \epsilon_n ( M ) + i \Gamma \epsilon_n '
( M ) - {\textstyle\frac12}\Gamma^2 \epsilon_n '' ( M ) + \dots   
\label{eq:pert} 
\end{eqnarray} 
The edges of the bound state gap and consequently the Dirac
eigenvalues wander into the complex-$\nu$ plane moving upwards or
downwards, depending on the sign of $\epsilon_n'(M)$.  The eigenvalues
fulfilling $ \epsilon_n ' (M) =0 $ stay stationary on the real
axis. According to Fig.~(\ref{fig:valence}), we have $ \epsilon_0
''(M) < 0 $ for any $M>0$.  The question is whether or not we should
also deform the contour $\gamma$ in the evaluation of the propagator
(\ref{Sspec}). On the one hand, the eigenvalues lying in the complex
plane should not cross the contour, as this would lead to
discontinuities, thus some deformation must be done. On the other
hand, if $ \epsilon_0' (M) =0 $ we move along the real axis in the
positive direction by the amount $- \frac12 \Gamma^2 \epsilon_0 '' ( M
)$, Eq.~(\ref{eq:pert}), hence crossing the contour only in the case
where also $ \epsilon_0 (M) =0 $. According to
Fig.~(\ref{fig:valence}) this situation never happens unless $M=0$ or
$M$ is much larger than the position of the maximum (which, as we will
see, is of no relevance to our construction). Hence, we deform the
contour in such a way that it continues to pass through the $\nu=0$
point, but never allows a complex eigenvalue to intersect.  Such a
contour yields
\begin{eqnarray} 
&& \int_\gamma \frac{d \nu}{2\pi i } \frac{ e^{i \nu (t-t') } }{\nu -
\epsilon_n (\w) } = \left[ \theta(t-t')\theta({\rm Re} \,
\epsilon_n(\w) ) \right . \nonumber \\ && \qquad
-\left. \theta(t'-t)\theta(-{\rm Re} \, \epsilon_n (\w) ) \right]
e^{-i \epsilon_n (\w) (t-t')}.
\end{eqnarray} 
We note that the large imaginary time evolution is damped, as we might
have expected. In the positive $t-t' > 0$ branch, where a baryon (and
not an antibaryon) propagates, we are thus left with the integral
\begin{eqnarray} 
\int_{C'} d \w \rho(\w) \theta({\rm Re} \, \epsilon_n(\w) ) e^{-\tau \epsilon_n(\w)}. \label{eqqq} 
\end{eqnarray} 
for the imaginary time $\tau \to \infty $. Note that we only have the
positive section of the contour, as implied by the condition ${\rm Re}
\, \epsilon_n(\w) >0$.  Recall that we wish to occupy the lowest $0^+$
orbit with the eigenvalue $\epsilon_0(\w)$.  We proceed by using the
{\em saddle-point} method. First we have to locate stationary points
of $\epsilon_0 (\w)$, {\em i.e.} the points $\w_m$ where
\begin{eqnarray} 
\epsilon_0 ' (\w_m) =0. \label{sadcon}
\end{eqnarray}
From Fig.~\ref{fig:valence} we find, that for any value of $R$ there
exists a saddle located at the maximum of the curve
$\epsilon_0(M)$. We denote the position of the saddle as
$M_0$\footnote{A priori we should admit multiple saddles $\w_m$. We
  choose the branch, denoted $\w_0$, with the lowest possible real
  part of the eigenvalue. In addition, the complex reflexion property
  implies that if $\w_0$ is a saddle, then $\w_0^*$ is also a
  saddle. As a consequence, for complex saddles the large Euclidean
  time behavior would not be purely exponential, but also some
  oscillations would set in. This obviously contradicts the spectral
  decomposition (\ref{eq:spec_th}), which allows only for real
  stationary points. Therefore, although complex solutions may a
  priori exist, they should be considered a spurious result of the
  model. Similar problems occur also in the local models
  \cite{Broniowski:1995yq}.}.  We call $M_0$ the {\em saddle mass}.
As we move left from the saddle, decreasing $M$, we enter at some
critical value $M_c$ into the upper continuum. On the basis of the toy
model results of Appendix~\ref{sec:toy}, we expect that
$\epsilon_0(\w)$ has a branch cut at $\w=M_c$ running downwards. We
observe from Fig.~\ref{fig:valence} that the saddles are located right
of $M_c$ for any value of the soliton size $R$.
\begin{figure}[tb]
\includegraphics[width=8.5cm]{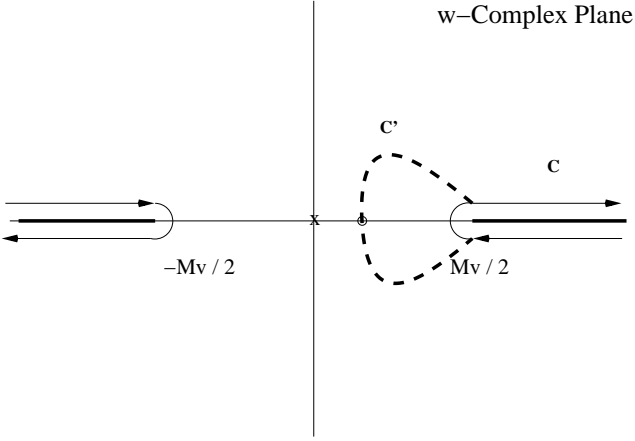} 
\caption{The deformed contour $C'$ in the complex $\omega$-plane for the saddle-point evaluation of the
valence contribution.  
\label{fig:contour}}
\end{figure}

In order to compute the integral in Eq.~(\ref{eqqq}) we deform the
original contour $C$ used in the vacuum sector into the contour $C'$
shown in Fig.~\ref{fig:contour}.  It contains the segment $C'-C$
(dashed line), which runs across $\w=M_0$ parallel to the imaginary
axis.
Note that it is not necessary to change the
path globally but only in the vicinity of the stationary
point. Making the
change of variables $ \w = M_0 + i \Gamma $, where $ M_0 $ is the 
saddle mass, $ \epsilon' (M_0) =0 $ we get
\begin{eqnarray} 
&& \int_{C'-C} d \w \rho(\w) \theta({\rm Re} \, \epsilon_n(\w) ) e^{-\tau
\epsilon_n (\w) } \nonumber \\  && \to  \int_{-\infty}^\infty i d \Gamma
\rho( M_0) e^{-\tau \epsilon (M_0) } e^{\frac12 \tau \epsilon_n''
(M_0) \Gamma^2 } \nonumber \\  && \sim i \rho( M_0) \sqrt{\frac{2 \pi }{-\epsilon_0 ''
(M_0) \tau }} \, e^{- \tau \epsilon (M_0)} \nonumber \\ 
&=& Z_0 e^{-\tau \epsilon (M_0)} \, 
\label{eq:sad}
\end{eqnarray} 
where the wave function renormalization factor $Z_0$ has been
included. Finally, we compare Eq.~(\ref{eq:sad}) to Eq.~(\ref{barbar})
and read off the valence contribution to the energy as
\begin{eqnarray}
E_{\rm val} = N_c \epsilon_0 (M_0). \label{eq:pres}
\end{eqnarray} 
The necessary conditions are $\epsilon_0' (M_0) =0$, $\epsilon_0 (M_0)
> 0$, and $\epsilon_0'' (M_0) < 0$, all satisfied in our case, as shown in
Fig.~\ref{fig:valence}. In addition, the solution is only admissible
with the conditions $M_c < M_0 < M_V/2$, which guarantee a real
residue. As we have already discussed, $M_c < M_0$ holds for all
values of $R$.  For $M_0 > M_V/2$ we would have a complex residue,
contradicting the spectral decomposition of the propagator,
Eq.~(\ref{eq:spec_th}).  According to Fig.~\ref{fig:valence}, for the
ansatz (\ref{eq:exp-prof}) this occurs for $R < 0.6~{\rm fm}$, where
no bound-state nucleon can be constructed. 
Thus, in the large imaginary time limit $t \to - i\infty $ and $t' \to
+ i\infty $ we get ( see e.g. Eq.~(\ref{eq:sad}))
\begin{eqnarray}
S(x',x) \to Z_0 \Psi_0  (x) \bar \Psi_0 (x') e^{-i (t-t') \epsilon_0 }, 
\label{eq:prop-val}
\end{eqnarray} 
where $\Psi_0$ is the valence Dirac spinor.

\begin{figure}[tb]
\includegraphics[width=8.5cm]{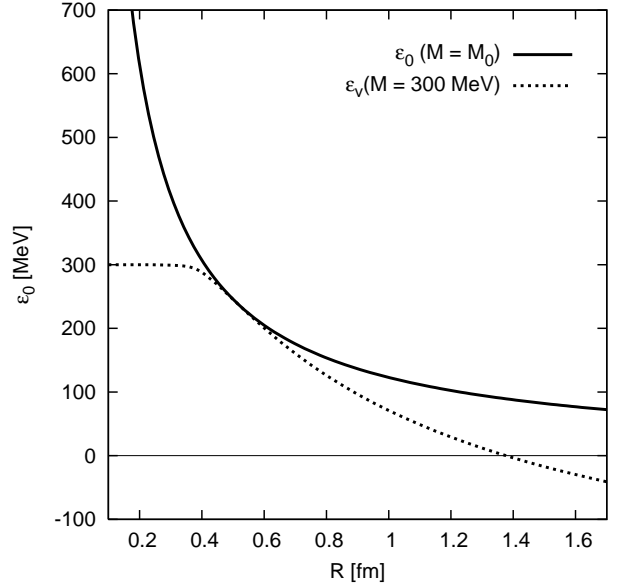} 
\caption{Solid line: the valence eigenvalue $ \epsilon_0$
and  the saddle mass, $M_0$, for the $0^+$
state (maxima of Fig.~\ref{fig:valence}), plotted as a function of the
soliton size $R$. Dotted line: the $0^+$ eigenvalue $\epsilon$ at
the fixed value of $M=300~{\rm MeV}$, as used in traditional
fixed-mass models.  We note that the SQM prescription has
$\epsilon_0 = const/R$, whereas $\epsilon \to const$ at $R \to 0$
and becomes negative at $R \to \infty$.
\label{fig:valR}}
\end{figure}

To summarize, our prescription for the valence orbit consists of the following
very simple steps:
\begin{enumerate}
\item Obtain (numerically) the spectrum for the valence orbital as a
function of the real positive mass $M$.
\item Look for the saddle mass, {\em i.e.}, 
a maximum with respect to $M$, with the help of conditions $\epsilon_0'
(M_0) =0$ and $\epsilon_0'' (M_0) < 0$.  From analyticity, this is the
saddle point which corresponds to the saddle mass $M_0$.
\item If $M_c \le M_0 < M_V/2$, the valence contribution to the soliton
mass is $N_c \epsilon_0 (M_0)$, otherwise there is no baryon state in
the model.
\end{enumerate} 
There is an essential difference between the standard valence
prescription based on the Fock-space decomposition, used in local
quark models, and our prescription for SQM presented above. In the
first case the size of the profile, $R$, and the quark mass, $M$, can
be fixed independently in the Dirac operator, whereas in the present
case we actually find a correlation between them. In the quark
propagator we have a superposition of masses $\omega$ which are
integrated over\footnote{This reminds of the familiar distinction
between the phase and group velocities of wave packets made out of a
continuous superposition of plane waves.}, thus no additional
independent scale is present in the problem.  An immediate consequence
can be seen in Fig.~\ref{fig:valR}, where we compare the valence
eigenvalue for the profile function (\ref{eq:spec_th}) along the
manifold where $\epsilon_0'(M)=0$ (for a given $R$) to the fixed-mass
result of the local chiral quark model, obtained with $M_Q=300~{\rm
MeV}$.  From the scaled Dirac equation (\ref{eq:scaled}) and our
valence prescription it follows that
\begin{eqnarray}
\epsilon_0 = {k}/R, \label{eps1R}
\end{eqnarray}
where $k$ is a constant depending on the particular profile.
In particular, for $R\to 0$ in SQM one has $\epsilon_0 \to 1/R$ (a
signature of a ``repulsive force''), as opposed to the behavior
$\epsilon_0 \to M_Q$ of a free particle at rest in the case of the
fixed-mass model.  This means in practice that the bound state never
becomes unbound in the presence of a chiral field background (see the
discussion of Sect.~\ref{sec:conf}) and reflects the absence of
on-shell quarks in the vacuum (analytic confinement) in SQM. Another
feature of SQM with our valence prescription is that the valence
eigenvalue never dives into the negative part of the spectrum. The
prescription for fixed-mass models allows the valence contribution to
become negative, and this happens with a finite slope which originates from
non-analyticity. This entering of the valence level into
the negative part of the spectrum was interpreted within the chiral
quark soliton model as entering the ``Skyrme'' model regime. In our
approach such a regime never arises. However, the $1/R$ dependence of
the valence contribution behaves very much like the Skyrme stabilizing
term.

Similarly, the saddle mass scales as 
\begin{eqnarray}
M_0 = {k'}/R. \label{M1R}
\end{eqnarray}
Numerically, for the exponential profile (\ref{eq:exp-prof}) one gets 
$k = 123$~ MeV~fm and $k' = 151$~ MeV~fm, 
for the linear profile (\ref{eq:lin-prof})
$k = 234$~ MeV~fm and $k' = 257$~ MeV~fm, and for 
the arctan profile (\ref{eq:atan-prof}) $k = 100$~ MeV~fm and $k' = 119$~ MeV~fm.

\subsection{The soliton baryon density and the baryon number}

\begin{figure}[tb]
\includegraphics[width=8.5cm]{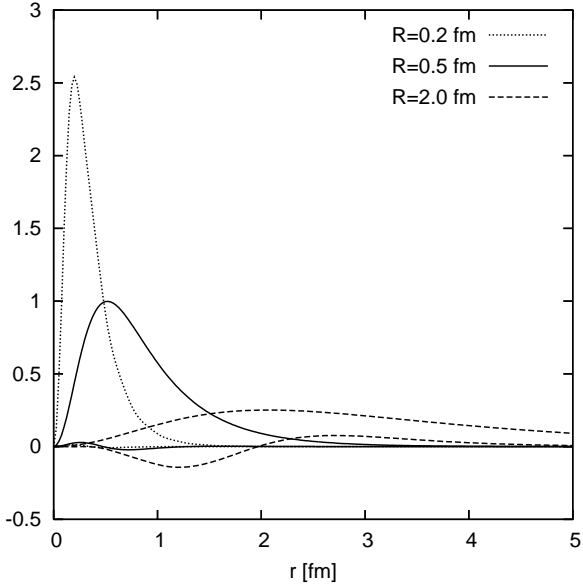} 
\caption{Sea and valence contribution to the radial baryon density for several 
soliton radii in the SQM (the valence parts have positive sign at all $r$). 
\label{fig:densiR}}
\end{figure}

A crucial point is to show that our soliton corresponds to a system
with baryon number equal to one. This requires the separation
between the valence and sea contributions to the baryon number.  To
analyze this point we have to compute the following time ordered
product,
\begin{eqnarray} 
\Pi_B^\mu (x,x',y) &=& \langle 0 | T \Bigl\{ B(x) \bar q (y) \gamma^\mu q(y)
\bar B(x') \Bigl\} | 0 \rangle \nonumber \\
&=& \frac{\delta }{\delta v_\mu (y)} \Pi_B^v (x,x') \Big|_0  
\end{eqnarray} 
with $B(x)$ denoting the interpolating baryonic operator,
Eq.~(\ref{eq:baryo-field}) and $\Pi_B^v (x,x') $ the baryon correlator
(\ref{eq:baryon-corr}) in the {\it presence} of a external vector
field $v^\mu(y)$ for which the result (\ref{eq:sol-corr}) follows if
the external source is included in the propagator 
\begin{eqnarray}
S_{a a'}^v (x,x') = \int_{C'} d \w \rho(\w) \langle x|
\left(i\slashchar{\partial} - \slashchar{v} - \omega
U^5\right)^{-1}_{a a'}|x'\rangle .
\label{eq:eff_prop_v} 
\end{eqnarray}  
For the open line propagator the functional derivative can be readily
evaluated yielding
\begin{eqnarray}
\frac{\delta S^v (x',x) }{\delta v_\mu (y)}\Big|_0  &=& \int_C d \w \rho(\w) \nonumber \\ &\times& 
\langle x| \frac1{i\slashchar{\partial} - \omega U^5 } |y\rangle \gamma^\mu \langle y|
\frac1{i\slashchar{\partial} - \omega U^5 } |x'\rangle
\nonumber \\ & \to & \bar \Psi_0 (y) \gamma^\mu \Psi_0 (y) S (x,x')
\end{eqnarray} 
where in the last line the limit $t \to - i\infty $ and $t' \to +
i\infty $ has been taken along the line of reasoning developed in
Sec.~\ref{sec:val} (see Eq.~\ref{eq:prop-val}). The determinant
contribution to the baryon current can be deduced from
Eq.~(\ref{eq:bc-sea}). Collecting all results we obtain for the
three-point correlator in the asymptotic limit the factorized form
\begin{eqnarray} 
\Pi_B^\mu (x,x',y) &\to & B^\mu (y) \Pi_B (x,x' ) 
\end{eqnarray} 
with the total baryon current given by 
\begin{eqnarray} 
B^\mu (x) & = & \bar \Psi_0 (x) \gamma^\mu \Psi_0 (x) \nonumber \\ &+&
\int_C d \w \rho(\w) \tr \left[ \gamma^\mu \langle x | \frac{-i}{
i\slashchar{\partial} - \omega U^5 } |x \rangle \right]
\label{eq:b-val+sea} 
\end{eqnarray} 
which is our result for the baryon current and in general for any
observable based on a one-body operator. In Fig.~\ref{fig:densiR} the
sea and the valence contributions to the baryon density for several
soliton radii in the SQM are displayed. As we see, the Dirac sea contribution
gives a vanishing contribution to the baryon number; only the valence
quarks contribute to the total normalization. This is consistent with
our finding that the valence level never dives into the negative
energy region.

It is worth mentioning that the previous calculation can be
extended mutatis mutandis to any one body observable and in particular
to the energy-momentum tensor. In the following section we analyze the
sea contribution as it arises from Eq.~(\ref{barbar}) onwards. A non
trivial check which results from the one body consistency condition is
that such a calculation coincides with the one based on the energy-momentum tensor in the soliton background.

\subsection{The sea contribution}

To identify the sea contribution we compare, as done for the valence
case, the spectral decomposition of the two point correlator a large
Euclidean times, Eq.~(\ref{barbar}), with the path integral
representation, Eq.~(\ref{eq:sol-corr}). Using Eq.~(\ref{eq:eff_ac_0})
for stationary configurations, the Dirac sea contribution to the energy
is obtained by the spectral integration along the original contour $C$
of Fig.~\ref{contour0},
\begin{eqnarray}
E_{\rm sea} &=& \frac{\rm i}{T} N_c \int_C d\w \rho(\w) {\rm Tr}\log
\left ( i\slashchar{\partial} - \omega U^5 \right ) \nonumber \\ &=& -
N_c \int_C d\w \rho(\w) \frac{1}{2} \sum_n \sqrt{\epsilon_n(\w)^2}
\nonumber \\ &=& N_c \int_C d\w \rho(\w) \sum_{i \in {\rm sea}}
\epsilon_i(\w) ,
\end{eqnarray}
where $T$ is the time and $n$ indicates all (positive- and
negative-energy) states of the Dirac Hamiltonian $H$ defined in
Eq.~(\ref{eq:Hdirac}) and the frequency integral has been carried out.
In the last line Eq.~(\ref{Hp}) has been used and summation $i$ runs
over negative energy (sea) states only. A full-fledged evaluation of this
quantity is presented in Sect.~\ref{sec:self}.  To understand the
general trend we will present an estimate based on the so-called {\em
two point function approximation} (TPA) proposed originally in
Ref.~\cite{Diakonov:1987ty} and further exploited in
Ref.~\cite{Diakonov:1988mg,Adjali:1991bi}. This approximation has the
virtue of reproducing both the limit of large and small soliton sizes
and is based in the identity for the normal parity contribution to the
action
\begin{eqnarray}
&& \int_C d\w \rho(\w) {\rm Tr}\log \left ( i\slashchar{\partial} -
\omega U^5 \right ) |_{n.p.}  \nonumber \\ &&= \frac12 \int_C d\w \rho(\w) {\rm
Tr}\log \left ( \partial^2 + i \omega\slashchar{\partial} U^5
+ \omega^2 \right)
\end{eqnarray}
(deduced by commuting the $\gamma_5 $ matrix across the Dirac operator)
and further expanding the logarithm to second order in the field
$U^5$, yielding
\begin{eqnarray}
E_{\rm sea}^{\rm TPA} &=& \frac{-{ i}N_c}{4 T} \int_C d\w \rho(\w)
{\rm Tr} \left\{ \left[ \frac1{\partial^2+ \w^2 }i
\omega\slashchar{\partial} U^5 \right]^2 \right\} \nonumber \\ &=&
\frac{-{ i}N_c}{T} \int_C d\w \rho(\w) \int \frac{d^4 q}{(2\pi)^4}
q^2 I(q^2,\w) \w^2 \langle |U(q)|^2 \rangle \nonumber \\
\end{eqnarray}
where the functional trace and Dirac traces have been evaluated,
$\langle . \rangle $ indicates the isospin trace, and $U(q)$ is the
Fourier transform of the chiral field $U(x)$ (see Appendix
\ref{sec:eff2} for notation). The one-loop integral $I(q^2, \w)$ is
introduced and calculated in Appendix~\ref{sec:integrals}. A
straightforward calculation in the meson-dominance version of the SQM
for static fields yields
\begin{eqnarray}
E_{\rm sea}^{\rm TPA}= \frac14 f_\pi^2 \int \frac{d^3  {q}}{(2 \pi)^3}
\,  {\vec q}^2 \langle | U( {\vec q}) |^2 \rangle \frac{M_V^2}{{\vec q}^2+M_V^2},
\label{eq:TPA}
\end{eqnarray} 
where 
\begin{eqnarray}
U( {\vec q})=\int d^3  { x} U({\vec x}) e^{i {\vec q} \cdot{\vec x}}.    
\end{eqnarray}
The factor appearing in Eq.~(\ref{eq:TPA}) is the pion form factor,
Eq.~(\ref{eq:pion-formfac}) in the space-like region, $q^2 = - {\bf
q}^2 < 0$.  Actually, this is a general feature, static soliton
profiles probe the Euclidean region of mesonic correlation functions.
At large values of the soliton size $R$, small values of $q$ dominate
in Eq.~(\ref{eq:TPA}) and one gets 
\begin{eqnarray}
E \sim a f_\pi^2 R, 
\end{eqnarray} 
where $a$ is a numerical constant. For the exponential profile
(\ref{eq:exp-prof}) one gets numerically $a=30.99$. In the opposite
limit of small soliton sizes, large $q$ values dominate and
Eq.~(\ref{eq:TPA}) yields
\begin{eqnarray}
E \sim b f_\pi^2 R (M_V R)^2,  
\end{eqnarray} 
with $b=28.11$.
This behavior is different from the $R^3 \log R$ short distance
behavior documented in Ref.~\cite{Diakonov:1987ty} for the proper-time regularized
fermion determinant; it is also free
of the $ R \log R $ behavior reported in
Ref.~\cite{Ripka:1987ne,Soni:1987zj} for the renormalized sea energy,
which generated a vacuum Landau instability.

\subsection{Existence of absolute minima \label{sec:conf}}

\begin{figure}[]
\begin{center}
\includegraphics[width=8.5cm]{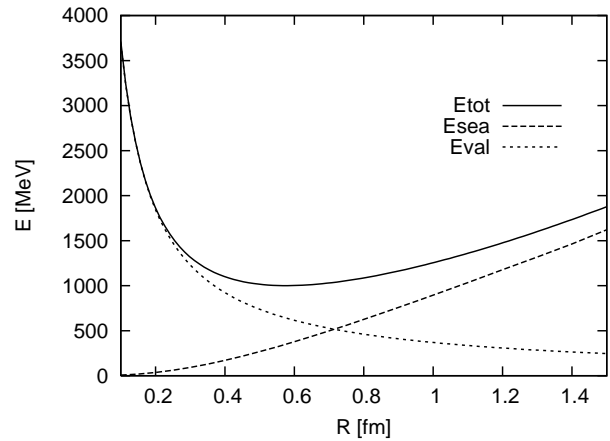}
\end{center}
\caption{Total energy of the soliton (solid line) and its valence and sea
components for the profile (\ref{eq:exp-prof}), plotted as functions of $R$.
An absolute minimum exists.
\label{fig:ER}}
\end{figure}

The total soliton energy is the valence plus the sea contribution as
\begin{eqnarray}
E_B = E_{\rm val} + E_{\rm sea}.   
\end{eqnarray} 
At small radii the valence contribution (\ref{eps1R}) dominates and $E
\sim 1/R$, while at large radii the Dirac sea contribution dominates
and $E \sim R$.  Since the function $E(R)$ is continuous, on these
simple grounds we prove to have a minimum. The behavior is illustrated
in Fig.~(\ref{fig:ER}), where we show the energy dependence as a
function of the profile size $R$ for the profile
(\ref{eq:exp-prof}). For this restricted configuration one clearly
sees the occurrence of an {\em absolute minimum} as a consequence of
the fact that the valence quarks never become unbound.  In contrast,
in the standard approach to chiral quark solitons the valence quarks
become unbound at small $R$ and one has instead a {\em local minimum},
which becomes unstable when the total soliton energy exceeds that of
$N_c$ free quarks at rest, $E > N_c M $. This instability has been an
obvious cause of concern since this situation corresponds to bound but
not confined solitons. In practice, it would not be a problem if
valence states where deeply bound. However, most calculations have
produced solutions which lie on the unstable branch. The fact that our
solutions correspond to an absolutely stable state is a remarkable
property of SQM together with our construction of the valence state.
Results for other profiles of the chiral field are qualitatively
similar to the case of Fig.~\ref{fig:ER}. 

Another interesting feature that can be seen from a direct comparison of
Figs.~\ref{fig:ER} and \ref{fig:valR} is that the minimum takes place
at a soliton size $R$ where the valence state for our model and the
fixed-mass models produce a similar dependence. This suggests that the
constituent fixed mass in soliton models may indeed correspond to a
given saddle mass in the spectral construction and that the shallow
(and unstable) minima found there may indeed be identified as the
absolute and stable minima obtained here.

\section{Results for the self-consistent soliton}
\label{sec:self}

In this section we describe the numerical 
properties of self-consistent chiral solitons in SQM, as well as
the corresponding observables.

\subsection{Evaluation of observables}

As is well known, observables in hedgehog solitons fall into two
categories: independent of and dependent on {\em cranking}
\cite{Cohen:1986va}.  The second case is much more complicated, hence
in this paper we deal only with the quantities that do not involve
cranking.  Observables independent of cranking can be written in terms
of single spectral sums of the form
\begin{equation}
A=\sum_i A_i, \label{obs}
\end{equation}
where $A$ is the value of the observable in the soliton, and $A_i$ is
the single-particle contribution which includes a valence orbit and
the Dirac sea. In the spectral approach this is generalized, in the
sense that an extra integration over the spectral variable $\omega$ is
present. With our method of treating the valenceness we also have a
separated contribution from $N_c$ occupied valence states,
\begin{eqnarray}
A &=& A_{\rm val}+A_{\rm sea}, \nonumber \\
A_{\rm val} &=& N_c A(\omega_{\rm val}), \nonumber \\
A_{\rm sea} &=& \int_C d \omega \rho(\omega) A_{\rm sea} (\omega), \label{obsom}
\end{eqnarray}
where ${\rm val}$ and ${\rm sea}$ indicate valence and sea, $\omega_{\rm
val}=M_0$ is the saddle mass, $\rho(\omega)$ is the
spectral function consisting of the vector and scalar parts given in
Eq.~(\ref{rhos}), and $A_{\rm sea}(\omega) = \sum_{i \in {\rm
sea}}A_i(\omega)$ with $A_i(\omega)$ denoting the $\omega$-dependent
single-particle value of the observable in the orbit $i$.  Due to the
hedgehog symmetries for the observables in question one may replace
the sum over the negative energy states, $\sum_{i \in {\rm sea}}$, with
$1/2$ of the sum over all states, $\frac{1}{2}\sum_{n \in {\rm all}}$.

The contour $C$ is invariant under the transformation 
$\omega \to - \omega$, see Fig.~\ref{contour0}. Therefore 
\begin{eqnarray}
A_{\rm sea}&=&\frac{1}{2} \left [\int_C d \omega \rho(\omega) A_{\rm sea}(\omega)- 
\int_C d \omega \rho(-\omega) A_{\rm sea}(-\omega) \right ] , \nonumber \\ 
&=& \int_C d \omega \rho_V(\omega) \frac{A_{\rm sea}(\omega)+A_{\rm sea}(-\omega)}{2}
\nonumber \\ &+& 
\int_C d \omega \rho_S(\omega) \frac{A_{\rm sea}(\omega)-A_{\rm sea}(-\omega)}{2},
\label{obsompm}
\end{eqnarray}
where $\rho_V$ and $\rho_S$ are the odd and even parts of the spectral 
function $\rho$,
respectively, see Eq.~(\ref{rhos}).  As we have stressed before, 
the contour $C$ in Fig.~\ref{contour0} is complex. This is a complication, since in a
numerical calculation we obviously do not have access to $A(\omega)$
in the complex $\omega$ plane. For that reason we use a method which
allows us to carry on the calculation along the real axis in the
$\omega$ plane, as explained in Appendix \ref{app:trick}.
According to the formalism of Sect.~\ref{sec:formal}, the the energy
of the soliton is
\begin{eqnarray}
&& E=N_c E_{\rm val}+ \left . \frac{d^2}{du^2}\bar [E_V(u)+E_S(u)] \right |_{u=1/M_V^2}, \nonumber \\
&& \bar E_{V,S}(u)=  2\int_{1/(2 \sqrt{u})}^\infty 
d \omega {\rm disc} [ \bar \rho_{V,S}(\omega) ] \nonumber \\ && \times
\sum_i \frac{\epsilon_i(\omega,m) \pm \epsilon_i(-\omega,m)}{2}. \label{E}
\end{eqnarray}
With the help of Eq. (\ref{trans}) we may also write
\begin{eqnarray}
\bar E_{V,S}(u)&=&  2\int_{1/(2 \sqrt{u})}^\infty 
d \omega {\rm disc} [ \bar \rho_{V,S}(\omega) ] \nonumber \\ && \times
\sum_i \frac{\epsilon_i(\omega,m)\pm \epsilon_i(\omega,-m)}{2}. \label{Etr}
\end{eqnarray}
In the chiral limit
\begin{eqnarray}
\bar E_V(u)&=&  2\int_{1/(2 \sqrt{u})}^\infty {\rm disc} [ \bar \rho_V(\omega) ] 
\sum_i {\epsilon_i(\omega,0)}, \nonumber \\ \bar E_S(u)&=&  0. \label{Etrchir}
\end{eqnarray}
All observables not involving cranking can be evaluated with the
technique described above and with the help of ``standard'' formulas
at each $\omega$.  The derivatives with respect to $u$ are carried out
numerically.

\subsection{Self-consistent equations}

Now we present some details of our self-consistent procedure. 
We use the linear representation of the hedgehog field, $U=s+i \hat r
\cdot \vec{\tau} p$, 
and impose the 
nonlinear constraint $s^2+p^2=1$ by renormalizing the fields after each 
numerical iteration. The Euler-Lagrange equations for the radial $s$ and $p$ fields 
are obtained from Eq.~(\ref{linear}) and have the form 
\begin{eqnarray}
\frac{s(r)}{g^2}&=& N_c \bar \psi_0({\vec x}, M_0)  \psi_0({\vec x}, M_0)
\nonumber \\ 
&+& \int_C d \omega \rho(\omega) \sum_i \bar \psi_i({\vec x}, \w)  
\psi_i({\vec x}, \w), \nonumber \\
\frac{p(r)}{g^2}&=& N_c \bar \psi_0({\vec x}, M_0) 
i \gamma_5 \vec \tau \cdot \hat r  \psi_0({\vec x}, M_0) \label{eq:ELe}\\ 
&+& \int_C d \omega \rho(\omega) \sum_i \bar \psi_i({\vec x}, \w)
i \gamma_5 \vec \tau \cdot \hat r \psi_i({\vec x}, \w), \nonumber
\end{eqnarray}
where $g^2$ is treated as a Lagrange multiplier. The first terms 
on the r.h.s. are the valence contributions,
evaluated according to the SQM prescription. The second terms 
are the Dirac sea contributions, 
where the sum over the negative-energy states $i$ is carried out, or, equivalently, 
it can be replaced by $1/2$ of the sum over all states.. With the help of formulas 
(\ref{disczv},\ref{disczs}) the spectral integration 
in (\ref{eq:ELe}) can be performed as a real-valued integral.
The spinors $\psi_i({\vec x}, \w)$
are obtained by solving the Dirac equation in the background of the fields 
$s(r)$ and $p(r)$ at a given value of $\w$.

The code used to find numerically the self consistent solutions 
is a modification of the method used in solving different versions
of chiral-quark models.
The quark orbits are calculated by diagonalizing $H$
for each value of $\omega$ in the discrete Kahana-Ripka basis \cite{Kahana:1984dx}.
The Euler-Lagrange equations are solved by iteration.
The numerical effort involved in the calculation is similar 
to that in the case of solitons in non-local models 
\cite{Golli:1998rf,Broniowski:2001cx}. 

\subsection{The self-consistent solution}

For the sake of obtaining physical properties of the soliton, we use two
versions of the model. The first one (model I) is just the standard 
meson-dominance SQM, with the single vector meson mass, $M_V=m_\rho=769~{\rm MeV}$.
This model (in the chiral limit) has only one scale, 
thus all observables are proportional to the value 
of $M_V$ in appropriate power. In particular, $f_\pi^2=M_V^2/(8\pi^2)$. 
The second model (model II) 
includes also the excited $\rho$ state, $\rho'(1465)$. The spectral function 
is taken to be the weighted sum of Eq.~(\ref{rhos}), containing 90\% of the
ground-state $\rho$, and 10\% of excited $\rho$. All quantities are 
distributive over the spectral density, {\em e.g.}  
$f_\pi^2=0.9m^2_\rho/(8\pi^2)+0.1m^{\prime 2}_\rho/(8\pi^2)$. 
Model II contains two scales, and produces somewhat heavier and more compact
solitons, as expected on simple scaling grounds. We work with 
the physical pion, $m_\pi=139.6~{\rm MeV}$, 
and the current quark mass is adjusted such that the 
GMOR relation is fulfilled, $m \langle \bar q q \rangle = m_\pi^2 f_\pi^2$.  

\begin{figure}[tb]
\begin{center}
\includegraphics[width=8.5cm]{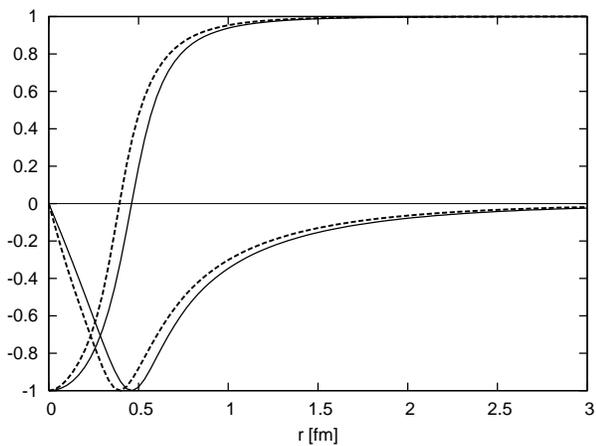}
\end{center}
\caption{The $s\equiv\cos\theta$ and $p\equiv\sin\theta$ 
fields as a function of the radius $r$ for the two
self-consistent solutions of Table~\ref{energies}:
the solid lines for the model I and the dashed lines for model II.
\label{fields}}
\end{figure}

The results of the self-consistent calculation for the case with
physical pion mass are displayed in Table~\ref{energies} and in
Fig.~\ref{fields}.  The chiral corrections are small and do not alter
our basic conclusions.  We note that for both models the value of the
saddle mass $M_0$ is well below the critical value ${1\over2}M_V$
discussed in Sect.~\ref{sec:formal}.

Comparing the soliton energy to the experimental nucleon mass we
should be aware that our soliton is a mean-field solution with grand
spin 0 and should not be identified with the nucleon, but rather with
the average of the mass of the nucleon and the $\Delta(1232)$ isobar
$\frac{1}{2}(M_N+M_\Delta)=1174~{\rm MeV}$.  Moreover, quantization of
the collective coordinates, or projection of the soliton wave function
on the subspace with good quantum numbers
\cite{Golli:1986er,Birse:1986qc}, reduces its energy by eliminating
the spurious rotational and translational energy.  We note that the
contribution to the energy arising from the scalar spectral density,
$E_{{\rm sea}, S}$, is much smaller than the vector part, $E_{{\rm
sea},V}$. Furthermore, we notice that the model gives approximate
equipartition of energy between each valence quark and the Dirac sea,
in agreement to the $1/R$ behavior of the valence contribution and
the approximate $ \sim R$ dependence of the sea part. 

\begin{table}[ht]
\begin{center}
\begin{tabular}{lrr}
\hline
 & model I & model II \\
\hline
$f_\pi$ [MeV]                              &   86.5  &  97.3  \\
$m_\pi$ [MeV]                              &  139.6  & 139.6  \\
$(-\langle \bar q q \rangle)^{1/3}$ [MeV]  &  243    &  243   \\ 
$m$ [MeV]                                  &  5.04   &  6.37  \\
$M_0$ [MeV]                                &   267   & 304    \\
$\varepsilon_0$ [MeV]                      &   233   & 263    \\
\hline
$E_\mathrm{sea,V}$ [MeV]                   & 285     & 351   \\
$E_\mathrm{sea,S}$ [MeV]                   &  36     & 35    \\
$E_\mathrm{total}$ [MeV]                   & 1019    & 1174  \\
\hline
\end{tabular}
\end{center}
\caption{Model parameters and the soliton energy in the self-consistent calculation
with the physical pion mass.
\label{energies}}
\end{table}

\subsection{Results for observables}

\begin{table}[ht]
\begin{center}
\begin{tabular}{lccc}
\hline
 & model I & model II & experiment \\
\hline
$\sqrt{\langle r^2\rangle_{I=0,\mathrm{val}}}$  [fm]  & 1.23  & 1.07 & - \\
$\sqrt{\langle r^2\rangle_{I=0,\mathrm{total}}}$ [fm] & 1.24  & 1.08 & 0.79 \\
\hline
$g_A$(val)                                 &  0.79   & 0.79  & -   \\
$g_A$(total)                               &  0.93   & 0.95  & 1.26 \\
\hline
$\mu_{I=1}$(val)                           &  2.40   & 2.07  &  -  \\
$\mu_{I=1}$(total)                         &  2.96   & 2.67  &  4.71 \\
\hline
$\sigma_{\pi\mathrm{N}}$ [MeV]             &  36     & 33    & $45\pm 8$ \\ 
\hline
\end{tabular}
\end{center}
\caption{Properties of the self-consistent solution of Table~\ref{energies}:
the rms isoscalar charge radius,
the axial-vector charge $g_A$, 
the isovector magnetic moment $\mu_{I=1}$,
and the sigma commutator $\sigma_{\pi\mathrm{N}}$.
The valence contribution (val) is given separately.
The contribution from the sea (not given explicitly) is dominated
by the vector part, similarly as in the case of the energy.
The experimental value of $\sigma_{\pi\mathrm{N}}$ is taken 
from Ref.~\cite{Gasser:1990ce}.
\label{observables}}
\end{table}

\begin{figure}[tb]
\begin{center}
\includegraphics[width=8.5cm]{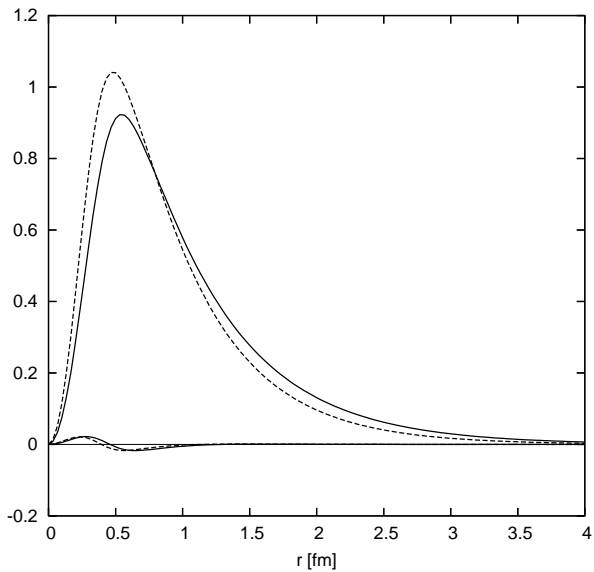}
\end{center}
\caption{The valence and the sea contributions to the
baryon density (multiplied by $4\pi r^2$) for the self-consistent 
solutions:the solid lines for model I and the dashed lines for model II.
\label{densities}}
\end{figure}

As discussed in Sec.~\ref{sec:self}, in the present approach we deal only
with those observables that do not involve cranking. Inclusion of cranking,
which requires a linear-response calculation, is numerically involved and 
is outside of the scope of this study.
In Table~\ref{observables} we display some characteristic observables
of the self consistent solution for the choice of parameters of
of Table~\ref{energies}.
Firstly, we note that the isoscalar rms charge radius of the soliton is too large 
as compared to the experiment.  
This quantity is dominated by the valence
contribution, as can also be seen from Fig.~\ref{densities}.
The quark spinors of the valence wave function exhibit a long tail
which is a feature related to our prescription for constructing the
valence orbit. The large $r$ behavior is of the form 
$\exp(-\sqrt{M_0^2-\epsilon_0^2}\,r)$, 
and in our case the value of $\kappa = \sqrt{M_0^2-\epsilon_0^2}$ is small, with 
$\kappa=1/(1.5~{\rm fm})$ in model I and $\kappa=1/(1.3~{\rm fm})$ in model II.
The large soliton radius can perhaps be reduced by subtracting the 
spurious center-of-mass motion of the soliton.
Including higher vector mesons in the model tends to reduce 
the soliton size, as expected by scaling arguments,
however, on the expense of increasing $f_\pi$ and $E_{\rm total}$.
 
In the evaluation\footnote{Explicit expressions for these quantities
may be found, {\em e.g.}, in Ref.~\cite{NJL:rev,Alkofer:rev}.} 
of $g_A$ and the isovector magnetic moment
$\mu_{I=1}$ we use
the semiclassical projection coefficients as obtained in the large-$N_c$ 
limit \cite{Cohen:1986va}.
Both values are lower than the corresponding experimental values.
But since our solution is dominated by the valence state we could
as well, instead of the semiclassical projection, 
use the quark model wave functions for evaluation of 
the valence parts of observables.
Then the expectation value of the $\sigma\tau$
operator for the proton and the neutron 
yields the coefficient ${5\over9}$ instead of ${1\over3}$.
That would result in higher values, in better agreement
with the experiment. 
The value of the sigma commutator, which in our model is equal to
\begin{equation}
 \sigma_{\pi\mathrm{N}} = -m_\pi^2 f_\pi^2\int d^3 r(s(r)-1)  
\label{sigmacom}
\end{equation}
is reasonably well reproduced, assuming values somewhat smaller than
the experimental number, showing that the spatial extent of our chiral profile 
is not too large.

The results for the observables turn out to be quite similar 
to those obtained in other chiral models. In particular, they are
strikingly close to the predictions of the
model with the non-local regulators \cite{Golli:1998rf,Broniowski:2001cx}.

\section{Conclusion}
\label{sec:concl}

For many years it has been accepted that baryons arise as solitons of
a chiral Lagrangean in the large-$N_c$-limit of QCD. However, the
precise realization of this attractive and fruitful idea is still
unclear. Many model Lagrangeans have been proposed, emphasizing
different aspects of the problem and incorporating as many known
features of the underlying quark-gluon dynamics as possible.  While on
the one hand the Skyrme soliton models incorporate confinement but not
the spin-1/2 partons, on the other hand, chiral quark soliton models
account for spin-1/2 partons but lack the confinement property. This
unsatisfactory situation has been calling for improvement.  As a first
and hopefully useful step, we have considered a model where quark
poles on the real axis are absent, be it in the vacuum or in the
soliton. Unfortunately, confinement does not hold since basic large-$N_c$ 
requirements on the analytic structure of meson correlators
are violated, similarly to other effective quark models. 
On the other hand the Spectral Quark Model provides a
framework with spin-1/2 partons in the vacuum which simultaneously
allows for baryonic soliton solutions corresponding to absolute minima
of the action and hence cannot decay into three free quarks, unlike
previous local chiral quark soliton models. This is due to the rather
peculiar features of the model characterized by a continuous
superposition of masses in the complex plane with a suitable spectral
function. As a necessary complementary study we have examined an
instance of an involved complex-mass relativistic system.

The solitons we find fit nicely into the phenomenological expectations
of more standard chiral quark solitons with full inclusion of the
polarized Dirac sea. In contrast to constituent chiral quark
models where one is allowed to tune the constituent quark mass as a free
parameter, the Spectral Quark Model does not have this freedom. In our
case the vector meson mass determines one unique solution in the
chiral limit and thus all properties scale with this mass. Actually,
due to the  superposition of complex spectral masses $\omega$, in the soliton
calculation there appears a saddle mass, $M_0$, defined as a
stationary point of the valence quark eigenvalue as a function of
$\omega$ in a fixed soliton profile. In the chiral limit $M_0$ also
scales with the vector meson mass and thus is a fixed number. To some
extent the saddle mass behaves as a constituent quark mass determined
uniquely by the soliton, and its numerical value comes out within the
expected range $M_0 \sim 300 {\rm MeV}$ for the typical solitons
minimizing the total energy.  Moreover, the close resemblance of the
valence contribution to the energy around the minimum suggests that
the phenomenology and results of previous calculations within
constituent chiral quark models with a constant mass are
reproduced, at least for the low-lying baryon states. The new aspect
unveiled by our calculation is the possibility of studying the monopole
vibrations and their quantization which are the traditional candidate
for describing excited baryonic states such as, {\em e.g.}, the Roper
resonance. While studies of this sort are routinely carried out in the
Skyrme model, the lack of confinement in the traditional chiral quark
soliton model has prevented, as a matter of principle, calculations of
excited states.

One distinguishing feature of the Spectral Quark Model construction is
the verification of an important consistency relation involving quark
two-point functions which were not fulfilled for fixed constituent
mass chiral quark models. This feature, in addition to the uniform
treatment of regularization removes theoretical doubts on the proper
computation of the high-energy processes as regards their partonic
interpretation and normalization. As we have mentioned, one motivation
for using chiral quark soliton models is the spin 1/2 nature of the
constituents, which makes the calculation of partonic distributions in
the nucleon possible as a matter of principle. The results found in
the present paper pave the way for such a calculation in the soliton
picture where the interplay between the analyticity enforced by the
lack of on-shell quarks and the chiral symmetry may simultaneously be
tested for the low-lying baryon states.

\begin{acknowledgments}

This research is supported by the Polish Ministry of Education and
Science, grants 2P03B~02828, by the Spanish Ministerio de Asuntos
Exteriores and the Polish Ministry of Education and Science, project
4990/R04/05, by the Spanish DGI and FEDER funds with grant
no. FIS2005-00810, Junta de Andaluc{\'\i}a grants no.  FQM225-05, EU
Integrated Infrastructure Initiative Hadron Physics Project contract
no. RII3-CT-2004-506078 and by the Bilateral Program for Scientific
and Technological Cooperation of the Ministries of Science,
Technology, and Higher Education of Poland and Slovenia.

\end{acknowledgments}

\appendix

\section{The effective action to second order in the fields} \label{sec:eff2}

When using the path integral (\ref{eq:pathint}) with the effective
action (\ref{eq:eff_ac}) it is enough to expand the action to second
order in the external sources and the dynamical pion field, 
\begin{eqnarray}
U = 1 +  \frac{i}{f} \vec{\tau}\cdot \vec{\pi} \gamma_5 -
\frac1{2 f^2}  \vec{\pi}\cdot \vec{\pi}  + \dots .
\end{eqnarray} 
This can be done by noting that the Dirac operator in
Eq.~(\ref{eq:dirac_op}) can be decomposed as a free part plus a
perturbation,
\begin{eqnarray}
i {\bf D} &=& i\slashchar{\partial} - \omega - V, 
\label{eq:dirac_op_pert} 
\end{eqnarray} 
where 
\begin{eqnarray}
V &=& \omega \left( \frac{i}{f} \vec{\tau}\cdot \vec{\pi} \gamma_5 -
\frac1{2 f^2}  \vec{\pi}\cdot \vec{\pi} + \dots \right) \nonumber \\ 
&+&  {\hat m_0} - \left(
\slashchar{v} + \slashchar{a} \gamma_5 - s - i \gamma^5 p \right) .
\end{eqnarray}
Note that {\it both} the free propagator and the potential $V$ may depend
on the spectral mass $\w$.  Then the perturbative expansion of the fermion
determinant can be readily done, yielding 
\begin{eqnarray}
 i \Gamma &=& N_c\int_C d \omega \rho(\omega) {\rm Tr} \log (
i\slashchar{\partial} - \omega ) \nonumber \\ &-& N_c\int_C d \omega
\rho(\omega) \sum_{n=1}^\infty \frac1{n} {\rm Tr} \left[ \frac1{
i\slashchar{\partial} - \omega } V \right]^n \nonumber \\ &=& i
\Gamma^{(0)} + i \Gamma^{(1)} + i \Gamma^{(2)} + \dots,
\label{gamma012}
\end{eqnarray} 
which can be classified according to the number of external sources as
well as dynamical fields.

The zeroth order contribution yields the vacuum energy,
\begin{eqnarray}
i \Gamma^{(0)} &=& N_c\int_C d \omega \rho(\omega) {\rm Tr} \log (
i\slashchar{\partial} - \omega ) \nonumber \\ &=& N_c N_f \int d^4 x \int_C d
\omega \rho( \w) \int \frac{d^4 p}{(2\pi)^4} {\rm tr} \log (
\slashchar{p} - \omega ) \nonumber \\ &=& \int d^4 x  \left\{ -\epsilon
+ N_c N_f \int \frac{d^4 p}{(2\pi)^4} {\rm tr} \log (
\slashchar{p} )  \right\},
\end{eqnarray}  
where the vacuum energy density, $\epsilon$, defined relative to
the case of massless quarks, has been explicitly separated. This
quantity was computed in Ref.~\cite{RuizArriola:2003bs} and in Sec.~\ref{sec:vac} of the present paper 
through the use of the
energy momentum tensor. It is instructive to check the calculation
in the effective action formalism. We use the reflection symmetry $p \to -p $
and the standard identity for matrices, ${\rm tr} \log A + {\rm tr}
\log B = \log \det A + \log \det B = \log \det(AB)= {\rm tr} \log (AB)
$ for the commuting Dirac matrices $A= \slashchar{p} - \w$ and $B=
\slashchar{p} + \w$ (in four dimensions $\det A = \det (-A) $). The
vacuum energy density can be written as one half of any of these
contributions. The Dirac trace becomes trivial and after interchanging
the order in the momentum and spectral integrals one gets
\begin{eqnarray}
\epsilon &=& -2i  N_c N_f \int \frac{d^4 p}{(2\pi)^4}  \nonumber \\ &\times&
\int_C
d \w \rho(\w)  \left[ \log(p^2 - \w^2 ) -\log(p^2) \right] .
\end{eqnarray} 
To evaluate first the $\w$-integral we write 
\begin{eqnarray}
\log(p^2 - \w^2 ) = \int^{p^2} \frac{dk^2}{k^2- \w^2}  . 
\end{eqnarray} 
Thus, we readily get from direct application of Eq.~(\ref{AB}), 
\begin{eqnarray}
\epsilon &=&-2i N_c N_f \int \frac{d^4 p}{(2\pi)^4}
\int^{p^2} dk^2 \left[A(k^2) -\frac1{k^2} \right] \nonumber \\ &=&
i N_c N_f \int \frac{d^4 p}{(2\pi)^4} \left[p^2
A(p^2) -1\right],
\end{eqnarray} 
after going to the Euclidean space and integrating by parts.
The surface term has been discarded, as the condition $\lim_{s \to \infty} s^3 A(-s) =0$, 
holds in the meson dominance realization Eq.~(\ref{AB}). 
Thus the final result is finite, 
\begin{eqnarray}
\epsilon &=& - \frac{ N_c N_f M_V^4}{192 \pi^2}  = -(0.2 {\rm GeV})^4 \qquad
(N_f=3) \nonumber \\ 
&=& -\frac{3 N_f \pi^2 f^4}{N_c} ,
\end{eqnarray} 
in agreement with Eq.~(\ref{enden})
based on the energy-momentum tensor. 

Returning to Eq.~(\ref{gamma012}), the first non-vanishing correction
involves only the scalar source, yielding
\begin{eqnarray} 
i \Gamma^{(1)} &=& N_c\int_C d \omega \rho(\omega) {\rm Tr} \left\{
\frac1{ i\slashchar{\partial} - \omega } s \right\}  \\ 
&=& 4 N_c \int \frac{d^4 p}{(2\pi)^4} \int_C d \omega 
\frac{\rho(\omega) }{p^2 - \omega^2 }  \int d^4 x  \langle s (x) \rangle 
\nonumber
\end{eqnarray} 
($\langle . \rangle $ means the flavor trace), whence the quark condensate
may be obtained as
\begin{eqnarray}
N_f \langle \bar q q \rangle = - 4 i N_c N_f \int \frac{d^4
p}{(2\pi)^4} B(p^2),
\end{eqnarray} 
which becomes an identity according to Eq.~(\ref{AB}). 

Finally, the second order contribution can be written as 
\begin{eqnarray}
\Gamma^{(2)} &=& \Gamma_{SS}+ \Gamma_{PP}+ \Gamma_{VV}+ \Gamma_{AA}+
\Gamma_{\pi\pi} \nonumber \\ &+& \Gamma_{\pi P}+ \Gamma_{\pi A} + \Gamma_{PA} .
\end{eqnarray} 
The calculation is straightforward. For the bilinears in the fields,
generically written as $\varphi(x) = \varphi_A(x) \Gamma^A $ with $A$
a denoting Lorentz-flavor index, we have
\begin{eqnarray}
i \Gamma^{(2)} &=& - \frac12 N_c \int d\omega \rho(\omega) {\rm Tr}
\left[ \varphi_A \Gamma^A {1\over i \slashchar{\partial}- \omega}
\varphi_B \Gamma^B {1\over i \slashchar{\partial}- \omega} \right]
\nonumber \\ &=&  \frac{i}2 \int \frac{d^4 q}{(2\pi)^4} \bar \varphi_A
(q) \bar \varphi_B (-q) K_{AB} (q^2),
\end{eqnarray} 
where 
\begin{eqnarray}
(-i) K_{AB} (q) &=& - N_c \int_C d\omega \rho(\omega) \int {d^4 p \over
(2\pi)^4 } \nonumber \\ 
&\times& 
{\rm Tr} \left[ \Gamma_A {i\over \slashchar{p}-\slashchar{q} -
\omega} \, \Gamma_B {i\over \slashchar{p} - \omega} \right].
\end{eqnarray} 
The Fourier-transformed fields are defined through the relation 
\begin{eqnarray}
\varphi (x) = \int \frac{d^4 q }{(2\pi)^4} e^{i q \cdot x} \bar \varphi (q).
\end{eqnarray} 
The flavor trace is trivial, $\langle \lambda_a \lambda_b \rangle =
\delta_{ab} /2 $.  After using the vanishing condition of the positive
moments, Eq.~(\ref{eq:moments}),  the calculation of the Dirac trace and
momentum integration yields
\begin{widetext} 
\begin{eqnarray}
(-i) K_{SS}^{ab} (q) &=& - N_c \int_C d\omega \rho(\omega) \int {d^4 p
\over (2\pi)^4 } {\rm Tr} \left[ \frac{\lambda^a}{2}{i\over
\slashchar{p}-\slashchar{q} - \omega} \frac{\lambda^b}{2} \, {i\over
\slashchar{p} - \omega} \right] \\ &=& \delta^{ab} N_c \int_C d \w
\rho(\w) \left[ 2 I ( 0 , \w ) \w^2 + I (q^2, \w ) (4 \w^2 -q^2 )
\right], \nonumber \\
(-i) K_{PP}^{ab} (q) &=& - N_c \int d\omega \rho(\omega) \int {d^4 p
\over (2\pi)^4 } {\rm Tr} \left[ {i\over \slashchar{p}-\slashchar{q} -
\omega} \, i \gamma_5 \frac{\lambda^a}{2} \, {i\over \slashchar{p} -
\omega} \, i \gamma_5 \frac{\lambda^b}{2} \, \right] \\ &=&
\delta^{ab} N_c \int_C d \w \rho(\w) \left[ 2I (0, \w ) \w^2 - I (q^2, \w ) q^2 \right] , \nonumber \\
(-i) K_{VV}^{a \mu; b \nu} (q) &=& - N_c \int d\omega \rho(\omega)
\int {d^4 p \over (2\pi)^4 } {\rm Tr} \left[ {i\over
\slashchar{p}-\slashchar{q} - \omega} \, \gamma_\mu
\frac{\lambda^a}{2}\, {i\over \slashchar{p} - \omega} \, \gamma_\nu
\frac{\lambda^b}{2} \, \right] \\ &=& \frac13\delta^{ab}N_c \left(
-g^{\mu\nu } + { q^\mu q^\nu \over q^2} \right) \int_C d \w \rho(\w)
\left[ 4 I (0,\w ) \w^2 - \frac{q^2}{24\pi^2}- 2 I (q^2 , \w ) (2 \w^2
+ q^2 ) \right], \nonumber \\
(-i) K_{AA}^{a \mu; b \nu} (q) &=& - N_c \int d\omega \rho(\omega)
\int {d^4 p \over (2\pi)^4 } {\rm Tr} \left[ {i\over
\slashchar{p}-\slashchar{q} - \omega} \, \gamma_\mu \gamma_5
\frac{\lambda^a}{2} \, {i\over \slashchar{p} - \omega} \, \gamma_\nu
\gamma_5 \frac{\lambda^b}{2} \, \right] \\ &=& \frac13 \delta^{ab}N_c
\left( -g^{\mu\nu } + { q^\mu q^\nu \over q^2} \right) \int_C d \w
\rho(\w) \left[ 4 I ( 0 , \w ) \w^2 - \frac{q^2}{24\pi^2}+ 2I (q^2, \w
) (4 \w^2 - q^2 ) \right] \nonumber \\ &-& \delta^{ab}4N_c { q^\mu
q^\nu \over q^2} \int_C d \w \rho(\w) I (q^2, \w ) \w^2, \nonumber\\
(-i) K_{AP}^{a \mu ; b} (q) &=& 
- N_c \int d\omega \rho(\omega) \int
{d^4 p \over (2\pi)^4 } {\rm Tr} \left[ {i\over
\slashchar{p}-\slashchar{q} - \omega} \, \gamma_\mu \gamma_5
\frac{\lambda^a}{2} \, {i\over \slashchar{p} - \omega} \, i \gamma_5
\frac{\lambda^b}{2} \, \right] \\ &=& \delta^{ab} 2i N_c q^\mu \int_C
d \w \rho(\w)\w I (q^2, \w ), \nonumber \\
(-i) K_{\pi\pi}^{ab} (q) &=& - N_c \int d\omega \rho(\omega) \int {d^4
p \over (2\pi)^4 } {\rm Tr} \left[ \frac{i \w}{ f^2} \delta^{ab} {i\over
\slashchar{p} - \omega} + {i\over \slashchar{p}-\slashchar{q} -
\omega} \frac{\w}{f} \gamma_5 \lambda^a{i\over
\slashchar{p} - \omega} \frac{\w}{f} \gamma_5 \lambda^b
\right] \\ &=& -\delta^{ab} 4 N_c \frac{q^2}{f^2} \int_C d \w \rho(\w)\w^2
I (q^2, \w ), \nonumber \\
(-i) K_{A\pi}^{a \mu ; b} (q) &=& - N_c \int d\omega \rho(\omega) \int
{d^4 p \over (2\pi)^4 } {\rm Tr} \left[ {i\over
\slashchar{p}-\slashchar{q} - \omega} \, \gamma_\mu \gamma_5
\frac{\lambda^a}{2}\, {i\over \slashchar{p} - \omega} \, i \gamma_5
\lambda^b \frac{\w}{f} \, \right] \\ &=& -\delta^{ab} 4 i N_c \frac{q^\mu}{f} \int_C d \w \rho(\w) 
I (q^2, \w ) \w^2 ,\nonumber \\
(-i) K_{ P \pi }^{ab} (q) &=& - N_c \int d\omega \rho(\omega) \int {d^4 p \over
(2\pi)^4 } {\rm Tr} \left[ {i\over \slashchar{p}-\slashchar{q} -
\omega} \, i \gamma_5 \frac{\lambda^a}{2} \, {i\over \slashchar{p} - \omega} \, i
\gamma_5 \lambda^b \frac{\w}{f} \, \right] \\ &=& \delta^{ab} 2 N_c \frac1{f}
\int_C d \w \rho(\w) \left[ 2 I (0, \w ) \w^3 - I(q^2,\w) q^2 \w \right] . \nonumber
\end{eqnarray} 
\end{widetext}
where the basic one-loop two-point integral $I(q^2, \w) $ is defined
in Eq.~(\ref{eq:I2qw}). 

To compute the correlation functions as functional derivatives we must
take into account the contributions from the pion pole. This is
accomplished in a standard way by eliminating the pion field at the
mean field level through the equations of motion, $\delta \Gamma /
\delta \pi^a(x)=0$, yielding at lowest order in the field
\begin{eqnarray}
\pi^a (q) K_{\pi \pi}^{ab} (q) + p^a (q) K_{\pi P}^{ab} (q)+ a^{a,\mu} (q)
K_{A \pi}^{a, \mu; b} (q)  =0 \nonumber \\ 
\end{eqnarray}  
and reinserting the pion field into the effective action
$\Gamma^{(2)}$. This contribution exactly cancels the non-transverse
piece of the $AA$ correlator, reproducing the gauge technique result of
Eq.~(\ref{eq:VV+AA}) and an additional pion pole contribution to the
$PP$ correlator yielding Eq.~(\ref{eq:SS+PP}).


\section{Useful Integrals}
\label{sec:integrals} 

The basic two-point integral is given by\footnote{There is a typo in
Eq.~(A4) of Ref.~\cite{RuizArriola:2003bs}.}
\begin{eqnarray}
I(q^2, \w) &=& \frac1{i} \int {d^4 k \over (2\pi)^4} {1\over k^2 -
\omega^2 } {1\over (q-k)^2 - \omega^2 } \label{eq:I2qw} \\ &=& \frac1{i}
\int {d^4 k \over (2\pi)^4} \int_0^1 \frac{dx}{\left[k^2 -\w^2 + q^2 x
(1-x) \right]^2} \nonumber \\ 
&=& - \frac1{(4\pi)^2} \int_0^1 d x \, \log \left[\w^2 - x(1-x)q^2 \right]. \nonumber
\end{eqnarray} 
For the meson dominance case one can use the moments method, based on
expansion in powers of $q^2$ and resummation of the series. This
method provides a useful check to our computations.  However, for the
purpose of illustrating the issues of quark unitarity with our
unconventional propagator it is enlightening to provide general
formulas in terms of the vector and scalar components of the quark
propagator. The $\w$ integrals can be evaluated first by using
Eqs.~(\ref{AB}).  We get the useful identities,
\begin{eqnarray}
A'(p^2) &=& -\int_C d\w \frac{\rho (\w)}{[p^2-\w^2]^2} ,
\nonumber \\ 
B'(p^2) &=& -\int_C d\w \frac{\w \rho (\w)}{[p^2-\w^2]^2} ,
\nonumber \\ 
A(p^2) + p^2 A'(p^2) &=& -\int_C d\w \frac{\w^2 \rho (\w)}{[p^2-\w^2]^2} ,
\nonumber \\ 
B(p^2) + p^2 B'(p^2) &=& -\int_C d\w \frac{\w^3 \rho (\w)}{[p^2-\w^2]^2} .
\end{eqnarray}  
The normalization of the pion field yields 
\begin{eqnarray}
  f^2  = 4 N_c i  \int \frac{d^4 k}{(2\pi)^4} \frac{d}{dk^2}
  \left[k^2 A(k^2) \right] = \frac{M_V^2 N_c }{24 \pi^2}.
\end{eqnarray} 
We also have 
\begin{eqnarray}
&& \int_C d \w \rho (\w) I(q^2 ,\w ) = \nonumber \\ && \;\; i \int_0^1 dx 
 \int \frac{d^4 k}{(2\pi)^4} A'\left(k^2 +x(1-x)q^2 \right), \nonumber \\ 
&& \int_C d \w \rho (\w) I(q^2 ,\w ) \w = \nonumber \\ && \;\;i \int_0^1 dx 
 \int \frac{d^4 k}{(2\pi)^4} B'\left(k^2 +x(1-x)q^2 \right),
\end{eqnarray}
and so on.  The integrals can be evaluated by passing to the Euclidean
space $k^2 \to - k_E^2 $ and $ (-i) d^4 k \to \pi^2 k_E^2 d k_E^2
= \pi^2 s ds $. We may then shift the integration variable in order to get, {\em e.g.},  
\begin{eqnarray}
J(q^2)&=&-i \int_0^1 dx \int \frac{d^4 k}{(2\pi)^4} F\left(k^2 +x(1-x)q^2
 \right) \\ &=& \frac{1}{16 \pi^2} \int_0^\infty s ds
 \int_0^1 dx F\left(-s +x(1-x)q^2 \right) \nonumber \\ &=& \frac{1}{16
 \pi^2} \int_0^\infty dS F(-S) \int_0^1 dx \left[S+x(1-x)q^2 \right]_+,
 \nonumber 
\end{eqnarray} 
where we have introduced the distribution $[x]_+ = x \theta (x) $. Let
us assume that for definiteness $q^2 < 0 $. Then, the argument of the
step function vanishes for $x<x_-$ and $x>x_+$, where
\begin{eqnarray} 
x_\pm = \frac12 \pm \sqrt{1 + \frac{4 S}{q^2}}.
\end{eqnarray} 
We have  
\begin{eqnarray}
&& \int_0^1 dx  \left[S+x(1-x)q^2 \right]_+ = S+ \frac{q^2}6 \nonumber  \\ 
&& \;\;\;\; - \theta \left( -q^2-4 S \right) \sqrt{1+\frac{4S}{q^2}} (q^2 + 4 S ).
\end{eqnarray}
Thus
\begin{eqnarray}
J(q^2) &=& -\frac{1}{16\pi^2} \Big\{ \int_{0}^\infty dS F(-S) \left(
S+ \frac{q^2}6 \right) \\ &-& \int_0^{-q^2/4} dS F(-S)
\sqrt{1+\frac{4S}{q^2}} (q^2 + 4 S ) \Big\}. \nonumber
\end{eqnarray} 
This formula can be analyticly continued to any complex $q$, and
the second integral becomes a line integral in the complex $S$-plane.
Clearly, if the function $F(-S)$ has a cut at say $S=-M_V^2/4$, the
result of the line integral becomes path dependent and will develop an
imaginary part discontinuity for $q^2 > M_V^2 $. This is the way how a
quark propagator with no poles generates the unitarity cuts in the two
point correlators. 

We list the final results: 
\begin{eqnarray}
&&\int_C d \w \rho (\w) I(q^2 ,\w ) = \int_C d \w \rho (\w) I(0 ,\w ) \nonumber \\ 
&& \;\; + \frac1{(4\pi)^2} \Big[ -\log \left( 1- \frac{q^2}{M_V^2} \right) + 
\frac23 \frac{q^2}{M_V^2-q^2} \Big], \nonumber  \\ 
&& \int_C d \w \rho (\w) I(q^2 ,\w ) \w = -\frac{\langle \bar q q \rangle }{2 N_c } \frac1{M_S^2 -q^2}, \nonumber \\
&& \int_C d \w \rho (\w) I(q^2 ,\w ) \w^2 = \frac{M_V^2}{96\pi^2}\frac{M_V^2}{M_V^2 -q^2}, \nonumber \\ 
&& \int_C d \w \rho (\w) I(q^2 ,\w ) \w^3 = \nonumber \\ && \;\;\; - \frac{\langle \bar q q \rangle}{8 N_c} \Big[ \frac{M_S^2}{M_S^2
-q^2} - 3 \frac{M_S}{q} \tanh^{-1} \frac{q}{M_S} \Big].
\end{eqnarray}

\section{Quantum-mechanical examples of complex-mass systems \label{qm}} 

Establishing analytic properties of the Dirac operator eigenvalues for
a complex mass is an involved mathematical problem. To gain some
insight and develop some intuition, in this Appendix we consider a few
cases of non-relativistic quantum mechanical system with complex
coupling potentials. Our aim is to define what we mean by the {\it
energy eigenvalues for a complex coupling}, assuming that we know the
definition for the real coupling, and to determine its analytic
properties in the complex plane. The most natural and obvious way to
do so is in terms of analytic continuation in the coupling from the
real case.

As a first example let us consider the harmonic oscillator which has
the reduced potential $U = 2m V = m^2 \w^2 x^2 $ and the ground state
energy is given by $ E_0 (\w) = \w /2 $ for real $\w> 0$. The problem
is obviously invariant under the change $ \w \to -\w $, and we should
write $ E_0 (\w) = |\w| /2 $ for real $\w$. Clearly, we have a branch
cut at $ \w=0$ as a function of the coupling $\w^2$, hence the right
way to write the energy for a complex coupling is $ E_0(\w) =
\sqrt{\w^2} /2 $. Written this way the bound state wave function is
given by $ \psi_0 = C e^{-\sqrt{\w^2} x^2 /2 }$.  The analytic
continuation of the decreasing exponential becomes an increasing
exponential on the second Riemann sheet, $\w^2 \to e^{2\pi i} \w^2
$. For negative $\w^2$ we have a negative energy. So, the energies on
the first and second Riemann sheets differ only in the sign.

As a second example let us consider the hydrogen atom where we have $
V = - Z /r $ with $ Z > 0$ and the energy is $ E_0 ( Z) = -Z^2 /2
$. This suggests an analytic behavior in $Z$, and in particular having
a continued bound state solution for repulsive potentials ($ Z < 0 $).
Again, the bound state function $ u(r) = r e^{-Z r} $ transforms into
a positive exponential for negative $Z$. So, we have a cut at $
Z=0$. For negative $Z$ we have a real energy but it is on the second
Riemann sheet. In this case the energy on the first and the second
Riemann sheets coincide.

As a final example, more directly related to the more complicated case
of the toy model for the Dirac equation in Appendix~\ref{sec:toy} , we
analyze the complex square well potential for which we have $U(r)=
-U_0 \Theta ( R-r) $, with $U_0$ complex. For real $U_0$ the s-wave
bound state solution is given by (we take $2m=1$)
\begin{eqnarray} 
u(r) = \Theta (R-r) A e^{-\kappa r} + \Theta (r-R) B \sin K r 
\end{eqnarray} 
where 
\begin{eqnarray} 
\kappa = \sqrt{-E} \quad , \qquad K = \sqrt{U_0 - E}
\end{eqnarray} 
The continuity condition for the logarithmic derivative yields the
bound state relation 
\begin{eqnarray} 
-\kappa = K \cot KR
\end{eqnarray} 
which defines an implicit function $ E_0 ( U_0 ) $ which we want to
extend to complex $U_0$. For real $U_0$ there is a critical value 
$U_{0,c}= (\pi/2 R)^2 $ above which the equation has real
solutions. Actually, close to the threshold we have for $ U_0 >
U_{0,c} $ the ground state
\begin{eqnarray}
E_0 ( U_0 ) &=& - \frac14 R^2 \left( U_0 - U_{0,c}\right)^2 \nonumber \\
&+& \frac18 R^4 \left(1 - \frac{4}{\pi^2} \right) \left( U_0 - U_{0,c}
\right)^3 + \dots,
\end{eqnarray} 
which suggests an analytic behavior of the energy. Again, it is the
wave function at large distances where we see that $ \kappa \to
-\kappa$ if we loop once about the critical point $ U_{0,c}= (\pi/2
R)^2$ in the complex $U_0$ plane. In this particular case, this
corresponds to the well known fact that a square well potential with a
subcritical coupling has a virtual state and no bound state. As $U_0
\to 0^+ $ one gets $ \kappa \to -\infty $ or $ E_0 \to -\infty $ on
the second Riemann sheet. Similar features should appear also for
excited states, $E_n$, with the corresponding critical values. Thus,
we may define the continuous function $E_0 ( U_0 ) $ for all real
values of $U_0$. For $ U_0 > U_{0,c} $ it corresponds to a bound
state, whereas for $ U_0 < U_{0,c} $ it describes a virtual state.

If we go now to the complex plane in $U_0$ we get the implicit
function defined through
\begin{eqnarray}
A ( k , U ) =0  
\end{eqnarray} 
If we have a solution $A(k_0 , U_0) =0 $ and go close to it, then by
Taylor expanding we get
\begin{eqnarray}
0 &=& A ( k_0 + \Delta k , U_0 + \Delta U ) \nonumber \\ 
&=& A (k_0, U_0 ) +
\frac{\partial A ( k_c , U_c )}{\partial k} \Delta k + \frac{\partial
A ( k_c , U_c )}{\partial U} \Delta U \nonumber \\ 
\end{eqnarray} 
This way we can define a one to one relation unless either derivative
vanishes at a critical point 
\begin{eqnarray}
\frac{\partial A ( k_c , U_c )}{\partial k} =0 \qquad {\rm or} \qquad 
\frac{\partial A ( k_c , U_c )}{\partial U} =0 
\end{eqnarray} 
If this is the case we have a square root branch point assuming 
\begin{eqnarray}
\frac{\partial^2 A ( k_c , U_c )}{\partial k^2} \neq 0 
\end{eqnarray} 
In our case we have the critical points located at
\begin{eqnarray}
-\kappa &=& K \cot (KR) \\ 
-1 &=&  \frac{\kappa R}{\sin(KR)^2} + \frac{\kappa}{K} \cot (KR)   
\end{eqnarray} 
Combining both equations we get 
\begin{eqnarray}
\kappa_c &=& - \frac1{R} \qquad E_c = -\frac1{R^2} \\ 1 &=& K R \cot
(KR) \qquad U_c = \frac{1 + x_n^2}{R^2}
\end{eqnarray} 
with $ x_n = 0 , \pm 4.49, \pm 7.72 , \pm 10.9 , \dots $.
Moreover, we have 
\begin{eqnarray}
\frac{\partial^2 A ( k_c , U_c )}{\partial k^2} = \frac{R^3 U_c}{R^2 U_c -1} 
\end{eqnarray} 
which diverges for $ U_c = 1/R^2 $. Note that the critical points are
located in the second Riemann sheet.  They generate branch points of
second order, {\em i.e.} looping twice around the point the function
returns to its original value. The Riemann surface is obtained by
joining all critical points with a line.  Since they are infinitely
many, the cut divides the complex $U$ plane into two disjoint pieces.
The complete and extensive study of Riemann sheets of this particular
problem can be looked up at Ref.~\cite{Gramma}.

The main outcome is that the analytic structure of a the
analyticly continued eigenvalues of a complex coupling potential can
be determined by the study of the critical points which generate
branch cuts. Otherwise, the function is analytic in the complex
coupling plane. This situation appears also in Appendix~\ref{sec:toy}
for the Dirac operator with a complex mass.

\section{Analytic properties of the valence eigenstate in a toy model} 
\label{sec:toy} 

The discussion in Sect.~\ref{sec:formal} made explicit use of the fact
that an integration path in the complex mass plane of the Dirac
Hamiltonian can be deformed without pinching any
singularities. Although we cannot prove this in general, the
complex-mass coupling Dirac systems may have unusual properties.  In
this appendix we investigate a model baring similarity to the full
chiral model where our assumptions are verified.  The main issue is
both to define the meaning of an eigenvalue as a {\em function of the
complex mass} $\w$ as well as to determine its analytic properties in
the complex-mass plane. To our knowledge this topic is not discussed
in the literature at all. However, strong similarities are found with
the analytic properties of the energy eigenvalues of complex
potentials, which have been motivated in the context of optical
models~\cite{Kok:1980fw,Badalian:1981xj} and for review purposes
Appendix~\ref{qm} includes some warm-up quantum-mechanical problems
which may be helpful in the understanding of analyticity properties
that arise in studies of our type. The prescription for valenceness
derived in Sect.~\ref{sec:formal} holds under specific conditions.  In
the chiral soliton model we have no precise knowledge of the analytic
properties, hence it is not mathematically proven that the
prescription can actually be used.  The model of this section is much
simpler and solvable semi-analyticly.  It shows that the desired
analytic properties are fulfilled, providing support for the method in
application to the chiral soliton model described in this paper.

Let us consider the Dirac equation for the state with $0$ grand-spin 
and positive parity, $G^P = 0^+ $, with the upper component $u$ and the lower
component $v$.
It has the form
\begin{eqnarray} 
u' &=& - u M \sin \theta -v ( \epsilon + M \cos \theta ) \nonumber \\
v' &=& u ( \epsilon - M \cos \theta ) + v ( -\frac2{r}+ M \sin \theta
) \label{eq:toym}
\end{eqnarray}  
with the usual boundary conditions for a normalizable state
\begin{eqnarray}
u'(0) =0, \; v(0) =0, \; u(\infty) =0, \; v(\infty) =0.
\end{eqnarray} 
Following Ref.~\cite{Zhao:1988in} we look for a solution with the
profile function 
\begin{eqnarray}
\theta (r) = \pi \Theta (R-r), \label{eq:th} 
\end{eqnarray} 
where $\Theta (x)$ 
the standard Heaviside step function. The profile (\ref{eq:th})
has the winding number $ (\theta (0) - \theta (\infty ))/ \pi=1$. The
analytic solution is
\begin{eqnarray}
u(r) &=& N\left (\Theta (R-r) \frac{\kappa^3}{\epsilon+M}\frac{\sinh \kappa
r}{r} \right . \nonumber \\ &+&  \left . \Theta(r-R) \frac{\kappa^3}{M-\epsilon} \frac{e^{-\kappa
r}}{r} \right ), \nonumber \\ v(r) &=& N \left ( \Theta (R-r) \frac{\kappa}{r^2} ( \kappa r \cosh
(\kappa r) - \sinh ( \kappa r) ) \right . \nonumber \\ &+& \left . \Theta (r-R) \kappa ( \kappa r +1)
\frac{e^{-\kappa r}}{r^2} \right ),
\end{eqnarray} 
where $\kappa = \sqrt{M^2 - \epsilon^2}$ and $N$ is the normalization factor 
chosen such that $N^2 \int d^3 x(u^2+v^2)=1$. Using the matching
condition expressing the continuity of $ u(r)/v(r)$ at $r=R$ we get the
eigenvalue equation
\begin{eqnarray}
\kappa R \coth (\kappa R ) = \frac{\kappa R ( M-\epsilon ) + 2 M
}{\epsilon + M}
\end{eqnarray} 
which can be written it terms of dimensionless variables $x= \kappa R$ and $ x_0 = MR $ as
\begin{eqnarray}
x \coth (x) = \frac{x ( x_0  \pm \sqrt{x_0^2 - x^2}) + 2 x_0 
}{\pm \sqrt{x_0^2-x^2} + x_0},  \label{eq:tra}
\end{eqnarray} 
where $\pm$ corresponds to the positive (negative) 
energy states, $\epsilon >0$ ($\epsilon < 0$ ).  
The value $x=0$ is formally always a solution of Eq.~(\ref{eq:tra}), 
but it does not correspond to a
zero momentum state $ \epsilon_0 =M $ since the solution is 
\begin{eqnarray}
u(r) &=& \frac{1}{R^2}\Theta (R-r) + \frac{2 M}r \Theta(r-R) \\ v(r)
&=& \Theta (R-r) \frac{2Mr}{3 R^2} + \Theta (r-R) \frac1{r^2},
\end{eqnarray} 
which are regular both at $r=0$ and $r=\infty$ but fail to fulfill the
matching condition since $ v / u = 2Mr/3 $ for $r > R $ whereas $ v/u
= 2 M r $ for $ r> R$. 

\begin{figure}[ttt]
\includegraphics[width=8.5cm]{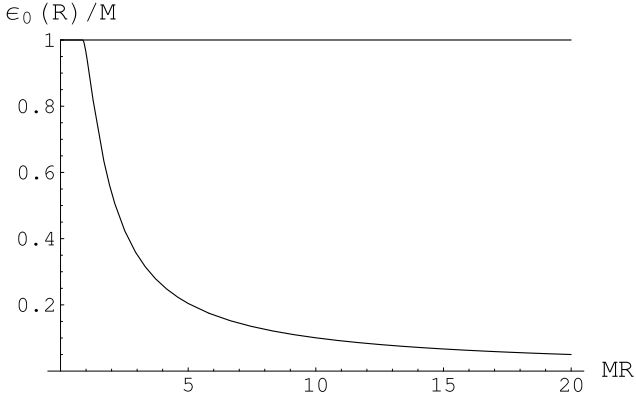} 
\caption{Valence eigenvalue $ \epsilon_0$ for the toy model
(\ref{eq:toym},\ref{eq:th}) as a function of the soliton size $R$ in
units of $1/M$.
\label{fig:eps-real}}
\end{figure}

The numerical dependence of the eigenvalue as a function of the
soliton size for the $0^+$ state in the toy model is depicted in
Fig.~\ref{fig:eps-real}. For large values of the soliton size at fixed
mass we get
\begin{eqnarray}
\epsilon_0 (R) \to \frac{1}{R} 
\end{eqnarray}
which means according to the scaling property that this behavior also
holds true for large mass and fixed radius. The solution is always in
the positive energy region and never dives into the sea (despite
having topological number 1 -- this was actually the point of
Ref.~\cite{Zhao:1988in}). For $R\to \infty $ one gets $\epsilon_0
(R)\to 0^+ $. The reason is related to the fact that always $\sin
\theta =0$ for the profile, and we have a scalar coupling for which $
H \gamma_0 \gamma_5 = - \gamma_0 \gamma_5 H $ and the spectrum is
symmetric.

The threshold value for having a bound state is $ (M R)^2 > 3/4$.
Note that since $ \kappa
R = x $ changes sign across the threshold value, the bound state
becomes virtual (exponentially growing). Thus, the analytic
continuation of the unbound valence state for $ R^2 M^2 < 3 / 4 $ is
an exponentially growing state.

Equation~(\ref{eq:tra})  defines implicitly a relation between $x$
and $x_0$. Actually, we can compute the inverse function
\begin{eqnarray}
x_0^2 = \frac{x^4\left( \coth(x)+1\right)^2}{4
(x+1)\left( x \coth(x)-1\right)} \label{eq:squared}
\end{eqnarray} 
For $x > 0 $ the function on the r.h.s. is a monotonously increasing
function and thus the inverse is uniquely defined. The minimum takes
place at $x=0$ and assumes the value of $3/4$. Hence there are no
positive energy bound state solutions for $x_0^2 \le 3/4$. For $ x_0^2
= 3/4 $ we have a negative energy solution $ \kappa R = x \simeq
-0.848 $ and $ \epsilon \simeq -0.175 /R $, of no concern here. For
$0.512 \le x_0^2 \le 3/4 $ we have a positive energy state with $
\kappa < 0 $. For $0.512 > x_0^2 $ the value of $ \kappa $ becomes
complex. For $ x_0 \to 0 $ we get $ x \to - \infty $. 

\begin{figure}[]
\includegraphics[width=6.5cm]{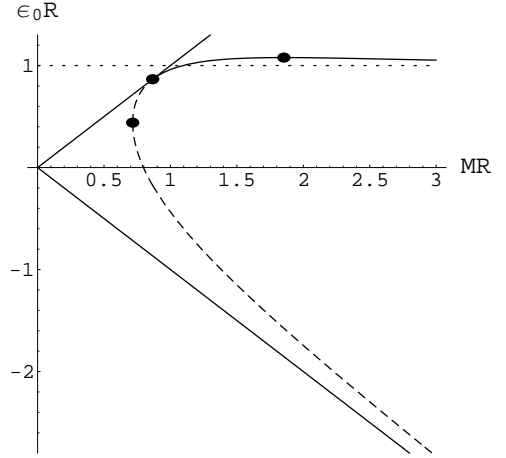} 
\caption{Solid straight lines are $\epsilon=\pm M$, the solid curve is
the bound state, the dashed curve is the virtual state. Blobs indicate
the position of the saddle, the emergence of a bound state, and the
bifurcation point of two real virtual states into two complex
states. For the bound state $\epsilon_0 R$ approaches unity at $M R
\to \infty$, while for the virtual branch it approaches the negative
continuum.
\label{fig:toybif}}
\end{figure}

In summary, for the real-mass case we have
\begin{itemize} 
\item For $ 0.866 < x_0$ we have one bound state solution.  
\item For $ 0.797 < x_0 < 0.866 $ we have one virtual state solution.
\item For $ 0.712 < x_0< 0.797 $ we have two virtual states. For $x_0=
0.7968 $ one at $ x=-0.797 $ and the other one at $ x= -0.196 $, the
higher one being a continuation of the previous case.
\item For $ x_0 < 0.716 $ the two solutions collide at $ x=-0.574$
and bifurcate into the complex plane.
\end{itemize} 
So clearly for $x_0 < 0.797 $ we have a cut in the $\kappa$ plane.
The situation for the real mass case has been displayed in
Fig.~\ref{fig:toybif}. This plot should be compared with the
exponential soliton profile of Fig.~\ref{fig:valence}.

\begin{figure}[]
\includegraphics[width=8.5cm]{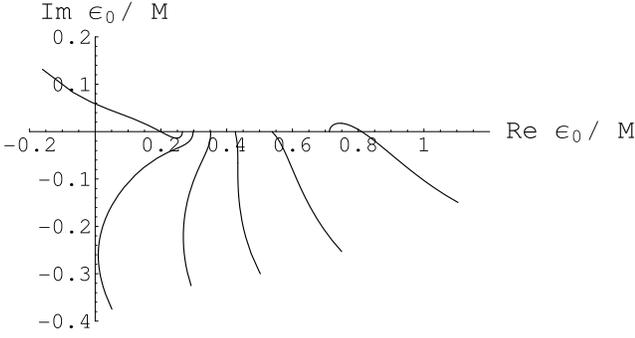} 
\caption{analyticly continued complex valence eigenvalue $
\epsilon_0 (R,w)$ in the toy model plotted as a function of the complex
mass $ \w=e^{i \alpha}$ for $\alpha$ in the first quadrant, $0 \le
\alpha \le \pi /2$. The axes represent ${\rm Re}(\epsilon_0)$ and
${\rm Im}(\epsilon_0)$. Subsequent curves, from right to left,
correspond to increasing values of $R = 2,2.5,3,3.5,3.9 $. The points
on the real axis are for $\alpha=0$, and the value of $\alpha$
parameterizes the curves, which evolve away from the real axis.  The
other quadrants are obtained by reflection about the real axis.
\label{fig:eps-complex}}
\end{figure}

We can now proceed to the complex plane in $\w$. The obvious way to
define the eigenvalue $\epsilon_0 (\w)$ as a function of the complex
variable $\w$ is by analytic continuation, since we want to follow
the evolution of the state in the complex-mass plane. For illustration
purposes we plot in Fig.~\ref{fig:eps-complex} the corresponding
analyticly continued eigenvalues when $\w=e^{i \alpha}$ for several
soliton sizes $R$. Equivalently, we consider the function of the
variable $ x_0^2$. Thus, we take the solutions which were defined for
real $M$ fulfilling the boundary conditions and arrive at the
transcendental equation (\ref{eq:tra}), which now defines implicitly
$\epsilon_0 (\w)$.\footnote{According to the implicit function theorem the
equation $y=f(x)$ can be inverted around the solution $y_0=f(x_0)$ if
$f'(x_0) \neq 0$. Critical points for which $f'(x_0)=0$ define
bifurcation branches if $f''(x_0) \neq 0$.} Thus, if we differentiate
with respect to $ x$ the implicit function we get the equation for the
critical points
\begin{eqnarray} 
\frac{2 t^4 +t^3 - 4 t^2 -2 t +1 + (2 t +3) \sqrt{1-t^2} }{t
\sqrt{1-t^2} (1+ \sqrt{1-t^2})^2} = 0,
\end{eqnarray} 
where for simplicity the variable $ t = x/x_0$ has been
introduced. This is an algebraic equation with solutions $t=t_n $. We
find numerically
\begin{eqnarray} 
t_n = x/x_0 &=& -1.2741 \pm i 0.1174 \, , \, 0.9092, \, \nonumber \\ 
&& 1.3898 \pm i 0.4832, \dots,
\end{eqnarray} 
which forms a bundle of solutions  
\begin{eqnarray}
t_n x_0 \coth (t_n x_0 ) = \frac{x_0 t_n ( 1 -\sqrt{1 - t_n^2}) + 2
t_n }{\sqrt{1-t_n^2} + 1}. 
\end{eqnarray}
We see that for any value of $t_n$ there are infinitely many solutions
for $x_0 $. Roughly, they are located along a straight line.  For
instance, for $ x/x_0= 0.9092 $ the critical points in $ x_0 $ are
located at about $ x_0 = 0.5 + i n \pi $ for large-$N$. So, the whole
critical points structure looks like five straight lines running in
all directions.
An alternative technique to analyze this kind of problems is by using the modular
landscape of the exponential as explained in detail in
Refs.~\cite{Kok:1980fw,Badalian:1981xj}.

The upshot of this study is that $ \epsilon (\w) $ has only critical
points on the second Riemann sheet. This means that whenever the bound
state becomes unbound we are crossing the branch cut, which happens
for sufficiently small values of $\w$. For the spectral integral that
would set the condition $ M_c < M_V/2 $ not to have cut intersection
between the spectral function and the energy
function.\footnote{Obviously, the smaller $M_c$ the better.  Now, it
would be interesting to see if there are chiral angles $ \theta(r) $
for which one always has a bound state. In such a case cuts would not
intersect. This is not such a strange situation, since e.g. in one
dimensional quantum mechanics, any tiny attractive potential generates
immediately a bound state. Also, in 1+1 dimensions the chiral angle $
\theta (x) = \alpha {\rm sign} (x) $ has always a bound state $
\epsilon = M \cos (\alpha) $~\cite{MacKenzie:1984mz}.}

\section{Computing the spectral integral  for the valence contribution first}  
\label{sec:spec-first}

In this appendix we show how our basic result for the valence
contribution, Eq.~(\ref{eq:pres}), 
can be reproduced by evaluating the spectral integral
first. In fact, experience with perturbative calculations of
quantities such as $ f_\pi $ or the pion form factor suggest that
there is no danger in computing the $\w$ integral first. At finite
temperature this is actually the only way to get analytic results
\cite{Megias:2004hj}. To compute the troublesome spectral integral
first in the soliton case, 
we transform
the scalar-pseudoscalar coupling into a vector-axial derivative
coupling and a constant mass. We write
\begin{eqnarray} 
\left( i\slashchar{\partial} - \omega U^5 \right) = u^5 \left(
i\slashchar{\partial} + u^{-5} i\slashchar{\partial} u^{-5} - \omega
\right) u^5
\end{eqnarray} 
where $ u^2 = U $. This is a field dependent axial
rotation and at the level of the determinant of the Dirac operator
generates the anomalous Wess-Zumino term in the SU(3) case. Then, we
get
\begin{eqnarray} 
&& S_{a a'} (x,x' ) = \int d \w \rho(\w) \\ &\times& \langle
x| \left[ u^{-5} \left( i\slashchar{\partial} + u^{-5}
i\slashchar{\partial} u^{-5} - \omega \right)^{-1} u^{-5} \right]_{a
a'}|x'\rangle \nonumber \\ &=& \left[ \langle x| u^{-5} \left\{ \int d \w
\frac{\rho(\w)}{ i\slashchar{\partial} + u^{-5} i\slashchar{\partial}
u^{-5} - \omega} \right\} u^{-5} |x'\rangle \right]_{a,a'} \nonumber \\ &=&
\left[ u(x)^{-5} \langle x| S \left( i\slashchar{\partial} + u^{-5}
i\slashchar{\partial} u^{-5} \right) |x'\rangle u(x')^{-5} 
\right]_{a,a'}. \nonumber 
\label{eq:eff_prop_2} 
\end{eqnarray}  
Now, in the meson dominance model the propagator function is given by
\begin{eqnarray}
S(\p) = \int_C d \w \frac{\rho(\w)}{\slashchar{p}-\w}= 2 \pi i \rho
(\slashchar{p}) + \frac1{\slashchar{p}}. \label{just}
\end{eqnarray} 
This relation is obtained for the closed contour under
the assumption that $\rho(\w)$ has a pole only at $\w=0$, 
which is clearly the case 
of the meson dominance model (\ref{rhos}). Equation (\ref{just}) is just a 
functional relation, and does not depend on the fact that the variable
$\slashchar{p}$ is a matrix. We have indeed computed the spectral
integral, but for this formula to be of any use, we have to look for
the eigenvalue problem of the chirally rotated Dirac operator
\begin{eqnarray} 
\left( i\slashchar{\partial} + u^{-5} i\slashchar{\partial} u^{-5}
\right) \Phi_n = M_n \Phi_n .
\end{eqnarray} 
This eigenvalue problem may be tricky in practice since the operator
is not normal, {\em i.e.} the adjoint operator does not commute with the
original operator even if we go to the Euclidean space (where the
derivative and the vector part of the coupling are anti self-adjoint
and the axial part is self adjoint). This means that $M_n $ may be
complex in general.

Undoing field dependent chiral rotation we get
\begin{eqnarray} 
\left( i\slashchar{\partial} - M_n U^5 \right) \Psi_n &=& 0 , 
\end{eqnarray} 
where 
\begin{eqnarray} 
\Psi_n = u^{-5}  \Phi_n \, , \qquad
\Psi_n^\dagger  =  \Phi_n^\dagger  u^{-5} .
\end{eqnarray} 
Thus, the eigenvalue problem corresponds to the search for the
(eigen)masses of the original Dirac operator which yield a zero mode of the
four dimensional Dirac operator. 
This equation looks as an on-shell condition for the bound
quarks. In the stationary field case we have, $ \Psi_n (\vec x, t ) =
e^{-i \epsilon t} \psi_n (\vec x) $. At this point we must make a
choice on the values of $\epsilon$. A natural condition would be to
request an exponential damping in the Euclidean space. Thus, we take $
\epsilon $ real and positive for forward propagating states, $ t> 0$,
and negative for backward propagating states. (Another possibility
that would be to take a contour with conditions on ${\rm Re}\, 
\epsilon $),
\begin{eqnarray} 
\left[ -i \alpha \cdot \nabla + M_n \beta U^5 \right] \psi_n = \epsilon
\psi_n .
\end{eqnarray}
Thus, here we encounter just the opposite problem to the case of the
spectral mass problem; for a given energy we have to look for the
eigenmasses, $M_n (\epsilon)$ (instead of $\epsilon_n (\w) $). 
Actually, this looks like an isospectral problem; given the energy
how many chiral couplings possess the same energy. Thus, the
eigenvectors of this problem are not exactly the eigenvectors of the
standard Dirac Hamiltonian problem.  Note also that because we are
searching for zero modes of the Dirac operator its eigenvectors are
also eigenvectors of the Dirac Hamiltonian, which is not the case for
non-zero eigenvalues, {\em i.e.} $ \gamma_0 (i \partial_t -H ) \Psi_n =
\lambda_n \Psi_n $ with $ \lambda_n \neq 0 $ cannot be solved by a
eigenstate of $H$ because of the $\gamma_0$. Clearly, multivaluedness
issues may become quite relevant when discussing the possible
equivalence of both problems. On the other hand, $H^\dagger = H $ if
and only if $M_n (\epsilon)$ is real.  For such a case if both $M_n $
and $\epsilon$ are real, then bound states happen for $ -M_n <
\epsilon < M_n $.  In general $M_n$ is complex, since even in the free
case, $U=1$, we have $ M_n (\epsilon) = \sqrt{\epsilon^2 - k_n^2}$,
with $k_n$ denoting the momentum quantized due to box boundary conditions.
Then, we are lead to
\begin{eqnarray} 
S_{a a'} (x,x' ) &=& u^{-5} (x) \langle x| S \left(
i\slashchar{\partial} + u^{-5} i\slashchar{\partial} u^{-5}
\right)|x'\rangle u^{-5} (x') \nonumber \\ &=& u^{-5} (x) \sum_n \Phi_n (x) S
\left( M_n \right) \Phi_n^\dagger (x') u^{-5} (x') \nonumber \\ &=& \sum_n
\Psi_n (x) S \left( M_n \right) \Psi_n^\dagger (x') \\ &=&
\int_{-\infty}^{\infty} \frac{d \epsilon}{2\pi} \sum_n \psi_n (x)
\left[ e^{i \epsilon (t-t')} \theta(\epsilon) \theta (t'-t) \right . \nonumber \\ &+&
\left .  e^{-i
\epsilon (t-t')} \theta(-\epsilon) \theta (t-t') \right] S \left( M_n (
\epsilon ) \right)  \psi_n^\dagger (x'). \nonumber
\label{eq:eff_prop_3} 
\end{eqnarray}  
In this representation, the states propagating forward in time have
been chosen to be those of positive energy and we are led to the basic
integral
\begin{eqnarray} 
\int_{0}^{\infty} \frac{d \epsilon}{2\pi} e^{- \tau \epsilon } S
\left( M_n ( \epsilon ) \right) = \int_{0}^{\infty} \frac{d
\epsilon}{2\pi} e^{- \tau \epsilon + \log S \left( M_n ( \epsilon )
\right)},
\end{eqnarray}  
with $\tau \to \infty $. The function $S(M)$ of Eq.~(\ref{just}) has no 
poles but branch
cuts starting at $M = \pm M_V / 2 $.  The stationary points are determined from
the equation 
\begin{eqnarray}
\tau = \frac{\partial \log S \left( M_n ( \epsilon ) \right) }{\partial
\epsilon} = M_n' (\epsilon) \frac{S' \left( M_n ( \epsilon ) \right) }{S
\left( M_n ( \epsilon ) \right) }.
\end{eqnarray} 
Note that a real $S(M)$ requires $M_n < M_V/2$, thus for $ M_n (\epsilon) \neq
M_V /2 $ we must have a divergent $ M_n ' (\epsilon) \to \pm \infty$. 
Obviously, the smallest positive $\epsilon$ (the valence orbit)
fulfilling $1/ M_n' (\epsilon)=0 $ dominates, and we get
the relation
\begin{eqnarray} 
\epsilon_{0} = {\rm min} \, \epsilon \, 
\Big|_{ M_n' (\epsilon) = \infty, \; \epsilon > 0 }\qquad , 
\end{eqnarray}  
in agreement with our result of Sect.~(\ref{sec:formal}), 
which is the result equivalent to (\ref{sadcon}) written in terms of the inverse function.

\section{Evaluation of the Dirac sea contribution to soliton 
observables \label{app:trick}}

Dirac sea contributions to observables involve the complex mass
integral. However, it is possible to rewrite this contribution as a real mass
distribution. 
The vector spectral density $\rho_V$ can be written as
\begin{equation}
\rho_V(\omega)=\left . \frac{d^2}{du^2} \bar \rho_V(\omega) \right
|_{u=1/M_V^2}, \label{rhov2} 
\end{equation}
where 
\begin{equation}
  \bar \rho_V(\omega) =\frac{1}{2 \pi i} \frac{1}{12 \omega^5}
\frac{1}{(1-4 u \omega^2)^{1/2}}. 
\end{equation}
The function $\bar \rho_V(\omega)$ has the property that its integrals
with smooth functions  vanish along the parts of the contour $C$
encircling the branch points of Fig. \ref{contour0}, as well as are
finite along the cut. Similarly, for the scalar part, $\rho_S$, we
have
\begin{equation}
\rho_S(\omega)=\left . \frac{d^2}{du^2} \bar \rho_S(\omega) \right
|_{u=1/M_V^2}, \;\;  
\end{equation}
with 
\begin{equation}
\bar \rho_S(\omega) =\frac{1}{2 \pi i}
\frac{1}{12 \omega^4}  \frac{12 \rho'_3}{M_S^4(1-4 u
\omega^2)^{1/2}}. \label{rhovs}
\end{equation}
We now define
\begin{eqnarray}
{\bar A}_{{\rm sea},V}(u)&=& 2 \int_{1/(2 \sqrt{u})}^\infty \!\!\!\!\!\!\! d \omega {\rm
disc} [ \bar \rho_V(\omega) ]  \frac{A_{\rm sea}(\omega)+A_{\rm sea}(-\omega)}{2},
\nonumber \\ {\rm disc}[ \bar \rho_V(\omega)] &=& \frac{1}{12 \pi
\omega^5 \sqrt{4u \omega^2 -1} }, \label{discv}
\end{eqnarray}
(the factor of 2 comes from the two sections of $C$) and, similarly,
\begin{eqnarray}
{\bar A}_{{\rm sea},S}(u)&=& 2 \int_{1/(2 \sqrt{u})}^\infty \!\!\!\!\!\!\! d \omega {\rm
disc} [ \bar \rho_S(\omega) ]  \frac{A_{\rm sea}(\omega)-A_{\rm s}(-\omega)}{2},
\nonumber \\ {\rm disc}[ \bar \rho_S(\omega)] &=& \frac{\rho'_3}{\pi
M_S^4 \omega^4 \sqrt{4u \omega^2 -1} }. \label{discs}
\end{eqnarray}
The Dirac-sea contributions to observables are now obtained from 
\begin{eqnarray}
A_{\rm sea}=\left . \frac{d^2}{du^2}[\bar A_{{\rm sea},V}(u)+
\bar A_{{\rm sea},S}(u)] \right |_{u=1/M_V^2}. \label{ph}
\end{eqnarray}
Thus, we have managed to ``put the model on the real axis'', at the
only expense of the need  of differentiation with respect to $u$ at
the end of the calculation. As a bonus we also get a very high 
degree of convergence at large $\omega$, 
since  ${\rm disc}[ \bar \rho_V(\omega)] \sim
1/\omega^6$ and ${\rm disc}[ \bar \rho_S(\omega)] \sim 1/\omega^5$. 
It is useful to get rid of the integrable singularities in
Eqs.~(\ref{discv},\ref{discs}) by means of introducing the  variable
\begin{eqnarray}
z=\sqrt{4 u \omega^2 -1}, \;\;\; \omega=\frac{\sqrt{z^2+1}}{2 \sqrt{u}} . \label{z}
\end{eqnarray} 
Then
\begin{eqnarray}
{\bar A}_{{\rm sea},V}(u)&=& \frac{8 u^2 }{3\pi}  
\int_0^\infty \frac{dz}{(z^2+1)^3} \\ 
&\times& \frac12  \left[
{A_{\rm sea}}\left( 
\frac{\sqrt{z^2+1}}{2 \sqrt{u}} \right )+ {A_{\rm sea}}\left( 
-\frac{\sqrt{z^2+1}}{2 \sqrt{u}} \right ) \right], \nonumber \label{disczv}
\end{eqnarray}
and
\begin{eqnarray}
{\bar A}_{{\rm sea},S}(u)&=& - \frac{8 \pi \langle {\bar q} q \rangle }{N_c M_V^4} (4u)^{3/2} 
\int_0^\infty \frac{dz} {(z^2+1)^{5/2}} \\ &\times& 
\frac12 \left [ {A_{\rm sea}}\left( 
\frac{\sqrt{z^2+1}}{2 \sqrt{u}} \right )- {A_{\rm sea}}\left( 
-\frac{\sqrt{z^2+1}}{2 \sqrt{u}} \right ) \right ]. \nonumber \label{disczs}
\end{eqnarray}
The Dirac Hamiltonian has the form
\begin{eqnarray} 
H(\omega,m)= -i {\alpha \cdot \nabla}+ \beta m+ \beta \omega U^5, \label{dirac}
\end{eqnarray}
with the corresponding Dirac equation
\begin{eqnarray}
H(\omega,m) |i;\omega,m \rangle= \epsilon_i(\omega,m)  |i;\omega,m \rangle . \label{diraceq}
\end{eqnarray}
It has the property 
\begin{eqnarray}
\gamma_5 H(\omega,m) \gamma_5&=& H(-\omega,-m), \nonumber \\
\gamma_5 |i;\omega,m \rangle &=& |i;-\omega,-m \rangle . \label{dirac5}
\end{eqnarray}
Thus $\gamma_5$ flips the sign of the current quark mass $m$ and $\omega$. 
Obviously, the spectrum of the operator is unchanged under
this similarity transformation. Nevertheless, it is convenient 
in the numerical work
to deal with positive values of $\omega$ only.
Thus, any one body observable at negative $\omega$ can be transformed according to 
\begin{eqnarray}
A_{\rm sea}(\omega)&=&\sum_i \langle i;\omega,m | a(\omega,m) |i;\omega,m \rangle 
\\ &=&
\sum_i \langle i;\omega,m | \gamma_5 \gamma_5 a(\omega,m) \gamma_5 \gamma_5  |i;\omega,m \rangle \nonumber \\
&=&\sum_i \langle i;-\omega,-m | \gamma_5 a(\omega,m) \gamma_5 
|i;-\omega,-m \rangle. \nonumber \label{trans}
\end{eqnarray}
Some examples of $\gamma_5 a(\omega, m) \gamma_5$ encountered in the 
evaluation of observables are 
$\gamma_5 H(\omega,m) \gamma_5=H(-\omega,-m)$, 
$\gamma_5 \beta \gamma_5=-\beta$, {\em etc.} 
The formula for the energy following from the above formulation 
is given in Eq.~(\ref{E},\ref{Etr}).

\section{Instability of the linear model \label{exlin}}

Although SQM is only constructed in the nonlinear case, it looks 
tempting to extend it in the spirit of the original bosonized NJL model 
to a linear version where the fields may depart 
from the chiral circle,
\begin{eqnarray}
I&=&i \int_C d \omega \rho(\omega) {\rm Tr} \log (i \slashchar{\partial}-m-
w (s + i \gamma_5 {\bm \tau} \cdot {\bm p}) ) \nonumber \\ 
&+& \frac{1}{2g^2} \int d^4 x (s^2+{\bm p}^2). \nonumber \\ \label{linear} 
\end{eqnarray}
Here $s$ and ${\bm p}$ denote the scalar-isoscalar and
pseudoscalar-isovector fields, and $g$ is a coupling constant.  The
meson fields $s$ and $p$ are {dimensionless}.  
In the chiral limit and in the vacuum ($s=1$, $p=0$) we find from the 
Euler-Lagrange equation for the $s$ field the condition 
\begin{eqnarray}
\hspace{-9mm} \frac{s}{g^2}&=&-i\int_C d \omega 
\rho(\omega) {\rm Tr}\frac{\omega}{i \slashchar{\partial}-\omega s} 
\nonumber \\ &=& -4i N_c N_f \int_C d \omega \rho(\omega) 
\int \frac{d^4 k}{(2\pi)^4} \frac{\omega^2 s}{k^2-(\omega s)^2}.
\nonumber \\ \label{svac}
\end{eqnarray}
With the explicit form of the meson-dominance spectral function (\ref{rhos})
and the techniques of Ref.~\cite{RuizArriola:2003bs} we find 
\begin{eqnarray}
\frac{1}{g^2}=
\frac{N_c N_f M_V^4 }{48 \pi^2}. \label{gcond}
\end{eqnarray}
The effective potential assumes the form 
\begin{eqnarray}
V=\frac{I}{\int d^4 x} = \frac{N_c N_f M_V^4 }{192 \pi^2}(2s^2-s^4), 
\end{eqnarray}
where the $s^4$ term originates form the first term in Eq.~(\ref{linear}).
This corresponds to an inverted Mexican Hat, and clearly displays instability. 
Therefore the linear version of the model (\ref{linear}) does not make sense.

It is worthwhile to mention that this feature should not be regarded as 
specific to the spectral regularization. 
The inverted potential also arises when renormalizing the fermion determinant
with the help of the $\zeta$-function regularization in the local NJL model.
Therefore one needs to use from the outset the non-linear realization of 
chiral symmetry on its own, and it cannot be treated as an approximation 
to the linear model.
In that regard we also note that the soliton instability in
linear NJL models has been found in Ref.~\cite{Sieber:1992uj}. 


\begin{thebibliography}{88}
\expandafter\ifx\csname natexlab\endcsname\relax\def\natexlab#1{#1}\fi
\expandafter\ifx\csname bibnamefont\endcsname\relax
  \def\bibnamefont#1{#1}\fi
\expandafter\ifx\csname bibfnamefont\endcsname\relax
  \def\bibfnamefont#1{#1}\fi
\expandafter\ifx\csname citenamefont\endcsname\relax
  \def\citenamefont#1{#1}\fi
\expandafter\ifx\csname url\endcsname\relax
  \def\url#1{\texttt{#1}}\fi
\expandafter\ifx\csname urlprefix\endcsname\relax\def\urlprefix{URL }\fi
\providecommand{\bibinfo}[2]{#2}
\providecommand{\eprint}[2][]{\url{#2}}

\bibitem[{\citenamefont{Skyrme}(1961)}]{Skyrme:1961vq}
\bibinfo{author}{\bibfnamefont{T.~H.~R.} \bibnamefont{Skyrme}},
  \bibinfo{journal}{Proc. Roy. Soc. Lond.} \textbf{\bibinfo{volume}{A260}},
  \bibinfo{pages}{127} (\bibinfo{year}{1961}).

\bibitem[{\citenamefont{'t~Hooft}(1974)}]{'tHooft:1973jz}
\bibinfo{author}{\bibfnamefont{G.}~\bibnamefont{'t~Hooft}},
  \bibinfo{journal}{Nucl. Phys.} \textbf{\bibinfo{volume}{B72}},
  \bibinfo{pages}{461} (\bibinfo{year}{1974}).

\bibitem[{\citenamefont{Witten}(1979)}]{Witten:1979kh}
\bibinfo{author}{\bibfnamefont{E.}~\bibnamefont{Witten}},
  \bibinfo{journal}{Nucl. Phys.} \textbf{\bibinfo{volume}{B160}},
  \bibinfo{pages}{57} (\bibinfo{year}{1979}).

\bibitem[{\citenamefont{Goldstone and Wilczek}(1981)}]{Goldstone:1981kk}
\bibinfo{author}{\bibfnamefont{J.}~\bibnamefont{Goldstone}} \bibnamefont{and}
  \bibinfo{author}{\bibfnamefont{F.}~\bibnamefont{Wilczek}},
  \bibinfo{journal}{Phys. Rev. Lett.} \textbf{\bibinfo{volume}{47}},
  \bibinfo{pages}{986} (\bibinfo{year}{1981}).

\bibitem[{\citenamefont{Witten}(1983)}]{Witten:1983tw}
\bibinfo{author}{\bibfnamefont{E.}~\bibnamefont{Witten}},
  \bibinfo{journal}{Nucl. Phys.} \textbf{\bibinfo{volume}{B223}},
  \bibinfo{pages}{422} (\bibinfo{year}{1983}).

\bibitem[{\citenamefont{Balachandran et~al.}(1983)\citenamefont{Balachandran,
  Nair, Rajeev, and Stern}}]{Balachandran:1982cb}
\bibinfo{author}{\bibfnamefont{A.~P.} \bibnamefont{Balachandran}},
  \bibinfo{author}{\bibfnamefont{V.~P.} \bibnamefont{Nair}},
  \bibinfo{author}{\bibfnamefont{S.~G.} \bibnamefont{Rajeev}},
  \bibnamefont{and} \bibinfo{author}{\bibfnamefont{A.}~\bibnamefont{Stern}},
  \bibinfo{journal}{Phys. Rev.} \textbf{\bibinfo{volume}{D27}},
  \bibinfo{pages}{1153} (\bibinfo{year}{1983}).

\bibitem[{\citenamefont{Adkins et~al.}(1983)\citenamefont{Adkins, Nappi, and
  Witten}}]{Adkins:1983ya}
\bibinfo{author}{\bibfnamefont{G.~S.} \bibnamefont{Adkins}},
  \bibinfo{author}{\bibfnamefont{C.~R.} \bibnamefont{Nappi}}, \bibnamefont{and}
  \bibinfo{author}{\bibfnamefont{E.}~\bibnamefont{Witten}},
  \bibinfo{journal}{Nucl. Phys.} \textbf{\bibinfo{volume}{B228}},
  \bibinfo{pages}{552} (\bibinfo{year}{1983}).

\bibitem[{\citenamefont{Holzwarth and Schwesinger}(1986)}]{Holzwarth:1985rb}
\bibinfo{author}{\bibfnamefont{G.}~\bibnamefont{Holzwarth}} \bibnamefont{and}
  \bibinfo{author}{\bibfnamefont{B.}~\bibnamefont{Schwesinger}},
  \bibinfo{journal}{Rept. Prog. Phys.} \textbf{\bibinfo{volume}{49}},
  \bibinfo{pages}{825} (\bibinfo{year}{1986}).

\bibitem[{\citenamefont{Zahed and Brown}(1986)}]{Zahed:1986qz}
\bibinfo{author}{\bibfnamefont{I.}~\bibnamefont{Zahed}} \bibnamefont{and}
  \bibinfo{author}{\bibfnamefont{G.~E.} \bibnamefont{Brown}},
  \bibinfo{journal}{Phys. Rept.} \textbf{\bibinfo{volume}{142}},
  \bibinfo{pages}{1} (\bibinfo{year}{1986}).

\bibitem[{\citenamefont{Weigel}(1996)}]{Weigel:1995cz}
\bibinfo{author}{\bibfnamefont{H.}~\bibnamefont{Weigel}},
  \bibinfo{journal}{Int. J. Mod. Phys.} \textbf{\bibinfo{volume}{A11}},
  \bibinfo{pages}{2419} (\bibinfo{year}{1996}).

\bibitem[{\citenamefont{Schechter and Weigel}(1999)}]{Schechter:1999hg}
\bibinfo{author}{\bibfnamefont{J.}~\bibnamefont{Schechter}} \bibnamefont{and}
  \bibinfo{author}{\bibfnamefont{H.}~\bibnamefont{Weigel}}
  (\bibinfo{year}{1999}), \eprint{hep-ph/9907554}.

\bibitem[{\citenamefont{Weigel}(2004)}]{Weigel:2004px}
\bibinfo{author}{\bibfnamefont{H.}~\bibnamefont{Weigel}},
  \bibinfo{journal}{Eur. Phys. J.} \textbf{\bibinfo{volume}{A21}},
  \bibinfo{pages}{133} (\bibinfo{year}{2004}).

\bibitem[{\citenamefont{Chemtob}(1987)}]{Chemtob:1987ut}
\bibinfo{author}{\bibfnamefont{M.}~\bibnamefont{Chemtob}},
  \bibinfo{journal}{Nucl. Phys.} \textbf{\bibinfo{volume}{A473}},
  \bibinfo{pages}{613} (\bibinfo{year}{1987}).

\bibitem[{\citenamefont{Birse and Banerjee}(1984)}]{Birse:1983gm}
\bibinfo{author}{\bibfnamefont{M.~C.} \bibnamefont{Birse}} \bibnamefont{and}
  \bibinfo{author}{\bibfnamefont{M.~K.} \bibnamefont{Banerjee}},
  \bibinfo{journal}{Phys. Lett.} \textbf{\bibinfo{volume}{B136}},
  \bibinfo{pages}{284} (\bibinfo{year}{1984}).

\bibitem[{\citenamefont{Birse and Banerjee}(1985)}]{Birse:1984js}
\bibinfo{author}{\bibfnamefont{M.~C.} \bibnamefont{Birse}} \bibnamefont{and}
  \bibinfo{author}{\bibfnamefont{M.~K.} \bibnamefont{Banerjee}},
  \bibinfo{journal}{Phys. Rev.} \textbf{\bibinfo{volume}{D31}},
  \bibinfo{pages}{118} (\bibinfo{year}{1985}).

\bibitem[{\citenamefont{Kahana et~al.}(1984)\citenamefont{Kahana, Ripka, and
  Soni}}]{Kahana:1984dx}
\bibinfo{author}{\bibfnamefont{S.}~\bibnamefont{Kahana}},
  \bibinfo{author}{\bibfnamefont{G.}~\bibnamefont{Ripka}}, \bibnamefont{and}
  \bibinfo{author}{\bibfnamefont{V.}~\bibnamefont{Soni}},
  \bibinfo{journal}{Nucl. Phys.} \textbf{\bibinfo{volume}{A415}},
  \bibinfo{pages}{351} (\bibinfo{year}{1984}).

\bibitem[{\citenamefont{Kaelbermann and Eisenberg}(1984)}]{Kaelbermann:1984xn}
\bibinfo{author}{\bibfnamefont{G.}~\bibnamefont{Kaelbermann}} \bibnamefont{and}
  \bibinfo{author}{\bibfnamefont{J.~M.} \bibnamefont{Eisenberg}},
  \bibinfo{journal}{Phys. Lett.} \textbf{\bibinfo{volume}{B139}},
  \bibinfo{pages}{337} (\bibinfo{year}{1984}).

\bibitem[{\citenamefont{Broniowski and Banerjee}(1985)}]{Broniowski:1984zd}
\bibinfo{author}{\bibfnamefont{W.}~\bibnamefont{Broniowski}} \bibnamefont{and}
  \bibinfo{author}{\bibfnamefont{M.~K.} \bibnamefont{Banerjee}},
  \bibinfo{journal}{Phys. Lett.} \textbf{\bibinfo{volume}{B158}},
  \bibinfo{pages}{335} (\bibinfo{year}{1985}).

\bibitem[{\citenamefont{Broniowski and Banerjee}(1986)}]{Broniowski:1985kj}
\bibinfo{author}{\bibfnamefont{W.}~\bibnamefont{Broniowski}} \bibnamefont{and}
  \bibinfo{author}{\bibfnamefont{M.~K.} \bibnamefont{Banerjee}},
  \bibinfo{journal}{Phys. Rev.} \textbf{\bibinfo{volume}{D34}},
  \bibinfo{pages}{849} (\bibinfo{year}{1986}).

\bibitem[{\citenamefont{Diakonov et~al.}(1988)\citenamefont{Diakonov, Petrov,
  and Pobylitsa}}]{Diakonov:1987ty}
\bibinfo{author}{\bibfnamefont{D.}~\bibnamefont{Diakonov}},
  \bibinfo{author}{\bibfnamefont{V.~Y.} \bibnamefont{Petrov}},
  \bibnamefont{and} \bibinfo{author}{\bibfnamefont{P.~V.}
  \bibnamefont{Pobylitsa}}, \bibinfo{journal}{Nucl. Phys.}
  \textbf{\bibinfo{volume}{B306}}, \bibinfo{pages}{809} (\bibinfo{year}{1988}).

\bibitem[{\citenamefont{Meissner et~al.}(1988)\citenamefont{Meissner,
  Ruiz~Arriola, Grummer, Mavromatis, and Goeke}}]{Meissner:1988bg}
\bibinfo{author}{\bibfnamefont{T.}~\bibnamefont{Meissner}},
  \bibinfo{author}{\bibfnamefont{E.}~\bibnamefont{Ruiz~Arriola}},
  \bibinfo{author}{\bibfnamefont{F.}~\bibnamefont{Grummer}},
  \bibinfo{author}{\bibfnamefont{H.}~\bibnamefont{Mavromatis}},
  \bibnamefont{and} \bibinfo{author}{\bibfnamefont{K.}~\bibnamefont{Goeke}},
  \bibinfo{journal}{Phys. Lett.} \textbf{\bibinfo{volume}{B214}},
  \bibinfo{pages}{312} (\bibinfo{year}{1988}).

\bibitem[{\citenamefont{Wakamatsu and Yoshiki}(1991)}]{Wakamatsu:1990ud}
\bibinfo{author}{\bibfnamefont{M.}~\bibnamefont{Wakamatsu}} \bibnamefont{and}
  \bibinfo{author}{\bibfnamefont{H.}~\bibnamefont{Yoshiki}},
  \bibinfo{journal}{Nucl. Phys.} \textbf{\bibinfo{volume}{A524}},
  \bibinfo{pages}{561} (\bibinfo{year}{1991}).

\bibitem[{\citenamefont{Weigel et~al.}(1999)\citenamefont{Weigel, Ruiz~Arriola,
  and Gamberg}}]{Weigel:1999pc}
\bibinfo{author}{\bibfnamefont{H.}~\bibnamefont{Weigel}},
  \bibinfo{author}{\bibfnamefont{E.}~\bibnamefont{Ruiz~Arriola}},
  \bibnamefont{and} \bibinfo{author}{\bibfnamefont{L.~P.}
  \bibnamefont{Gamberg}}, \bibinfo{journal}{Nucl. Phys.}
  \textbf{\bibinfo{volume}{B560}}, \bibinfo{pages}{383} (\bibinfo{year}{1999}).

\bibitem[{\citenamefont{Birse}(1990)}]{Birse:1991cx}
\bibinfo{author}{\bibfnamefont{M.~C.} \bibnamefont{Birse}},
  \bibinfo{journal}{Prog. Part. Nucl. Phys.} \textbf{\bibinfo{volume}{25}},
  \bibinfo{pages}{1} (\bibinfo{year}{1990}).

\bibitem[{\citenamefont{Diakonov}(1996)}]{Diakonov:1995zi}
\bibinfo{author}{\bibfnamefont{D.}~\bibnamefont{Diakonov}},
  \bibinfo{journal}{Prog. Part. Nucl. Phys.} \textbf{\bibinfo{volume}{36}},
  \bibinfo{pages}{1} (\bibinfo{year}{1996}).

\bibitem[{\citenamefont{{C. V. Christov, A. Blotz, H.-C. Kim, P. V. Pobylitsa,
  T. Watabe, Th. Meissner, E. Ruiz Arriola, and K. Goeke}}(1996)}]{NJL:rev}
\bibinfo{author}{\bibnamefont{{C. V. Christov, A. Blotz, H.-C. Kim, P. V.
  Pobylitsa, T. Watabe, Th. Meissner, E. Ruiz Arriola, and K. Goeke}}},
  \bibinfo{journal}{{Prog. Part. Nucl. Phys.}} \textbf{\bibinfo{volume}{37}},
  \bibinfo{pages}{1} (\bibinfo{year}{1996}).

\bibitem[{\citenamefont{Alkofer et~al.}(1996)\citenamefont{Alkofer, Reinhardt,
  and Weigel}}]{Alkofer:rev}
\bibinfo{author}{\bibfnamefont{R.}~\bibnamefont{Alkofer}},
  \bibinfo{author}{\bibfnamefont{H.}~\bibnamefont{Reinhardt}},
  \bibnamefont{and} \bibinfo{author}{\bibfnamefont{H.}~\bibnamefont{Weigel}},
  \bibinfo{journal}{{Phys. Rep.}} \textbf{\bibinfo{volume}{265}},
  \bibinfo{pages}{139} (\bibinfo{year}{1996}).

\bibitem[{\citenamefont{Ripka}(1997{\natexlab{a}})}]{ripka:book}
\bibinfo{author}{\bibfnamefont{G.}~\bibnamefont{Ripka}},
  \emph{\bibinfo{title}{Quarks Bound by Chiral Fields}}
  (\bibinfo{publisher}{Clarendon Press}, \bibinfo{address}{Oxford},
  \bibinfo{year}{1997}{\natexlab{a}}).

\bibitem[{\citenamefont{Diakonov and Petrov}(1986)}]{Diakonov86}
\bibinfo{author}{\bibfnamefont{D.~I.} \bibnamefont{Diakonov}} \bibnamefont{and}
  \bibinfo{author}{\bibfnamefont{V.~Y.} \bibnamefont{Petrov}},
  \bibinfo{journal}{Nucl. Phys.} \textbf{\bibinfo{volume}{B 272}},
  \bibinfo{pages}{457} (\bibinfo{year}{1986}).

\bibitem[{\citenamefont{Roberts and Williams}(1994)}]{Roberts94}
\bibinfo{author}{\bibfnamefont{C.~D.} \bibnamefont{Roberts}} \bibnamefont{and}
  \bibinfo{author}{\bibfnamefont{A.~G.} \bibnamefont{Williams}},
  \bibinfo{journal}{Prog. Part. and Nucl. Phys.} \textbf{\bibinfo{volume}{33}},
  \bibinfo{pages}{475} (\bibinfo{year}{1994}).

\bibitem[{\citenamefont{Ball}(1990)}]{Ball90}
\bibinfo{author}{\bibfnamefont{R.~D.} \bibnamefont{Ball}},
  \bibinfo{journal}{Int. Journ. Mod. Phys.} \textbf{\bibinfo{volume}{A 5}},
  \bibinfo{pages}{4391} (\bibinfo{year}{1990}).

\bibitem[{\citenamefont{Ripka}(1997{\natexlab{b}})}]{Ripka97}
\bibinfo{author}{\bibfnamefont{G.}~\bibnamefont{Ripka}},
  \emph{\bibinfo{title}{Quarks Bound by Chiral Fields}}
  (\bibinfo{publisher}{Oxford University Press}, \bibinfo{address}{Oxford},
  \bibinfo{year}{1997}{\natexlab{b}}).

\bibitem[{\citenamefont{Praschifka et~al.}(1987)\citenamefont{Praschifka,
  Roberts, and Cahill}}]{Cahill87}
\bibinfo{author}{\bibfnamefont{J.}~\bibnamefont{Praschifka}},
  \bibinfo{author}{\bibfnamefont{C.~D.} \bibnamefont{Roberts}},
  \bibnamefont{and} \bibinfo{author}{\bibfnamefont{R.~T.}
  \bibnamefont{Cahill}}, \bibinfo{journal}{Phys. Rev.}
  \textbf{\bibinfo{volume}{D 36}}, \bibinfo{pages}{209} (\bibinfo{year}{1987}).

\bibitem[{\citenamefont{Holdom et~al.}(1989)\citenamefont{Holdom, Terning, and
  K.Verbeek}}]{Holdom89}
\bibinfo{author}{\bibfnamefont{B.}~\bibnamefont{Holdom}},
  \bibinfo{author}{\bibfnamefont{J.}~\bibnamefont{Terning}}, \bibnamefont{and}
  \bibinfo{author}{\bibnamefont{K.Verbeek}}, \bibinfo{journal}{Phys. Lett.}
  \textbf{\bibinfo{volume}{B 232}}, \bibinfo{pages}{351}
  (\bibinfo{year}{1989}).

\bibitem[{\citenamefont{Buballa and Krewald}(1992)}]{Krewald92}
\bibinfo{author}{\bibfnamefont{M.}~\bibnamefont{Buballa}} \bibnamefont{and}
  \bibinfo{author}{\bibfnamefont{S.}~\bibnamefont{Krewald}},
  \bibinfo{journal}{Phys. Lett.} \textbf{\bibinfo{volume}{B 294}},
  \bibinfo{pages}{19} (\bibinfo{year}{1992}).

\bibitem[{\citenamefont{Ball and Ripka}(1993)}]{Ripka93}
\bibinfo{author}{\bibfnamefont{R.~D.} \bibnamefont{Ball}} \bibnamefont{and}
  \bibinfo{author}{\bibfnamefont{G.}~\bibnamefont{Ripka}}, in
  \emph{\bibinfo{booktitle}{Many Body Physics (Coimbra 1993)}}, edited by
  \bibinfo{editor}{\bibfnamefont{C.}~\bibnamefont{Fiolhais}},
  \bibinfo{editor}{\bibfnamefont{M.}~\bibnamefont{Fiolhais}},
  \bibinfo{editor}{\bibfnamefont{C.}~\bibnamefont{Sousa}}, \bibnamefont{and}
  \bibinfo{editor}{\bibfnamefont{J.~N.} \bibnamefont{Urbano}}
  (\bibinfo{publisher}{World Scientific}, \bibinfo{address}{Singapore},
  \bibinfo{year}{1993}).

\bibitem[{\citenamefont{Bowler and Birse}(1995)}]{BowlerB}
\bibinfo{author}{\bibfnamefont{R.~D.} \bibnamefont{Bowler}} \bibnamefont{and}
  \bibinfo{author}{\bibfnamefont{M.~C.} \bibnamefont{Birse}},
  \bibinfo{journal}{Nucl. Phys.} \textbf{\bibinfo{volume}{A582}},
  \bibinfo{pages}{655} (\bibinfo{year}{1995}).

\bibitem[{\citenamefont{Plant and Birse}(1998)}]{PlantB}
\bibinfo{author}{\bibfnamefont{R.~S.} \bibnamefont{Plant}} \bibnamefont{and}
  \bibinfo{author}{\bibfnamefont{M.~C.} \bibnamefont{Birse}},
  \bibinfo{journal}{Nucl. Phys.} \textbf{\bibinfo{volume}{A628}},
  \bibinfo{pages}{607} (\bibinfo{year}{1998}).

\bibitem[{\citenamefont{Broniowski}(1999{\natexlab{a}})}]{coim99wb}
\bibinfo{author}{\bibfnamefont{W.}~\bibnamefont{Broniowski}}, in
  \emph{\bibinfo{booktitle}{Hadron Physics: Effective theories of low-energy
  QCD, Coimbra, Portugal, September 1999, AIP Conference Proceedings}}, edited
  by \bibinfo{editor}{\bibfnamefont{A.~H.} \bibnamefont{Blin}},
  \bibinfo{editor}{\bibfnamefont{B.}~\bibnamefont{Hiller}},
  \bibinfo{editor}{\bibfnamefont{M.~C.} \bibnamefont{Ruivo}},
  \bibinfo{editor}{\bibfnamefont{C.~A.} \bibnamefont{Sousa}}, \bibnamefont{and}
  \bibinfo{editor}{\bibfnamefont{E.}~\bibnamefont{van Beveren}}
  (\bibinfo{publisher}{AIP}, \bibinfo{address}{Melville, New York},
  \bibinfo{year}{1999}{\natexlab{a}}), vol. \bibinfo{volume}{508}, p.
  \bibinfo{pages}{380}, \bibinfo{note}{nucl-th/9910057}.

\bibitem[{\citenamefont{Broniowski}(1999{\natexlab{b}})}]{bled99}
\bibinfo{author}{\bibfnamefont{W.}~\bibnamefont{Broniowski}},
  \bibinfo{type}{Tech. Rep.} \bibinfo{number}{1828/PH},
  \bibinfo{institution}{INP Cracow} (\bibinfo{year}{1999}{\natexlab{b}}),
  \bibinfo{note}{hep-ph/9909438, talk presented at the Mini-Workshop on {\em
  Hadrons as Solitons}, Bled, Slovenia, 6-17 July 1999}.

\bibitem[{\citenamefont{Arriola and Salcedo}(1999{\natexlab{a}})}]{Arrio}
\bibinfo{author}{\bibfnamefont{E.} \bibnamefont{Ruiz Arriola}} \bibnamefont{and}
  \bibinfo{author}{\bibfnamefont{L.~L.} \bibnamefont{Salcedo}},
  \bibinfo{journal}{Phys. Lett.} \textbf{\bibinfo{volume}{B450}},
  \bibinfo{pages}{225} (\bibinfo{year}{1999}{\natexlab{a}}).

\bibitem[{\citenamefont{Arriola and Salcedo}(1999{\natexlab{b}})}]{ArrSal}
\bibinfo{author}{\bibfnamefont{E.} \bibnamefont{Ruiz Arriola}} \bibnamefont{and}
  \bibinfo{author}{\bibfnamefont{L.~L.} \bibnamefont{Salcedo}}, in
  \emph{\bibinfo{booktitle}{Hadron Physics: Effective theories of low-energy
  QCD, Coimbra, Portugal, September 1999, AIP Conference Proceedings}}, edited
  by \bibinfo{editor}{\bibfnamefont{A.~H.} \bibnamefont{Blin}},
  \bibinfo{editor}{\bibfnamefont{B.}~\bibnamefont{Hiller}},
  \bibinfo{editor}{\bibfnamefont{M.~C.} \bibnamefont{Ruivo}},
  \bibinfo{editor}{\bibfnamefont{C.~A.} \bibnamefont{Sousa}}, \bibnamefont{and}
  \bibinfo{editor}{\bibfnamefont{E.}~\bibnamefont{van Beveren}}
  (\bibinfo{publisher}{AIP}, \bibinfo{address}{Melville, New York},
  \bibinfo{year}{1999}{\natexlab{b}}), vol. \bibinfo{volume}{508},
  \bibinfo{note}{nucl-th/9910230}.

\bibitem[{\citenamefont{Plant and Birse}(2000)}]{Plant00}
\bibinfo{author}{\bibfnamefont{R.~S.} \bibnamefont{Plant}} \bibnamefont{and}
  \bibinfo{author}{\bibfnamefont{M.~C.} \bibnamefont{Birse}}
  (\bibinfo{year}{2000}), \bibinfo{note}{hep-ph/0007340}.

\bibitem[{\citenamefont{Diakonov and Petrov}(2000)}]{Diak00}
\bibinfo{author}{\bibfnamefont{D.}~\bibnamefont{Diakonov}} \bibnamefont{and}
  \bibinfo{author}{\bibfnamefont{V.~Y.} \bibnamefont{Petrov}}
  (\bibinfo{year}{2000}), \bibinfo{note}{hep-ph/0009006}.

\bibitem[{\citenamefont{Gocke et~al.}(2001)\citenamefont{Gocke, Blaschke,
  Khalatyan, and Grigorian}}]{Gocke}
\bibinfo{author}{\bibfnamefont{C.}~\bibnamefont{Gocke}},
  \bibinfo{author}{\bibfnamefont{D.}~\bibnamefont{Blaschke}},
  \bibinfo{author}{\bibfnamefont{A.}~\bibnamefont{Khalatyan}},
  \bibnamefont{and} \bibinfo{author}{\bibfnamefont{H.}~\bibnamefont{Grigorian}}
  (\bibinfo{year}{2001}), \bibinfo{note}{hep-ph/0104183}.

\bibitem[{\citenamefont{Prasza\l{}owicz and Rostworowski}(2001)}]{Prasz}
\bibinfo{author}{\bibfnamefont{M.}~\bibnamefont{Prasza\l{}owicz}}
  \bibnamefont{and}
  \bibinfo{author}{\bibfnamefont{A.}~\bibnamefont{Rostworowski}}
  (\bibinfo{year}{2001}), \bibinfo{note}{hep-ph/0105188}.

\bibitem[{\citenamefont{Dorokhov and Broniowski}(2003)}]{Dorokhov:2003kf}
\bibinfo{author}{\bibfnamefont{A.~E.} \bibnamefont{Dorokhov}} \bibnamefont{and}
  \bibinfo{author}{\bibfnamefont{W.}~\bibnamefont{Broniowski}},
  \bibinfo{journal}{Eur. Phys. J.} \textbf{\bibinfo{volume}{C32}},
  \bibinfo{pages}{79} (\bibinfo{year}{2003}).

\bibitem[{\citenamefont{Golli et~al.}(1998)\citenamefont{Golli, Broniowski, and
  Ripka}}]{Golli:1998rf}
\bibinfo{author}{\bibfnamefont{B.}~\bibnamefont{Golli}},
  \bibinfo{author}{\bibfnamefont{W.}~\bibnamefont{Broniowski}},
  \bibnamefont{and} \bibinfo{author}{\bibfnamefont{G.}~\bibnamefont{Ripka}},
  \bibinfo{journal}{Phys. Lett.} \textbf{\bibinfo{volume}{B437}},
  \bibinfo{pages}{24} (\bibinfo{year}{1998}).

\bibitem[{\citenamefont{Broniowski et~al.}(2002)\citenamefont{Broniowski,
  Golli, and Ripka}}]{Broniowski:2001cx}
\bibinfo{author}{\bibfnamefont{W.}~\bibnamefont{Broniowski}},
  \bibinfo{author}{\bibfnamefont{B.}~\bibnamefont{Golli}}, \bibnamefont{and}
  \bibinfo{author}{\bibfnamefont{G.}~\bibnamefont{Ripka}},
  \bibinfo{journal}{Nucl. Phys.} \textbf{\bibinfo{volume}{A703}},
  \bibinfo{pages}{667} (\bibinfo{year}{2002}).

\bibitem[{\citenamefont{Praszalowicz et~al.}(1995)\citenamefont{Praszalowicz,
  Blotz, and Goeke}}]{Praszalowicz:1995vi}
\bibinfo{author}{\bibfnamefont{M.}~\bibnamefont{Praszalowicz}},
  \bibinfo{author}{\bibfnamefont{A.}~\bibnamefont{Blotz}}, \bibnamefont{and}
  \bibinfo{author}{\bibfnamefont{K.}~\bibnamefont{Goeke}},
  \bibinfo{journal}{Phys. Lett.} \textbf{\bibinfo{volume}{B354}},
  \bibinfo{pages}{415} (\bibinfo{year}{1995}).

\bibitem[{\citenamefont{Broniowski and Cohen}(1986)}]{BC86}
\bibinfo{author}{\bibfnamefont{W.}~\bibnamefont{Broniowski}} \bibnamefont{and}
  \bibinfo{author}{\bibfnamefont{T.~D.} \bibnamefont{Cohen}},
  \bibinfo{journal}{{Nucl. Phys.}} \textbf{\bibinfo{volume}{A458}},
  \bibinfo{pages}{652} (\bibinfo{year}{1986}).

\bibitem[{\citenamefont{Ruiz~Arriola}(2001)}]{RuizArriola:2001rr}
\bibinfo{author}{\bibfnamefont{E.}~\bibnamefont{Ruiz~Arriola}}
  (\bibinfo{year}{2001}), talk at {\em Workshop on Lepton Scattering, Hadrons and QCD}, 
Adelaide, Australia, 26 Mar - 6 Apr 2001, \eprint{hep-ph/0107087}.

\bibitem[{\citenamefont{Ruiz~Arriola and
  Broniowski}(2003{\natexlab{a}})}]{RuizArriola:2003bs}
\bibinfo{author}{\bibfnamefont{E.}~\bibnamefont{Ruiz~Arriola}}
  \bibnamefont{and}
  \bibinfo{author}{\bibfnamefont{W.}~\bibnamefont{Broniowski}},
  \bibinfo{journal}{Phys. Rev.} \textbf{\bibinfo{volume}{D67}},
  \bibinfo{pages}{074021} (\bibinfo{year}{2003}{\natexlab{a}}).

\bibitem[{\citenamefont{Ruiz~Arriola and
  Broniowski}(2003{\natexlab{b}})}]{RuizArriola:2003wi}
\bibinfo{author}{\bibfnamefont{E.}~\bibnamefont{Ruiz~Arriola}}
  \bibnamefont{and}
  \bibinfo{author}{\bibfnamefont{W.}~\bibnamefont{Broniowski}}
  (\bibinfo{year}{2003}{\natexlab{b}}), presented at {\em Light-Cone Workshop: Hadrons and Beyond (LC 03)}, 
Durham, England, 5-9 Aug 2003, \eprint{hep-ph/0310044}.

\bibitem[{\citenamefont{Megias et~al.}(2004)\citenamefont{Megias, Ruiz~Arriola,
  Salcedo, and Broniowski}}]{Megias:2004uj}
\bibinfo{author}{\bibfnamefont{E.}~\bibnamefont{Megias}},
  \bibinfo{author}{\bibfnamefont{E.}~\bibnamefont{Ruiz~Arriola}},
  \bibinfo{author}{\bibfnamefont{L.~L.} \bibnamefont{Salcedo}},
  \bibnamefont{and}
  \bibinfo{author}{\bibfnamefont{W.}~\bibnamefont{Broniowski}},
  \bibinfo{journal}{Phys. Rev.} \textbf{\bibinfo{volume}{D70}},
  \bibinfo{pages}{034031} (\bibinfo{year}{2004}).

\bibitem[{\citenamefont{Efimov and Ivanov}(1993)}]{Efimov:1993zg}
\bibinfo{author}{\bibfnamefont{G.~V.} \bibnamefont{Efimov}} \bibnamefont{and}
  \bibinfo{author}{\bibfnamefont{M.~A.} \bibnamefont{Ivanov}},
  \emph{\bibinfo{title}{The Quark confinement model of hadrons}}
  (\bibinfo{publisher}{IOP}, \bibinfo{address}{Bristol, UK},
  \bibinfo{year}{1993}).

\bibitem[{\citenamefont{Lee and Wick}(1969)}]{Lee:1969fy}
\bibinfo{author}{\bibfnamefont{T.~D.} \bibnamefont{Lee}} \bibnamefont{and}
  \bibinfo{author}{\bibfnamefont{G.~C.} \bibnamefont{Wick}},
  \bibinfo{journal}{Nucl. Phys.} \textbf{\bibinfo{volume}{B9}},
  \bibinfo{pages}{209} (\bibinfo{year}{1969}).

\bibitem[{\citenamefont{Cutkosky et~al.}(1969)\citenamefont{Cutkosky,
  Landshoff, Olive, and Polkinghorne}}]{Cutkosky:1969fq}
\bibinfo{author}{\bibfnamefont{R.~E.} \bibnamefont{Cutkosky}},
  \bibinfo{author}{\bibfnamefont{P.~V.} \bibnamefont{Landshoff}},
  \bibinfo{author}{\bibfnamefont{D.~I.} \bibnamefont{Olive}}, \bibnamefont{and}
  \bibinfo{author}{\bibfnamefont{J.~C.} \bibnamefont{Polkinghorne}},
  \bibinfo{journal}{Nucl. Phys.} \textbf{\bibinfo{volume}{B12}},
  \bibinfo{pages}{281} (\bibinfo{year}{1969}).

\bibitem[{\citenamefont{Nakanishi}(1972)}]{Nakanishi:1972pt}
\bibinfo{author}{\bibfnamefont{N.}~\bibnamefont{Nakanishi}},
  \bibinfo{journal}{Prog. Theor. Phys. Suppl.} \textbf{\bibinfo{volume}{51}},
  \bibinfo{pages}{1} (\bibinfo{year}{1972}).

\bibitem[{\citenamefont{Gribov}(2002)}]{Gribov:book}
\bibinfo{author}{\bibfnamefont{V.~N.} \bibnamefont{Gribov}},
  \emph{\bibinfo{title}{Gauge Theories and Quark Confinement}}
  (\bibinfo{publisher}{Phasis}, \bibinfo{address}{Moscow},
  \bibinfo{year}{2002}).

\bibitem[{\citenamefont{Kleefeld}(2004)}]{Kleefeld:2004jb}
\bibinfo{author}{\bibfnamefont{F.}~\bibnamefont{Kleefeld}}
  (\bibinfo{year}{2004}), \eprint{hep-th/0408028}.

\bibitem[{\citenamefont{Arriola}(2002)}]{Ru02}
\bibinfo{author}{\bibfnamefont{E.} \bibnamefont{Ruiz Arriola}},
  \bibinfo{journal}{Acta Phys. Pol.} \textbf{\bibinfo{volume}{B33}},
  \bibinfo{pages}{4443} (\bibinfo{year}{2002}).

\bibitem[{\citenamefont{Delbourgo and West}(1977)}]{Delbourgo:1977jc}
\bibinfo{author}{\bibfnamefont{R.}~\bibnamefont{Delbourgo}} \bibnamefont{and}
  \bibinfo{author}{\bibfnamefont{P.~C.} \bibnamefont{West}},
  \bibinfo{journal}{J. Phys.} \textbf{\bibinfo{volume}{A10}},
  \bibinfo{pages}{1049} (\bibinfo{year}{1977}).

\bibitem[{\citenamefont{Delbourgo}(1979)}]{Delbourgo:1978bu}
\bibinfo{author}{\bibfnamefont{R.}~\bibnamefont{Delbourgo}},
  \bibinfo{journal}{Nuovo Cim.} \textbf{\bibinfo{volume}{A49}},
  \bibinfo{pages}{484} (\bibinfo{year}{1979}).

\bibitem{Dorokhov:2006qm}
  A.~E.~Dorokhov, W.~Broniowski and E.~Ruiz Arriola,
  Phys.\ Rev.\  D {\bf 74}, 054023 (2006).
  
\bibitem{TDA} W. Broniowski and E. Ruiz Arriola, Phys. Lett. {\bf B649}, 49, 2007.   

\bibitem[{\citenamefont{Broniowski and Ruiz~Arriola}(2003)}]{Broniowski:2003rp}
\bibinfo{author}{\bibfnamefont{W.}~\bibnamefont{Broniowski}} \bibnamefont{and}
  \bibinfo{author}{\bibfnamefont{E.}~\bibnamefont{Ruiz~Arriola}},
  \bibinfo{journal}{Phys. Lett.} \textbf{\bibinfo{volume}{B574}},
  \bibinfo{pages}{57} (\bibinfo{year}{2003}).

\bibitem[{\citenamefont{Bowman et~al.}(2002)\citenamefont{Bowman, Heller, and
  Williams}}]{Bowman:2002bm}
\bibinfo{author}{\bibfnamefont{P.~O.} \bibnamefont{Bowman}},
  \bibinfo{author}{\bibfnamefont{U.~M.} \bibnamefont{Heller}},
  \bibnamefont{and} \bibinfo{author}{\bibfnamefont{A.~G.}
  \bibnamefont{Williams}}, \bibinfo{journal}{Phys. Rev.}
  \textbf{\bibinfo{volume}{D66}}, \bibinfo{pages}{014505}
  (\bibinfo{year}{2002}).

\bibitem[{\citenamefont{Broniowski et~al.}(1996)\citenamefont{Broniowski,
  Ripka, Nikolov, and Goeke}}]{Broniowski:1995yq}
\bibinfo{author}{\bibfnamefont{W.}~\bibnamefont{Broniowski}},
  \bibinfo{author}{\bibfnamefont{G.}~\bibnamefont{Ripka}},
  \bibinfo{author}{\bibfnamefont{E.}~\bibnamefont{Nikolov}}, \bibnamefont{and}
  \bibinfo{author}{\bibfnamefont{K.}~\bibnamefont{Goeke}}, \bibinfo{journal}{Z.
  Phys.} \textbf{\bibinfo{volume}{A354}}, \bibinfo{pages}{421}
  (\bibinfo{year}{1996}).

\bibitem[{\citenamefont{Arriola and Broniowski}(2006)}]{Arriola:2006sv}
\bibinfo{author}{\bibfnamefont{E.} \bibnamefont{Ruiz Arriola}} \bibnamefont{and}
  \bibinfo{author}{\bibfnamefont{W.}~\bibnamefont{Broniowski}},
to appear in proc. {\em 4th International Conference on Quarks and Nuclear Physics (QNP06)}, Madrid, 5-10 Jun 2006, \eprint{hep-ph/0609266}.

\bibitem{RuizArriola:2006gq}
  E.~Ruiz Arriola and W.~Broniowski,
  Phys.\ Rev.\  D {\bf 73}, 097502 (2006).


\bibitem[{\citenamefont{Ecker et~al.}(1989)\citenamefont{Ecker, Gasser, Pich,
  and de~Rafael}}]{Ecker:1988te}
\bibinfo{author}{\bibfnamefont{G.}~\bibnamefont{Ecker}},
  \bibinfo{author}{\bibfnamefont{J.}~\bibnamefont{Gasser}},
  \bibinfo{author}{\bibfnamefont{A.}~\bibnamefont{Pich}}, \bibnamefont{and}
  \bibinfo{author}{\bibfnamefont{E.}~\bibnamefont{de~Rafael}},
  \bibinfo{journal}{Nucl. Phys.} \textbf{\bibinfo{volume}{B321}},
  \bibinfo{pages}{311} (\bibinfo{year}{1989}).

\bibitem[{\citenamefont{Pich}(2002)}]{Pich:2002xy}
\bibinfo{author}{\bibfnamefont{A.}~\bibnamefont{Pich}} (\bibinfo{year}{2002}),
proc. {\em The Phenomenology of Large $N_c$ QCD}, Tempe, Arizona, 9-11 Jan 2002, ed. R.~Lebed, World Scientific, Singapore, 2002, 
 \eprint{hep-ph/0205030}.

\bibitem[{\citenamefont{Salcedo and Ruiz~Arriola}(1996)}]{Salcedo:1994qy}
\bibinfo{author}{\bibfnamefont{L.~L.} \bibnamefont{Salcedo}} \bibnamefont{and}
  \bibinfo{author}{\bibfnamefont{E.}~\bibnamefont{Ruiz~Arriola}},
  \bibinfo{journal}{Ann. Phys.} \textbf{\bibinfo{volume}{250}},
  \bibinfo{pages}{1} (\bibinfo{year}{1996}).

\bibitem[{\citenamefont{Diakonov et~al.}(1989)\citenamefont{Diakonov, Petrov,
  and Praszalowicz}}]{Diakonov:1988mg}
\bibinfo{author}{\bibfnamefont{D.}~\bibnamefont{Diakonov}},
  \bibinfo{author}{\bibfnamefont{V.~Y.} \bibnamefont{Petrov}},
  \bibnamefont{and}
  \bibinfo{author}{\bibfnamefont{M.}~\bibnamefont{Praszalowicz}},
  \bibinfo{journal}{Nucl. Phys.} \textbf{\bibinfo{volume}{B323}},
  \bibinfo{pages}{53} (\bibinfo{year}{1989}).

\bibitem[{\citenamefont{Adjali et~al.}(1992)\citenamefont{Adjali, Aitchison,
  and Zuk}}]{Adjali:1991bi}
\bibinfo{author}{\bibfnamefont{I.}~\bibnamefont{Adjali}},
  \bibinfo{author}{\bibfnamefont{I.~J.~R.} \bibnamefont{Aitchison}},
  \bibnamefont{and} \bibinfo{author}{\bibfnamefont{J.~A.} \bibnamefont{Zuk}},
  \bibinfo{journal}{Nucl. Phys.} \textbf{\bibinfo{volume}{A537}},
  \bibinfo{pages}{457} (\bibinfo{year}{1992}).

\bibitem[{\citenamefont{Ripka and Kahana}(1987)}]{Ripka:1987ne}
\bibinfo{author}{\bibfnamefont{G.}~\bibnamefont{Ripka}} \bibnamefont{and}
  \bibinfo{author}{\bibfnamefont{S.}~\bibnamefont{Kahana}},
  \bibinfo{journal}{Phys. Rev.} \textbf{\bibinfo{volume}{D36}},
  \bibinfo{pages}{1233} (\bibinfo{year}{1987}).

\bibitem[{\citenamefont{Soni}(1987)}]{Soni:1987zj}
\bibinfo{author}{\bibfnamefont{V.}~\bibnamefont{Soni}}, \bibinfo{journal}{Phys.
  Lett.} \textbf{\bibinfo{volume}{B195}}, \bibinfo{pages}{569}
  (\bibinfo{year}{1987}).

\bibitem[{\citenamefont{Cohen and Broniowski}(1986)}]{Cohen:1986va}
\bibinfo{author}{\bibfnamefont{T.~D.} \bibnamefont{Cohen}} \bibnamefont{and}
  \bibinfo{author}{\bibfnamefont{W.}~\bibnamefont{Broniowski}},
  \bibinfo{journal}{Phys. Rev.} \textbf{\bibinfo{volume}{D34}},
  \bibinfo{pages}{3472} (\bibinfo{year}{1986}).

\bibitem[{\citenamefont{Golli and Rosina}(1985)}]{Golli:1986er}
\bibinfo{author}{\bibfnamefont{B.}~\bibnamefont{Golli}} \bibnamefont{and}
  \bibinfo{author}{\bibfnamefont{M.}~\bibnamefont{Rosina}},
  \bibinfo{journal}{Phys. Lett.} \textbf{\bibinfo{volume}{B165}},
  \bibinfo{pages}{347} (\bibinfo{year}{1985}).

\bibitem[{\citenamefont{Birse}(1986)}]{Birse:1986qc}
\bibinfo{author}{\bibfnamefont{M.~C.} \bibnamefont{Birse}},
  \bibinfo{journal}{Phys. Rev.} \textbf{\bibinfo{volume}{D33}},
  \bibinfo{pages}{1934} (\bibinfo{year}{1986}).

\bibitem[{\citenamefont{Gasser et~al.}(1991)\citenamefont{Gasser, Leutwyler,
  and Sainio}}]{Gasser:1990ce}
\bibinfo{author}{\bibfnamefont{J.}~\bibnamefont{Gasser}},
  \bibinfo{author}{\bibfnamefont{H.}~\bibnamefont{Leutwyler}},
  \bibnamefont{and} \bibinfo{author}{\bibfnamefont{M.~E.}
  \bibnamefont{Sainio}}, \bibinfo{journal}{Phys. Lett.}
  \textbf{\bibinfo{volume}{B253}}, \bibinfo{pages}{252} (\bibinfo{year}{1991}).

\bibitem[{\citenamefont{Grama et~al.}(2000)\citenamefont{Grama, Grama, and
  Zamfirescu}}]{Gramma}
\bibinfo{author}{\bibfnamefont{C.}~\bibnamefont{Grama}},
  \bibinfo{author}{\bibfnamefont{N.}~\bibnamefont{Grama}}, \bibnamefont{and}
  \bibinfo{author}{\bibfnamefont{I.}~\bibnamefont{Zamfirescu}},
  \bibinfo{journal}{Phys. Rev.} \textbf{\bibinfo{volume}{A61}},
  \bibinfo{pages}{032716} (\bibinfo{year}{2000}).

\bibitem[{\citenamefont{Kok and van Haeringen}(1981)}]{Kok:1980fw}
\bibinfo{author}{\bibfnamefont{L.~P.} \bibnamefont{Kok}} \bibnamefont{and}
  \bibinfo{author}{\bibfnamefont{H.}~\bibnamefont{van Haeringen}},
  \bibinfo{journal}{Ann. Phys.} \textbf{\bibinfo{volume}{131}},
  \bibinfo{pages}{426} (\bibinfo{year}{1981}).

\bibitem[{\citenamefont{Badalian et~al.}(1982)\citenamefont{Badalian, Kok,
  Polikarpov, and Simonov}}]{Badalian:1981xj}
\bibinfo{author}{\bibfnamefont{A.~M.} \bibnamefont{Badalian}},
  \bibinfo{author}{\bibfnamefont{L.~P.} \bibnamefont{Kok}},
  \bibinfo{author}{\bibfnamefont{M.~I.} \bibnamefont{Polikarpov}},
  \bibnamefont{and} \bibinfo{author}{\bibfnamefont{Y.~A.}
  \bibnamefont{Simonov}}, \bibinfo{journal}{Phys. Rept.}
  \textbf{\bibinfo{volume}{82}}, \bibinfo{pages}{31} (\bibinfo{year}{1982}).

\bibitem[{\citenamefont{Zhao and Hiller}(1989)}]{Zhao:1988in}
\bibinfo{author}{\bibfnamefont{M.-s.} \bibnamefont{Zhao}} \bibnamefont{and}
  \bibinfo{author}{\bibfnamefont{J.~R.} \bibnamefont{Hiller}},
  \bibinfo{journal}{Phys. Rev.} \textbf{\bibinfo{volume}{D40}},
  \bibinfo{pages}{1329} (\bibinfo{year}{1989}).

\bibitem[{\citenamefont{MacKenzie and Wilczek}(1984)}]{MacKenzie:1984mz}
\bibinfo{author}{\bibfnamefont{R.}~\bibnamefont{MacKenzie}} \bibnamefont{and}
  \bibinfo{author}{\bibfnamefont{F.}~\bibnamefont{Wilczek}},
  \bibinfo{journal}{Phys. Rev.} \textbf{\bibinfo{volume}{D30}},
  \bibinfo{pages}{2194} (\bibinfo{year}{1984}).

\bibitem[{\citenamefont{Megias et~al.}(2006)\citenamefont{Megias, Ruiz~Arriola,
  and Salcedo}}]{Megias:2004hj}
\bibinfo{author}{\bibfnamefont{E.}~\bibnamefont{Megias}},
  \bibinfo{author}{\bibfnamefont{E.}~\bibnamefont{Ruiz~Arriola}},
  \bibnamefont{and} \bibinfo{author}{\bibfnamefont{L.~L.}
  \bibnamefont{Salcedo}}, \bibinfo{journal}{Phys. Rev.}
  \textbf{\bibinfo{volume}{D74}}, \bibinfo{pages}{065005}
  (\bibinfo{year}{2006}).

\bibitem[{\citenamefont{Sieber et~al.}(1992)\citenamefont{Sieber, Meissner,
  Gruemmer, and Goke}}]{Sieber:1992uj}
\bibinfo{author}{\bibfnamefont{P.}~\bibnamefont{Sieber}},
  \bibinfo{author}{\bibfnamefont{T.}~\bibnamefont{Meissner}},
  \bibinfo{author}{\bibfnamefont{F.}~\bibnamefont{Gruemmer}}, \bibnamefont{and}
  \bibinfo{author}{\bibfnamefont{K.}~\bibnamefont{Goke}},
  \bibinfo{journal}{Nucl. Phys.} \textbf{\bibinfo{volume}{A547}},
  \bibinfo{pages}{459} (\bibinfo{year}{1992}).

\end{thebibliography}

\end{document}